\documentclass[twocolumn]{aastex631}
\usepackage{booktabs}
\usepackage{graphicx}
\usepackage{enumitem}
\usepackage{textgreek}
\usepackage{amsmath}
\usepackage{comment}
\usepackage{tabularx}
\usepackage{multirow}
\definecolor{yscol}{rgb}{0.8, 0.6, 1}
\usepackage{xcolor}

\usepackage{verbatim}
\usepackage{mathptmx}
\usepackage{url}
\usepackage{numprint}
\usepackage{ulem}

\usepackage{afterpage}
\usepackage{tikz}

\usepackage{blindtext}
\usepackage{graphicx}
\usepackage{everypage}
\usepackage{environ}

\begin{document}

\title{The \textit{AGORA} High-resolution Galaxy Simulations Comparison Project. X: Formation and Evolution of Galaxies at the High-redshift Frontier
}

\correspondingauthor{Hyeonyong Kim, Ji-hoon Kim, Minyong Jung, Santi Roca-Fàbrega}

\author[0000-0002-7820-2281]{Hyeonyong Kim}
\altaffiliation{Code leaders}
\affiliation{Center for Theoretical Physics, Department of Physics and Astronomy, Seoul National University, Seoul 08826, Republic of Korea;
\rm{\href{mailto:gusdyd398@snu.ac.kr}{gusdyd398@snu.ac.kr}}}

\author[0000-0003-4464-1160]{Ji-hoon Kim}

\affiliation{Center for Theoretical Physics, Department of Physics and Astronomy, Seoul National University, Seoul 08826, Republic of Korea;
\rm{\href{mailto:mornkr@snu.ac.kr}{mornkr@snu.ac.kr}}}
\affiliation{Institute for Data Innovation in Science, Seoul National University, Seoul 08826, Republic of Korea}
\affiliation{Seoul National University Astronomy Research Center, Seoul 08826, Republic of Korea}
\author[0000-0002-9144-1383]{Minyong Jung}
\affiliation{Center for Theoretical Physics, Department of Physics and Astronomy, Seoul National University, Seoul 08826, Republic of Korea;
\rm{\href{mailto:wispedia@snu.ac.kr}{wispedia@snu.ac.kr}}}
\author[0000-0002-6299-152X]{Santi Roca-Fàbrega}
\altaffiliation{Code leaders}
\affiliation{Lund Observatory, Division of Astrophysics, Department of Physics, Lund University, Box 118, SE-221 00 Lund, Sweden;
\rm{\href{matiltoo:santi.roca_fabrega@fysik.lu.se}{santi.roca\_fabrega@fysik.lu.se}}}
\affiliation{Departamento de Física de la Tierra y Astrofísica, Facultad de Ciencias Físicas, Plaza Ciencias, 1, 28040 Madrid, Spain}

\author[0000-0002-8680-248X]{Daniel Ceverino}
\affiliation{Universidad Autónoma de Madrid, Ciudad Universitaria de Cantoblanco, E-28049 Madrid, Spain}
\affiliation{CIAFF, Facultad de Ciencias, Universidad Autónoma de Madrid, E-28049 Madrid, Spain}

\author[0009-0002-1398-6537]{Pablo Granizo}
\altaffiliation{Code leaders}
\affiliation{Universidad Autónoma de Madrid, Ciudad Universitaria de Cantoblanco, E-28049 Madrid, Spain}
\affiliation{Theoretical Astrophysics, Department of Earth and Space Science, Graduate School of Science, Osaka University, Toyonaka, Osaka, 560-0043, Japan}

\author[0000-0001-7457-8487]{Kentaro Nagamine}
\affiliation{Theoretical Astrophysics, Department of Earth and Space Science, Graduate School of Science, Osaka University, Toyonaka, Osaka, 560-0043, Japan}
\affiliation{Theoretical Joint Research, Forefront Research Center, Graduate School of Science, Osaka University, Toyonaka, Osaka 560-0043, Japan}
\affiliation{Kavli IPMU (WPI), University of Tokyo, 5-1-5 Kashiwanoha, Kashiwa, Chiba, 277-8583, Japan}
\affiliation{Department of Physics \& Astronomy, University of Nevada Las Vegas, Las Vegas, NV 89154, USA}
\affiliation{Nevada Center for Astrophysics, University of Nevada, Las Vegas, 4505 S. Maryland Pkwy, Las Vegas, NV 89154-4002, USA}

\author[0000-0001-5091-5098]{Joel R. Primack}
\affiliation{Department of Physics, University of California at Santa Cruz, Santa Cruz, CA 95064, USA}

\author{H\'{e}ctor Vel\'{a}zquez}
\altaffiliation{Code leaders}
\affil{Instituto de Astronom\'{i}a, Universidad Nacional Aut\'{o}noma de M\'{e}xico, A.P. 70-264, 04510, Mexico, D.F., Mexico}

\author[0000-0002-8638-1697]{Kirk S. S. Barrow}
\affiliation{Department of Astronomy, University of Illinois at Urbana-Champaign, Urbana, IL 61801, USA}

\author[0000-0002-1109-1919]{Robert Feldmann}
\affiliation{Department of Astrophysics, University of Zurich, Zurich CH-8057, Switzerland}
\author[0000-0002-5045-6052]{Keita Fukushima}
\affiliation{Institute for Data Innovation in Science, Seoul National University, Seoul 08826, Republic of Korea}

\author[0000-0002-7078-2074]{Lucio Mayer}
\affil{Department of Astrophysics, University of Zurich, Winterthurerstrasse 190, CH-8057 Zürich, Switzerland}

\author[0000-0003-4597-6739]{Boon Kiat Oh}
\affiliation{Department of Physics, University of Connecticut, U-3046, Storrs, CT 06269, USA}
\affiliation{School of Physics, Korea Institute for Advanced Study, 85 Hoegiro, Dongdaemun-gu, Seoul 02455, Republic of Korea}
\author[0000-0002-3764-2395]{Johnny W. Powell}
\affiliation{Department of Physics, Reed College, Portland, OR 97202, USA}

\author[0000-0002-5969-1251]{Tom Abel}
\affiliation{Kavli Institute for Particle Astrophysics and Cosmology, Stanford University, Stanford, CA 94305, USA}
\affiliation{Department of Physics, Stanford University, Stanford, CA 94305, USA}
\affiliation{SLAC National Accelerator Laboratory, Menlo Park, CA 94025, USA}

\author[0000-0002-4287-1088]{Oscar Agertz}
\affiliation{Lund Observatory, Division of Astrophysics, Department of Physics, Lund University, Box 118, SE-221 00 Lund, Sweden}

\author[0009-0009-6888-7967]{Chaerin Jeong}
\affiliation{Department of Astronomy \& Space Science, Kyung Hee University, 1732 Deogyeong-daero, Yongin-si, Gyeonggi-do 17104, Republic of Korea}

\author[0000-0001-6106-7821]{Alessandro Lupi}
\affiliation{Como Lake Center for Astrophysiscs, DiSAT, Universit\`a degli Studi dell'Insubria, via Valleggio 11, 22100, Como, Italy}
\affiliation{INAF, Osservatorio Astronomico di Bologna, Via Gobetti 93/3, I-40129 Bologna, Italy}

\author[0000-0002-5712-6865]{Yuri Oku}
\affil{Center for Cosmology and Computational Astrophysics, Institute for Advanced Study in Physics, Zhejiang University, Hangzhou 310027, People's Republic of China}

\author[0000-0001-5510-2803]{Thomas R. Quinn}
\affil{Department of Astronomy, University of Washington, Seattle, WA 98195, USA}

\author[0000-0002-6227-0108]{Yves Revaz}
\affiliation{Institute of Physics, Laboratoire d'Astrophysique, \'{E}cole Polytechnique F\'{e}d\'{e}rale de Lausanne (EPFL), CH-1015 Lausanne, Switzerland}

\author[0000-0002-9158-195X]{Ram\'{o}n Rodr\'{i}guez-Cardoso}
\affil{Departamento de Física de la Tierra y Astrofísica, Fac. de C.C. Físicas, Universidad Complutense de Madrid, E-28040 Madrid, Spain}
\affil{GMV, Space and Avionics Equipment, Isaac Newton, 11 Tres Cantos, E-28760 Madrid, Spain}
\affil{Instituto de Física de Partículas y del Cosmos, IPARCOS, Fac. C.C. Físicas, Universidad Complutense de Madrid, E-28040 Madrid, Spain}

\author{Ikkoh Shimizu}
\affiliation{Shikoku Gakuin University, 3-2-1 Bunkyocho, Zentsuji, Kagawa, 765-8505, Japan}

\author[0000-0001-7689-0933]{Romain Teyssier}
\affiliation{Department of Astrophysical Sciences, Princeton University, 4 Ivy Lane, Princeton, NJ 08540, USA}

\author{ \textit{AGORA} COLLABORATION}
\affiliation{\rm  \href{http://www.AGORAsimulations.org}{http://www.AGORAsimulations.org}}
\vspace{4mm}



\begin{abstract}

Recent observations from JWST have revealed unexpectedly luminous galaxies, exhibiting stellar masses and luminosities significantly higher than predicted by theoretical models at Cosmic Dawn. In this study, we present a suite of cosmological zoom-in simulations targeting high-redshift ($z \geq 10$) galaxies with dark matter halo masses in the range $10^{10} - 10^{11}\ {\rm M}_{\odot}$ at  $z=10$, using state-of-the-art galaxy formation simulation codes ({\sc Enzo}, {\sc Ramses}, {\sc Changa}, {\sc Gadget-3}, {\sc Gadget-4}, and {\sc Gizmo}). This study aims to evaluate the convergence of the participating codes and their reproducibility of high-redshift galaxies with the galaxy formation model calibrated at relatively low redshift, without additional physics for high-redshift environments. The subgrid physics follows the \textit{AGORA} \textit{CosmoRun} framework, with adjustments to resolution and initial conditions to emulate similar physical environments in the early universe. The participating codes show consistent results for key galaxy properties (e.g., stellar mass), but also reveal notable differences (e.g., metallicity), indicating that galaxy properties at high redshifts are highly sensitive to the feedback implementation of the simulation. Massive halos (${\rm M}_{\rm halo}\geq5\times10^{10}\,{\rm M}_{\odot}$ at $z=10$) succeed in reproducing observed stellar masses, metallicities, and UV luminosities at $10\leq z\leq12$ without requiring additional subgrid physics, but tend to underpredict those properties at higher redshift. We also find that varying the dust-to-metal ratio modestly affects UV luminosity of simulated galaxies, whereas the absence of dust significantly enhances it. In future work, higher-resolution simulations will be conducted to better understand the formation and evolution of galaxies at Cosmic Dawn.

\end{abstract}

\keywords{galaxies: high-redshift --- galaxies: formation --- galaxies: evolution --- method: numerical --- hydrodynamics}


\section{Introduction} \label{sec:intro}

Emerging with the first data from the James Webb Space Telescope (JWST), the Cosmic Dawn was unveiled, shedding light on the faintest galaxies ever observed. Although high-redshift objects around $z=8 \ \textendash\ 10$ were detected by the Hubble Space Telescope (HST) \citep[e.g.,][]{2013ApJ...763L...7E, 2016ApJ...819..129O, 2018MNRAS.479.1180M, 2022ApJ...928...52F}, galaxies beyond this redshift were too faint to be observed for HST. Since the launch of the JWST, however, numerous studies using the JWST NIRCam data have reported high-redshift galaxy candidates beyond $z=10$ \citep[e.g.,][]{2022ApJ...938L..15C, 2023MNRAS.518.4755A, 2023ApJ...954L..46L,2024ApJ...969L...2F, 2024Natur.633..318C, 2024ApJ...970...31R, 2025ApJ...992...63W,2025ApJ...991..179P}.

Although one of the ultra-high-redshift galaxy candidates suggested in early JWST studies \citep[e.g.,][]{2023MNRAS.518.6011D,2023MNRAS.523.1036B, 2023ApJ...946L..13F} turned out to be a low-redshift interloper \citep{2023Natur.622..707A, 2024ApJ...960...56H}, many other candidates have been confirmed through spectroscopic analyses \citep[e.g.,][]{2023A&A...677A..88B, 2023ApJ...952...74T,2023ApJ...957L..34W,2025ApJ...988...19S,2024ApJ...960...56H,2025ApJ...980..138H}. Observational efforts to detect even more distant galaxies continue to progress rapidly. Recent spectroscopic measurements have confirmed galaxies at redshifts of up to $z=14$ \citep[e.g.,][]{2024Natur.633..318C, 2024ApJ...970...31R, 2025ApJ...992...63W} and photometric surveys have identified galaxy candidates extending to $z = 15\ \textendash{}\ 25$ \citep[e.g.,][]{2025ApJ...991..179P, 2025A&A...704A.158C}.

JWST observations have also revealed that these high-z galaxies exhibit exceptionally high UV luminosities, creating a strong tension in standard galaxy formation theories within the $\Lambda$CDM cosmological framework. Initially, \cite{2023Natur.616..266L} reported six massive galaxy candidates with stellar masses greater than $10^{10}\,{\rm M}_{\odot}$ at $z=7.4\ \textendash{}\ 9.1$. These candidates were several orders of magnitude brighter and more massive than the $\Lambda$CDM cosmology prediction, implying baryon-to-stellar conversion ratios approaching unity \citep{2023MNRAS.518.2511L, 2023NatAs...7..731B}. This strong tension was partially resolved after the revision of the paper \citep{2023NatAs...7..731B} with calibration of the NIRCam and updated stellar masses. Still, the early universe unveiled by JWST seems to have more luminous galaxies than expected by $\Lambda$CDM cosmology.

To alleviate this tension, various explanations have been proposed. Some studies suggest a top-heavy initial mass function (IMF) characterized by a larger fraction of massive stars can enhance the UV luminosity of high-z galaxies \citep[e.g.,][]{2022ApJ...938L..10I, 2023ApJ...946L..13F, 2024MNRAS.527.5929Y}. This fraction could be achieved by the environments of the early universe \textemdash{} high cosmic microwave background (CMB) temperature, which can suppress fragmentation and facilitate the collapse of giant molecular clouds \citep{2022MNRAS.514.4639C}. Active galactic nuclei (AGN) are also considered as a possible source of UV photons. Several studies, however, suggest that AGNs are not dominant contributors to UV luminosity \citep[e.g.,][]{2023ApJ...951...72O, 2024MNRAS.529.3563T}, and their contribution to UV luminosity is still in debate.

Alternatively, \cite{2023MNRAS.523.3201D} and \cite{2024A&A...690A.108L} propose the feedback-free starburst (FFB) model. According to this model, the free-fall time of dense gas clouds can be shorter than the timescale for stellar wind or supernova feedback effects. Although this condition holds in dense gas environments in very massive halos, this scenario leads to significantly increased star formation efficiencies in early galaxies. Indeed, the FirstLight simulations \citep{2024A&A...689A.244C} have shown that the galaxy-averaged star formation efficiency (SFE) increases at these high redshifts due to higher ISM densities. The so-called attenuation-free model (AFM) has also been proposed to reconcile the tension \citep[e.g.,][]{2023MNRAS.526.4801T, 2025A&A...694A.286F}. In this model, the radiation outflows from bursty star formation push dust outward, resulting in significantly reduced attenuation, which could also enhance the overabundance of bright galaxies at high redshift.

While a variety of theoretical models to explain high-redshift galaxies are actively proposed and debated, recent numerical simulations suggest the tension may be less severe than initially thought. For example, \cite{2023OJAp....6E..47M} argues that the most massive galaxies in Renaissance simulations are consistent with the JWST survey data in stellar mass and star formation rate (SFR). Also, results from the FIRE-2 and $\rm FIREbox^{\it HR}$ simulations \citep[e.g.,][]{2023MNRAS.526.2665S, 2025MNRAS.536..988F} demonstrate that the observed abundance of UV-bright galaxies and the evolution of the UV luminosity density can natural ly be explained by the standard galaxy formation model without any additional subgrid physics.

In this paper, we introduce the \textit{AGORA High-z Run}, a suite of zoom-in simulations during Cosmic Dawn based on the \textit{AGORA CosmoRun} model. The \textit{AGORA} simulation comparison project began with the objective of enhancing the reliability and predictive power of numerical galaxy formation simulations \citep[][hereafter Papers I and II]{2014ApJS..210...14K, 2016ApJ...833..202K}. By comparing multiple state-of-the-art simulations from identical initial conditions and common subgrid physics, the project aims to establish robust and reliable astrophysical models independent of specific simulation codes. Starting from Papers III and IV \citep{2021ApJ...917...64R, 2024ApJ...968..125R}, the \textit{AGORA} project initiated the \textit{CosmoRun} series. \textit{CosmoRun} is a suite of simulations modeling the formation and evolution of Milky Way-size galaxies (halo mass of $10^{12} M_\odot$ at $z=0$) across multiple numerical codes. This series comprehensively examined properties of Milky Way-mass galaxy, satellite galaxy populations, circumgalactic medium (CGM) characteristics, satellite quenching mechanisms, and disk formation of Milky Way-mass galaxy across multiple numerical platforms \citep[][hereafter Papers V-VIII]{2024ApJ...964..123J,2024ApJ...962...29S,2025A&A...698A.303R,2025ApJ...994..245J}. The \textit{CosmoRun} simulation has not only demonstrated good agreement with observations in key galaxy properties such as stellar mass and the mass–metallicity relation (MZR), but also revealed similarities and differences in different gravity and hydrodynamics schemes. Now, the \textit{AGORA High-z Run} suite extends the \textit{AGORA} framework to the early universe, focusing on zoom-in simulations of overdense regions that host the earliest, most massive galaxies. By applying the galaxy formation model calibrated in \textit{CosmoRun} to these environments, the \textit{High-z Run} allows us to assess its ability to reproduce luminous high-redshift galaxies observed by JWST, while offering new insights into inter-code agreements and differences at Cosmic Dawn.

This paper is structured as follows. Section 2 provides an
overview of the \textit{CosmoRun} and details of the \textit{High-z Run}, including initial conditions, refinement strategies, analysis and mock observation methodologies, and a brief description of participating simulation codes. In Section 3, we analyze the properties and evolution of high-z galaxies from $z = 15$ to $z = 10$, focusing on gas properties, stellar mass, and metallicity of target halos, and compare these results with JWST spectroscopy data. We also present mock observations of the \textit{High-z Run} and examine the UV brightness of simulated galaxies. Section 4 discusses the star formation rate, efficiency and the role of dust attenuation in the \textit{High-z Run}. Finally, we summarize our findings and provide conclusions in Section 5.

\begin{table}[]
\centering
\vspace{2mm}
\begin{tabular}{ccccccc}
\hline
\multirow{2}{*}{Halo} & \multicolumn{6}{c}{Halo mass $(10^{9}\,{\rm M}_{\odot})$} \\\cline{2-7}
& $z=10$ & $z=11$ & $z=12$ & $z=13$ & $z=14$ &$z=15$ \\\hline\hline
{\sc Halo 1}  &  $15.1$ & $8.89$ & $5.79$ & $3.32$ & $1.83$ & $0.82$ \\
{\sc Halo 2}  &  $15.3$ & $7.76$ & $5.03$ & $3.31$ & $1.71$ & $0.87$ \\
{\sc Halo 3}  &  $46.8$ & $28.6$ & $5.02$ & $3.05$ & $2.09$ & $1.06$ \\
{\sc Halo 4}  &  $51.6$ & $33.4$ & $18.6$ & $8.78$ & $4.88$ & $1.79$ \\
{\sc Halo 5}  &  $81.2$ & $47.7$ & $25.3$ & $15.9$ & $5.96$ & $1.78$ \\\hline
\end{tabular}
\vspace{3mm}
\caption{The mass history of halos selected as the initial conditions. The values are the mean halo mass of six numerical codes.}
\vspace{-2mm}
\label{tab:halo_mass}
\end{table}

\begin{figure*}
    \vspace{0mm}
    \centering
    \includegraphics[width = 1\linewidth]{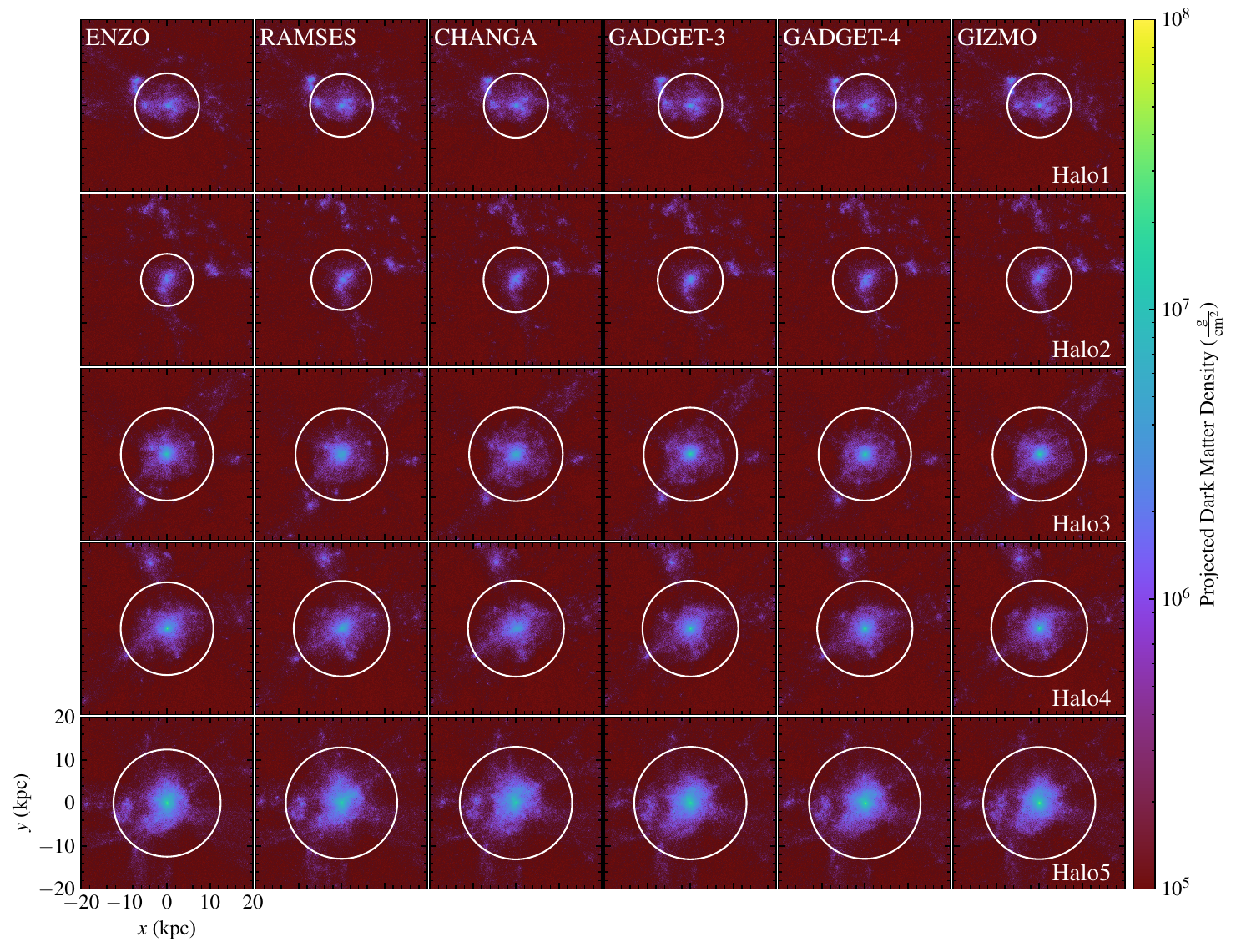}
    \vspace{0mm}
    \caption{The dark matter surface densities at $z=10$ for five massive halos across six different hydrodynamical simulations, ``\textit{High-z Run}''. The projections are taken through a 40 kpc slice and the target halo's virial radius $R_{\rm vir}$ identified by the ROCKSTAR halo finder \citep{2013ApJ...762..109B} is denoted with a white circle. The simulations are conducted by Santi Roca-Fàbrega (\textsc{Ramses}), Hyeonyong Kim (\textsc{Enzo}, \textsc{Gadget-3}, \textsc{Gizmo}), Héctor Velázquez (\textsc{Changa}), and Pablo Cuadrado (\textsc{Gadget-4}). 
    See Section \ref{sec:simulation} for detailed information including subgrid physics,  initial condition, and refinement schemes of these simulations.}
    \label{fig:darkmatter_plot}
\end{figure*}

\section{Simulation} \label{sec:simulation}
\subsection{The \textit{AGORA} “High-z Run” Simulation Suite}
The \textit{High-z Run} adopts the subgrid physics framework developed for the \textit{AGORA CosmoRun} simulations (Papers III and IV), which model the formation and evolution of Milky Way–mass galaxy's halo ($\sim10^{12} M_\odot$ at $z=0$) across multiple numerical codes. The suite involves various numerical methods, including adaptive mesh refinement (AMR), smoothed particle hydrodynamics (SPH), and hybrid techniques. Participating AMR codes include 
\textsc{Enzo} \citep{2014ApJS..211...19B} and \textsc{Ramses} \citep{2002A&A...385..337T}. SPH codes are represented by \textsc{Changa} \citep{2015ComAC...2....1M}, \textsc{Gadget-3} \citep{2005MNRAS.364.1105S}, \textsc{Gadget-4} \citep{2021MNRAS.506.2871S} and \textsc{Gear} \citep{2012A&A...538A..82R}. \textsc{Arepo} \citep{2010MNRAS.401..791S} and \textsc{Gizmo} \citep{2015MNRAS.450...53H} are classified as hybrid methods. We will refer to AMR codes as mesh-based codes and SPH and hybrid methods as particle-based codes for convenience.

While the gravity and hydrodynamics solvers differ across the participating codes, all simulations employ the {\sc Grackle} chemistry and cooling library \citep{2017MNRAS.466.2217S}. Based on the gas metallicity, cooling and heating rates are computed from the chemistry tables generated with the photoionization code {\sc Cloudy} \citep{2013RMxAA..49..137F} and a redshift-dependent cosmic UV background \citep{2012ApJ...746..125H} is applied starting at $z = 15$. Star formation occurs when the gas density exceeds the number density $n_{\rm H} = 1\mathrm{cm}^{-3}$. The local star formation rate follows the Schmidt law, $d\rho_{\star}/{dt} = \epsilon_\star{\rho_{\text{gas}}}/{t_{\text{ff}}}$, where $\epsilon_\star = 0.01$ is the formation efficiency and local free fall time, $t_{\text{ff}} = ({{3\pi}/{32G\rho_{\text{gas}}}})^{1/2}$. A metallicity floor of $10^{-4}\ Z_{\odot}$\footnote{ $Z_{\odot}$ = 0.02041 is used across all participating codes, following Paper II (see Section 2 of Paper II for details).} is applied to consider the metal enrichment by first stars that cannot be resolved. At the same time, each code follows its own implementation of stellar feedback, feedback energy, and metal diffusion scheme, reflecting the most widely used practice in the community. For more detailed information about the feedback description, we refer readers to Section 3.1 of Paper III and Section 4 of Paper IV.

\vspace{+4mm}

\subsection{Initial Conditions and Refinement Scheme} \label{sec:style}
We use {\sc Music}, a multi-scale initial condition generator code \citep{2011MNRAS.415.2101H}, to generate the initial conditions for the \textit{High-z Run}. We adopt cosmological parameters WMAP7/9+SNe+BAO: $\Omega_{\Lambda} = 0.728$, $\Omega_{\rm matter} = 0.272$, $\Omega_{\rm DM} = 0.227$, $\sigma_{8} = 0.807$, $n_{\rm s} = 0.961$, and $h = 0.702$. Since this paper does not study the population of high-redshift galaxies, we utilize the cosmological parameters from \textit{CosmoRun} for consistency, but plan to use updated cosmological parameters in future high-redshift simulation work. Starting from $z = 100$, we select five halos with $M_{\rm halo} > 10^{10}{\ \rm M}_{\odot}$ at $z = 10$, representative of observed high-redshift galaxies. Especially, the most massive target, Halo 5 ($8 \times 10^{10}{\ \rm M}_{\odot}$) has a number density of $\sim 2.5 \times 10^{-6} \ \rm cMpc^{-3}$. We estimate the number density by integrating the halo mass function from  \cite{2013MNRAS.433.1230W}. Such an object would be extremely rare both in the simulation box and in observations, given that typical effective volumes of JWST surveys are around $10^5$ to $10^6 \ \rm cMpc^3$ \citep[e.g.,][]{2024ApJ...969L...2F}.
Table \ref{tab:halo_mass} summarizes the average halo masses at selected epochs. Based on a $256^3$ ($512^3$ for Halo 5) root resolution in a 64 comoving $(\mathrm{Mpc}/h)^3$ box, we set an additional 4 (3 for Halo 5) levels to generate a zoom-in region of a $4096^3$ effective resolution. The most refined regions are set to enclose all the particles within $4R_{\rm vir}$ of the target halos at $z = 10$, and the masses of particles residing in the most refined region are $m_{\rm DM} = 3.41 \times 10^{5}{\ \rm M}_{\odot}$ and for particle-based codes, $m_{\rm gas, ic} = 6.86 \times 10^{4}{\ \rm M}_{\odot}$. Dark matter distributions of these runs at the final redshift ($z=10$) are shown in Figure \ref{fig:darkmatter_plot}.

\begin{figure*}
    \vspace{0mm}
    \centering
    \includegraphics[width = 1\linewidth]{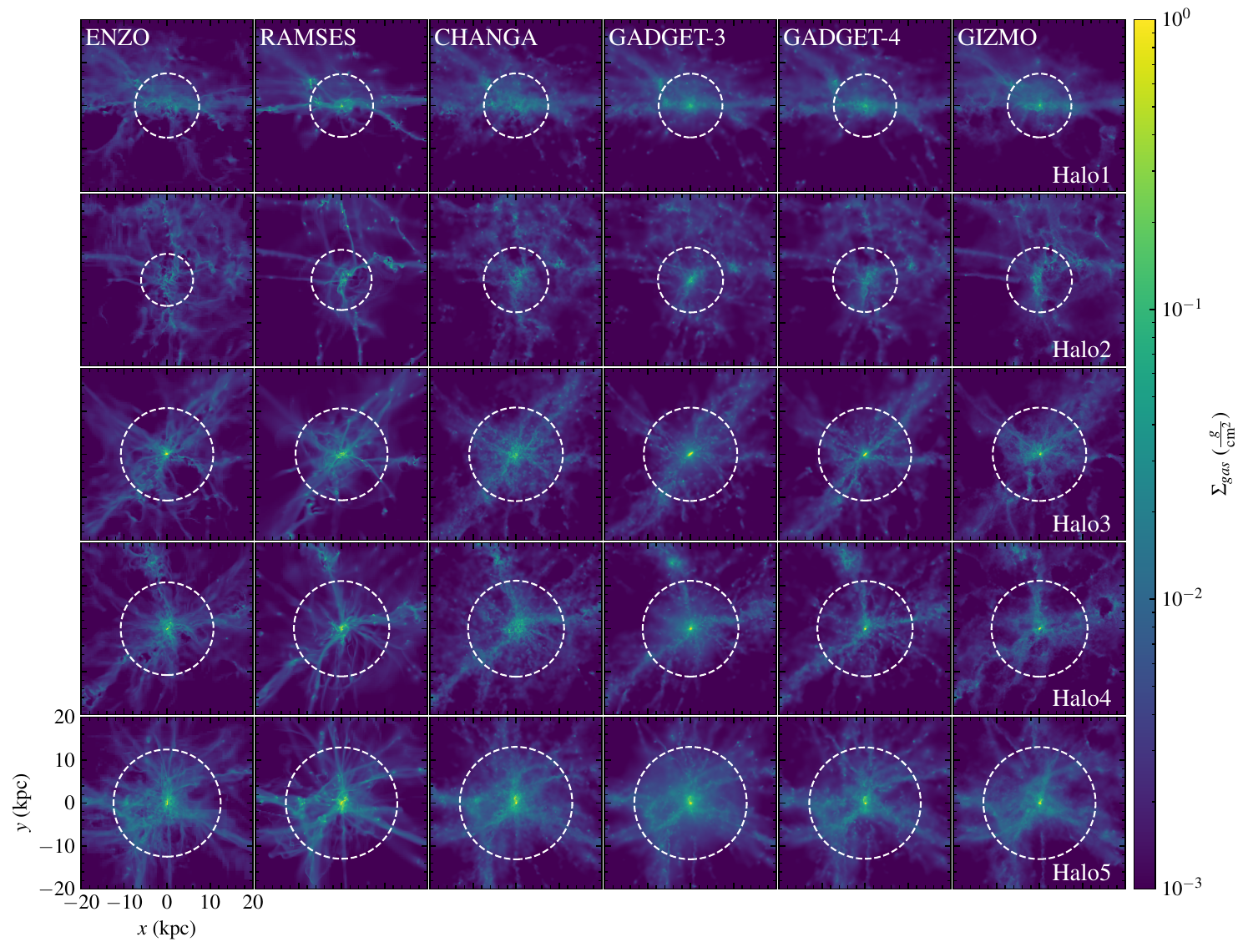}
    \vspace{-5mm}
    \caption{Gas density projections at $z=10$ from the \textit{High-z Run} for the five target halos. The virial radius of each target halo is indicated by a white dashed circle. While the participating codes show good overall agreement in gas distribution, some inter-code differences in the gas concentration in the circumgalactic region are notable. See Section \ref{sec:gas_properties} for more information.}
    \vspace{2mm}
    \label{fig:density_projection}
\end{figure*}

\begin{figure*}
    \vspace{0mm}
    \centering
    \includegraphics[width = 1\linewidth]{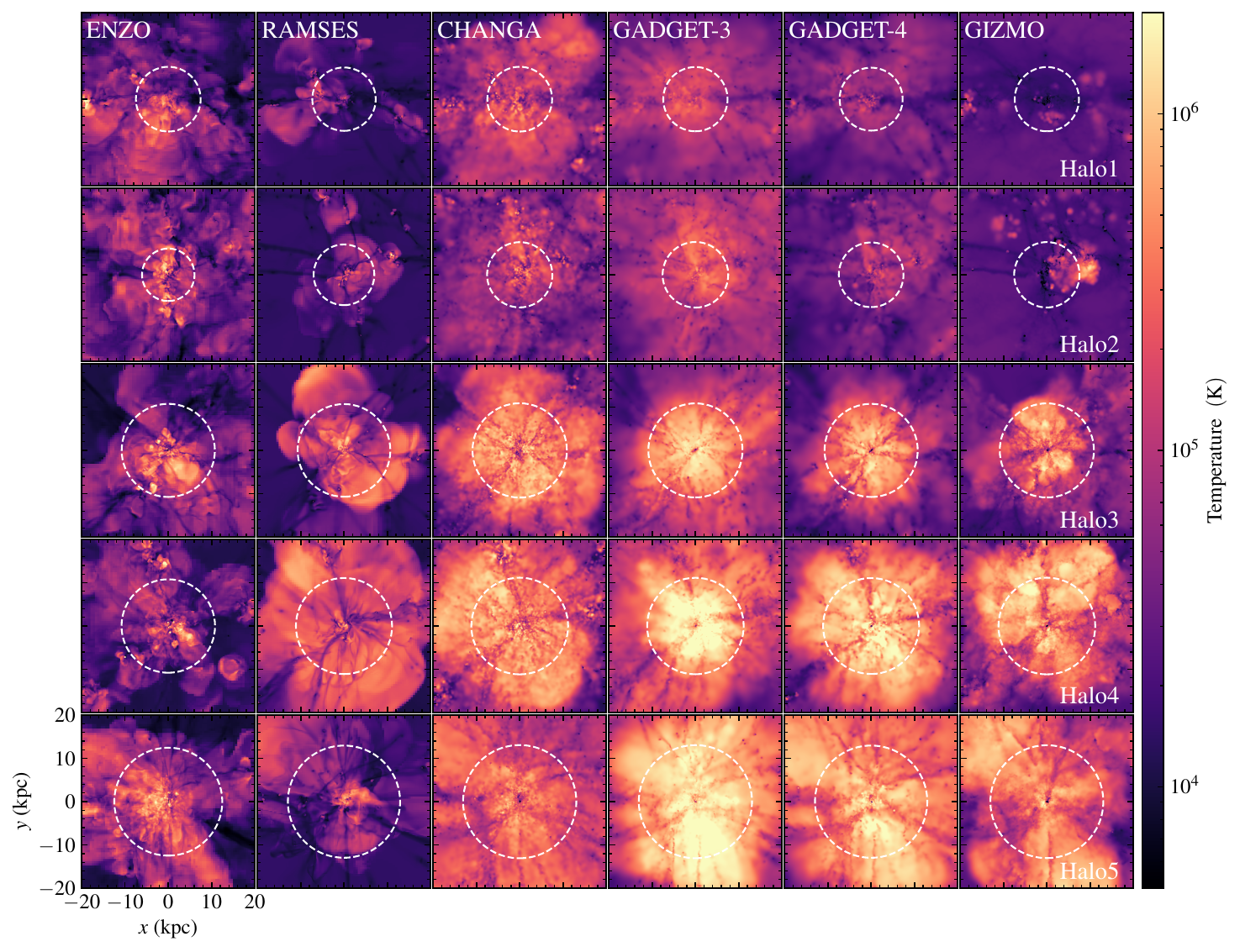}
    \vspace{-5mm}
    \caption{Density-weighted projections of gas temperature at $z=10$ from the \textit{High-z Run} for the five target halos. Similar to Figure \ref{fig:density_projection}, white circles indicate the virial radius of the halo. As discussed in Paper VI, the adaptive mesh refinement (AMR) codes \textsc{Enzo} and \textsc{Ramses} exhibit stronger contrast between the cold, high-density clouds and the hot, low-density bulk when compared to the smoothed particle hydrodynamics (SPH) codes in the circumgalactic medium (CGM). See Section \ref{sec:gas_properties} for more information.} 
    \vspace{2mm}
    \label{fig:temperature_projeciton}

\end{figure*}

We adopt a new refinement scheme to ensure consistent physical resolution between the \textit{High-z Run} and the \textit{CosmoRun} simulations. In the \textit{High-z Run}, mesh-based codes reach the finest cell size of 348 comoving pc, while particle-based codes use gravitational softening lengths of 1600 comoving pc until $z = 19$ where the first star emerges, and 80 physical pc thereafter. This strategy yields a uniform physical resolution of $22\ \textendash{} \ 32$ pc during the peak star formation epoch, a spatial resolution that corresponds to the epoch when star formation actively occurs in the \textit{CosmoRun}. The 1600 comoving pc for the particle-based code is set to be in accordance with the mesh-based codes, which have a comoving length twice as large compared to the \textit{CosmoRun} at $z\geq19$.


\begin{figure*}
    \vspace{0mm}
    \centering
    \includegraphics[width = 1\linewidth]{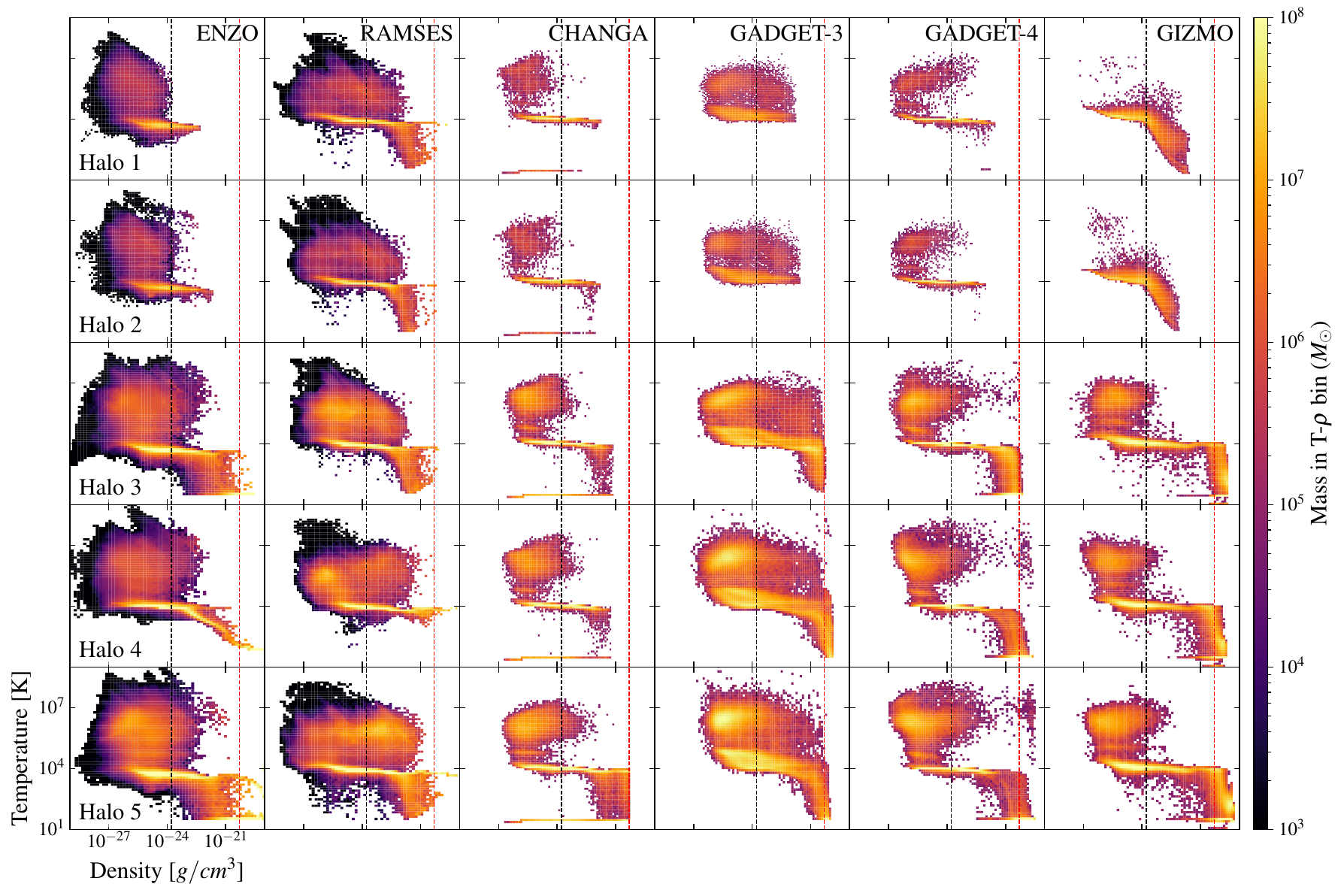}
    \vspace{-5mm}
    \caption{Two-dimensional probability distribution functions (PDFs) of gas density and temperature at $z=10$ in the \textit{High-z Run}. Gas within the $R_{\rm vir}$ of the halos is included, with the color representing the total gas mass in each bin. Star formation density threshold ($n_{\rm H} = 1 \,\rm cm^{-3}$) and feedback-free starbursts (FFB) density threshold ($n_{\rm H} = 3000 \,\rm cm^{-3}$) are annotated as vertical black and red dashed lines, respectively. Halos with virial masses above the FFB mass threshold ($\rm M_{\rm halo, z=10} = 3.5 \times 10^{10}\,{\rm M}_{\odot}$; Halos 3, 4, and 5) contain denser gas than halos below the threshold. \textsc{Ramses} consistently produces very dense gas regardless of halo mass, in contrast to other codes. See Section \ref{sec:gas_properties} for more information.}
    \vspace{2mm}

    \label{fig:pdf}
\end{figure*}
\begin{figure*}
    \vspace{5mm}
    \centering
    \includegraphics[width = 1\linewidth]{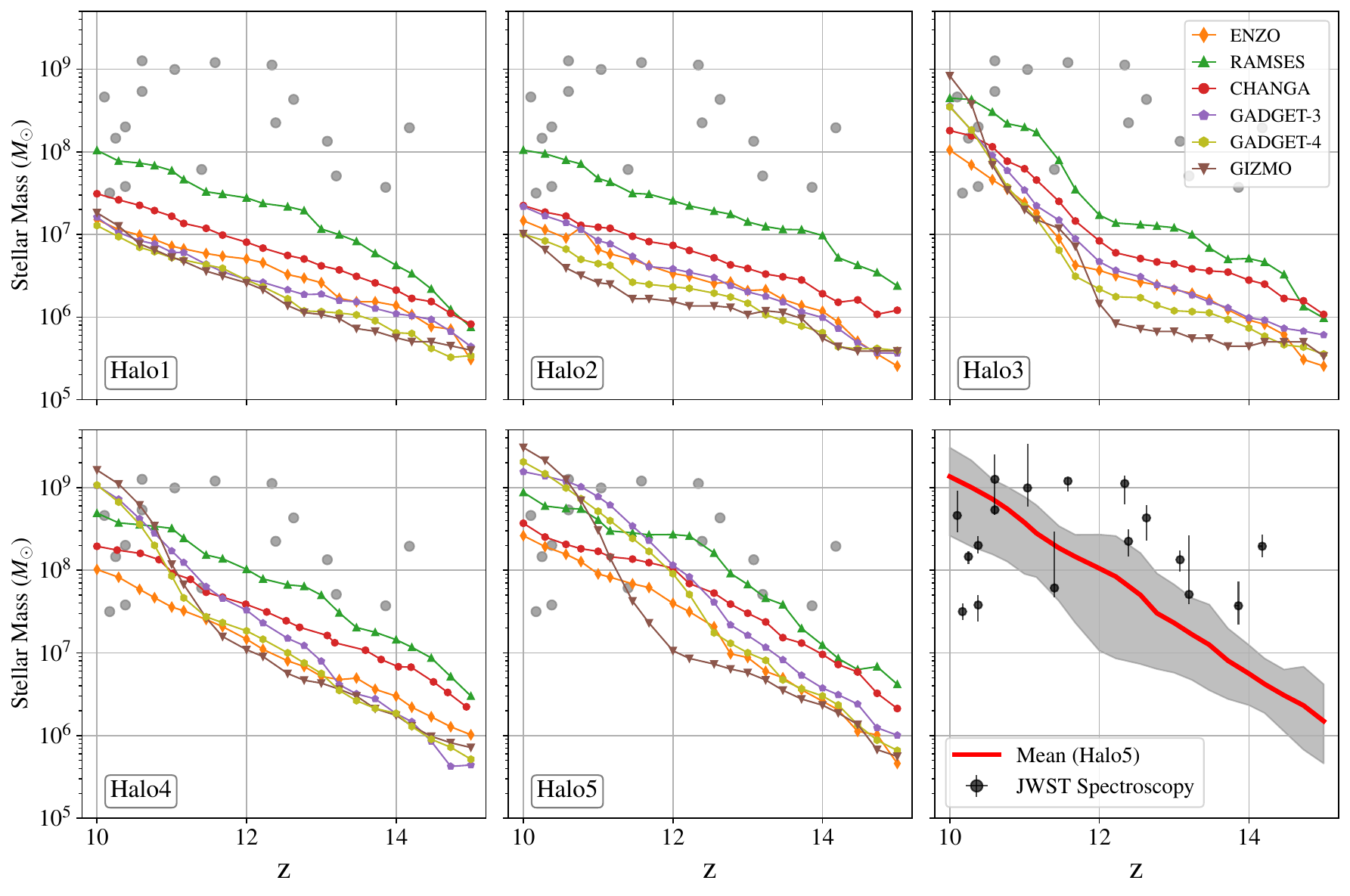}

    \caption{Evolution of stellar mass histories in the \textit{High-z Run}. Each panel shows the results for a different target halo. Stellar mass is calculated as the total stellar mass within $R_{\rm vir}$ of the halo, following the methodology adopted for Figure 4 in Paper IV. To compare with observations, in the bottom right panel we present the mean value (thick red line) and the minimum–maximum range (gray shaded region) of Halo 5 and JWST observation data (black dots with error bars) from the following works: \cite{2023A&A...677A..88B, 2023ApJ...952...74T, 2023ApJ...957L..34W, 2024ApJ...960...56H,2024ApJ...973....8H,2024ApJ...970...31R, 2025A&A...696A..87C, 2025ApJ...992..212W}. The observation data points are replicated in other panels as translucent gray dots. Despite some variations in stellar mass across simulations, the \textit{High-z Run} Halo 5 results show good agreement with observations, reproducing similar stellar masses except at $z \sim 14$. See Section \ref{sec:stellar} for more information.}
    \label{fig:stellar_mass}
    \vspace{2mm}
\end{figure*}

\subsection{SKIRT Radiative Transfer Code}
We utilize the radiative transfer code SKIRT \citep{2020A&C....3100381C} to generate mock images and estimate the UV luminosity of our simulated galaxies, accounting for both dust attenuation and emission.

Each galaxy's radiation sources are represented by star particles within the halo. For every stellar particle inside $R_{\rm vir}$, we imported the positions, initial masses, smoothing lengths, metallicities, and ages directly from the simulation outputs. The initial stellar masses were either derived from the stellar mass loss routines of the respective simulation codes or obtained from explicitly provided stellar mass data in each code. To determine the smoothing length for each star particle, we calculated the distance to its 32nd-nearest neighboring star particle with an upper limit of 80 pc. We divide stellar populations into two groups to include star-forming nebular emission, as nebular emission is an important indicator in measuring galaxy properties at high redshift in the limited available wavelength. For the older stellar population, we adopted the Binary Population and Spectral Synthesis (BPASS) model of \cite{2018MNRAS.479...75S}, utilizing a \cite{2003PASP..115..763C} initial mass function (IMF) to define the stellar emission properties. For young stellar particles (ages $\leq 30$ Myr), the TODDLERS library \citep{2023MNRAS.526.3871K} is employed. Coupled with a semi-analytical gas-cloud evolution model, the TODDLERS library produces a spectral energy distribution (SED) that includes stellar continuum, dust emission, nebular continuum, and line emission from both H II and molecular gas regions. In these models, we assumed a nebular density of $n_{\mathrm{H}} = 100\ \mathrm{cm}^{-3}$ following \cite{2019MNRAS.484.1366C} and a star formation efficiency of $\epsilon_{\mathrm{SF}} = 0.1$, higher than that in the local universe, allowing the disruption of clouds \citep{2023MNRAS.526.3871K}.

The interstellar medium (ISM) in our simulations is modeled according to the output format specific to each code: grid-based structure for \textsc{Enzo} and \textsc{Ramses}, and particle-based representations for \textsc{Changa}, \textsc{Gadget-3}, \textsc{Gadget-4}, and \textsc{Gizmo}. The dust distribution is proportional to the gas density, with the condition that dust formation occurs only in gas elements with temperatures below $T_{\text{max}} = 15000 \ \text{K}$, assuming that dust is destroyed at $T \geq 15000 \ \text{K}$ (Paper VIII). The dust density ($\rho_{\text{dust}}$) is calculated using the following relation:

\[
\rho_{\text{dust}} = 
\begin{cases}
f_{\text{dust}} \, Z_{\text{gas}} \, \rho_{\text{gas}} & \text{if } T < T_{\text{max}} \\
0 & \text{otherwise}
\end{cases}
\]

Where $\rho_{\text{dust}}$ and $\rho_{\text{gas}}$ are the densities of dust and gas, respectively. $Z_{\text{gas}}$ is the gas metallicity fraction and $f_{\text{dust}}$ is the dust-to-metal (DTM) ratio. Considering that dust production is inefficient at high redshift \citep{2018MNRAS.477..552B} and the decrease in the dust-to-metal ratio toward higher redshift \citep{2023A&A...679A..91H}, we adopt a value of $f_{\text{dust}} = 0.1$, which is relatively low compared to that in the local universe of 0.2 \textendash{} 0.5 \citep{2019A&A...623A...5D,2024A&A...681A..64K}.

In the observed spectra of high-redshift objects, the Gunn–Peterson trough \citep{1965ApJ...142.1633G} arises from absorption of Lyman-$\alpha$ photons by neutral hydrogen in the intergalactic medium (IGM), producing a broad region of zero flux at wavelengths shorter than the Lyman lines. Because SKIRT does not include IGM absorption along the line of sight, this effect is mimicked by setting the flux to zero at wavelengths shorter than the Lyman-$\alpha$ line corresponding to the source redshift.

\subsection{Participating Codes} \label{sec:style}
Here, we provide a brief introduction to the numerical codes that participate in the \textit{High-z Run} and their applications to high-redshift galaxy studies if available. It should be noted that the participating codes do not represent their entire code communities, as differences in parameter choices and subgrid physics can exist even within a single code. However, for clarity, we refer to each participating code by its code name. Previous \textit{AGORA} papers have presented comprehensive descriptions of each code: gravity solvers (Paper I), hydrodynamic solvers (Paper II), stellar feedback implementations (Papers III and IV). Therefore, we refer interested readers to these papers for full technical details. 

\subsubsection{ENZO}
\textsc{Enzo} \citep{2014ApJS..211...19B} is a block-structured AMR code. Originally developed by Greg Bryan and later released as a publicly available code, it is now actively maintained and extended by a broad collaboration of developers. \textsc{Enzo} employs the stochastic star formation prescription commonly adopted in the \textit{AGORA} project and models stellar feedback through thermal energy injection from Type II supernovae. Renaissance simulation \citep{2023OJAp....6E..47M} uses \textsc{Enzo} to model galaxy formation in the early universe, finding that the properties of the most massive galaxies in the simulations are consistent with JWST observations.
\newpage
\subsubsection{RAMSES}
\textsc{Ramses} \citep{2002A&A...385..337T} is also an AMR code that uses a tree-based grid structure to solve the gravity and hydrodynamic equations. In this paper, \textsc{Ramses} adopts the delayed-cooling thermal feedback model introduced by \cite{2015MNRAS.452.1502D}, which locally delays the gas cooling after the energy injection from thermal stellar feedback to prevent overcooling in the ISM. For every Type II supernova event, both thermal dump energy ($4\times 10^{51}{\rm erg}$) and non-thermal turbulent energy density are injected into the host cell. The turbulent energy decays gradually on a timescale set by paramter $t_{\rm delay}=10\ \rm Myr$, and radiative cooling is disabled as long as the turbulence remains above a given threshold $\mathrm{\sigma}_{\rm min} = 100\ \rm km/s$.
\cite{2025MNRAS.540.3350A} investigates the mechanisms that drive massive galaxy formation at Cosmic Dawn using \textsc{Ramses}, focusing on the star formation efficiency of galaxies.

\subsubsection{CHANGA}
\textsc{Changa} \citep{2015ComAC...2....1M} is an SPH code built on the Charm++ parallel programming framework. Reimplemented from the earlier SPH code \textsc{Gasoline} \citep{2017MNRAS.471.2357W}, it employs a k-th nearest neighbor algorithm with $N_{\rm ngb}=64$ for both gravitational and hydrodynamic calculations. For the \textit{High-z Run}, \textsc{Changa} adopts the “superbubble” feedback model, which introduces a small, hot region around supernova events to capture their nature \citep{2014MNRAS.442.3013K}. In this study, we follow the same stellar feedback prescription used in Paper III, but use $3.25 \times 10^{51}{\rm erg/SN}$ instead of $5 \times 10^{51}{\rm erg/SN}$, which is revised in Paper IV. Using \textsc{Changa}, \citet{2025MNRAS.543.2760V} conducts the \textsc{Phoebos} simulation, suggesting that a weak stellar feedback prescription can successfully reproduce galaxy properties at $z \geq 8$.

\subsubsection{GADGET-3}
\textsc{Gadget-3-Osaka} is a modified version of the SPH code \textsc{Gadget-3}, enhanced with a suite of subgrid physics modules for star formation and feedback developed by the Osaka group. Its prescriptions follow \citet{2017MNRAS.466..105A, 2019MNRAS.484.2632S}, implementing both thermal and kinetic feedback. Feedback energy is injected into the surrounding particles through a hot-bubble scheme with a time delay, where the delay time and bubble size depend on local gas density and pressure and the feedback energy. Chemical enrichment is modeled using metal yields from the chemical evolution library CELib \citep{2017AJ....153...85S}.
\cite{2025ApJ...980...10J} conducts cosmological hydrodynamic zoom-in simulations in a high redshift environment with a top-heavy IMF and star formation efficiency using \textsc{Gadget-3}.

\subsubsection{GADGET-4}
In the \textit{High-z Run}, we add the results of the \textsc{Gadget-4} \citep{2021MNRAS.506.2871S} code using the \textsc{Gadget-4-Osaka} model. \textsc{Gadget-4-Osaka} is also a modified version of the SPH code \textsc{Gadget-4}, incorporating an updated stellar feedback model \citep{2022ApJS..262....9O} and AGN physics developed by the Osaka group \citep{2024ApJ...975..183O}. It also supports delayed cooling and feedback implemented through two distinct channels, a local mechanical feedback and a large-scale hot galactic wind component. The mechanical feedback injects the terminal momentum of unresolved supernova remnants calibrated from high-resolution simulations, while the wind model captures the large-scale expulsion of hot gas. The detailed calibration steps of \textsc{Gadget-4} in the \textit{CosmoRun} are presented in the Appendix of Paper VIII.

\vspace{1mm}
\subsubsection{GIZMO}
\textsc{Gizmo} \citep{2015MNRAS.450...53H} is a hybrid code that bridges particle-based and grid-based methods through a fully Lagrangian hydrodynamics solver. It supports multiple schemes, including meshless finite mass (MFM) and meshless finite volume (MFV), allowing flexible treatment of fluid dynamics. In the \textit{High-z Run}, the MFM method is adopted. The stellar feedback model follows the FIRE-2 model \citet{2018MNRAS.477.1578H}, distributing both kinetic and thermal energy and metals to neighboring particles through kernel-weighted schemes. The FIRE-2 galaxy formation model \citet{2018MNRAS.477.1578H} has been implemented into \textsc{GIZMO} and applied to the study of high-redshift galaxies, including investigations of star formation and UV luminosity from the FIRE-2 and $\rm FIREbox^{\it HR}$ cosmological simulations \citep{2023MNRAS.526.2665S, 2025MNRAS.536..988F}. Please note that although the \textsc{Gizmo} in the \textit{High-z Run} adopts the FIRE-2 stellar feedback prescription, it does not represent the FIRE-2 model, as many other aspects (e.g., star formation model, cooling algorithm) differ significantly from those in the original FIRE-2 framework.

\begin{figure*}
    \vspace{0mm}
    \centering
    \includegraphics[width = 1\linewidth]{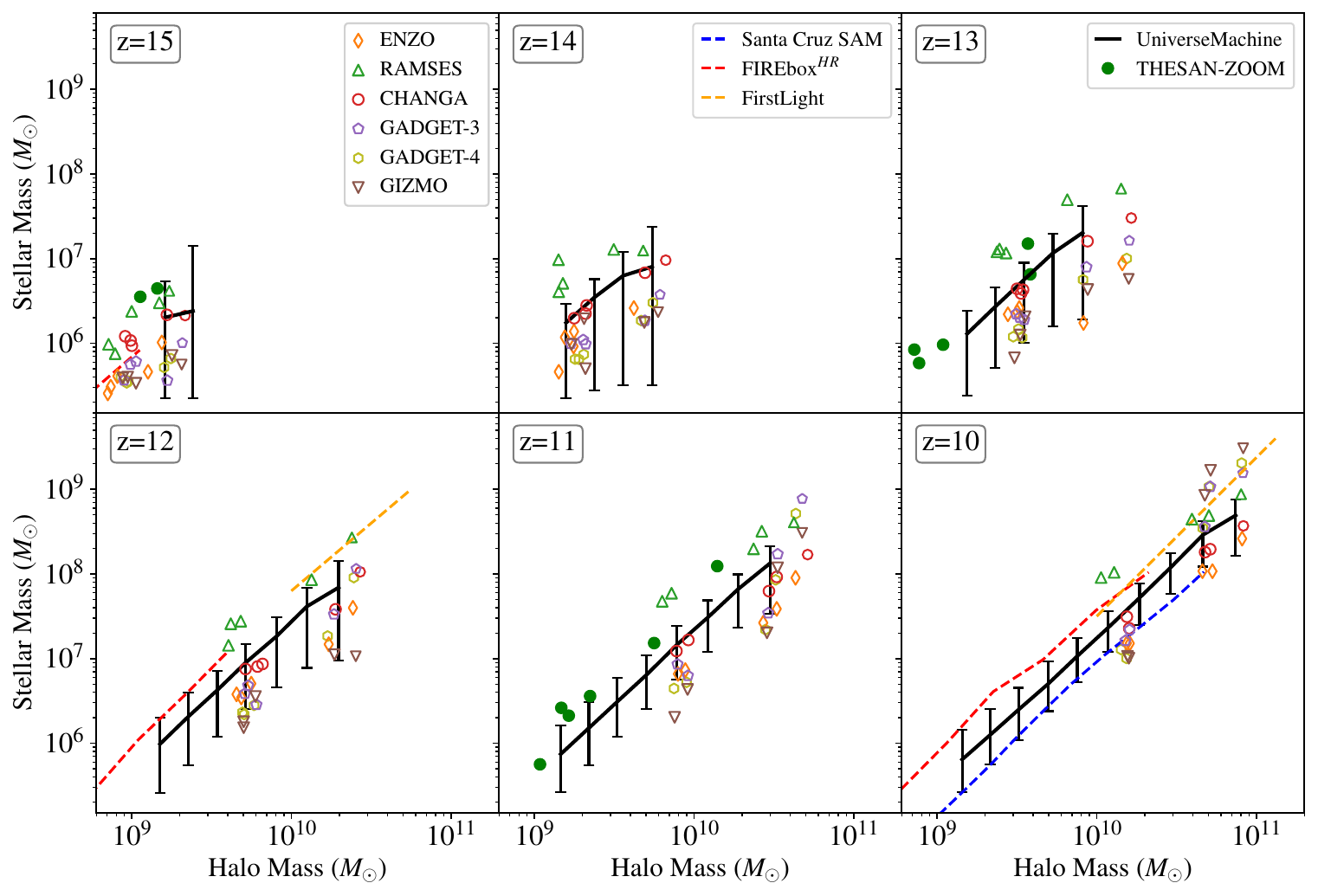}
    \caption{Evolution of the stellar mass-halo mass (SMHM) relation of the galaxies in the \textit{High-z Run}. The galaxies in the \textit{High-z Run} are marked with an open colored symbol at selected epochs. For comparison, we display the estimates from abundance matching techniques UniverseMachine \citep[][black solid line with error bar]{2020MNRAS.499.5702B} and semi-analytic model Santa Cruz SAM \citep[][blue dashed line at $z=10$]{2019MNRAS.490.2855Y}. We also present the SMHM relation in other cosmological simulations, $\rm FIREbox^{\it HR}$ \citep[][red dashed line at $z=10, 12$, and $15$]{2025MNRAS.536..988F}, FirstLight \citep[][orange dashed line at $z=10, 12$]{2024A&A...689A.244C} and THESAN-ZOOM \citep[][green dots at $z=11, 13$, and $15$]{2025OJAp....8E.153K}. Most of the \textit{High-z Run} galaxies align with the previous works, but some galaxies have slightly higher stellar mass at $z=10$. See Section \ref{sec:stellar} for more information.}
    \vspace{2mm}
    \label{fig:shmr}
\end{figure*}

\begin{figure*}
    \vspace{0mm}
    \centering
    \includegraphics[width = 1\linewidth]{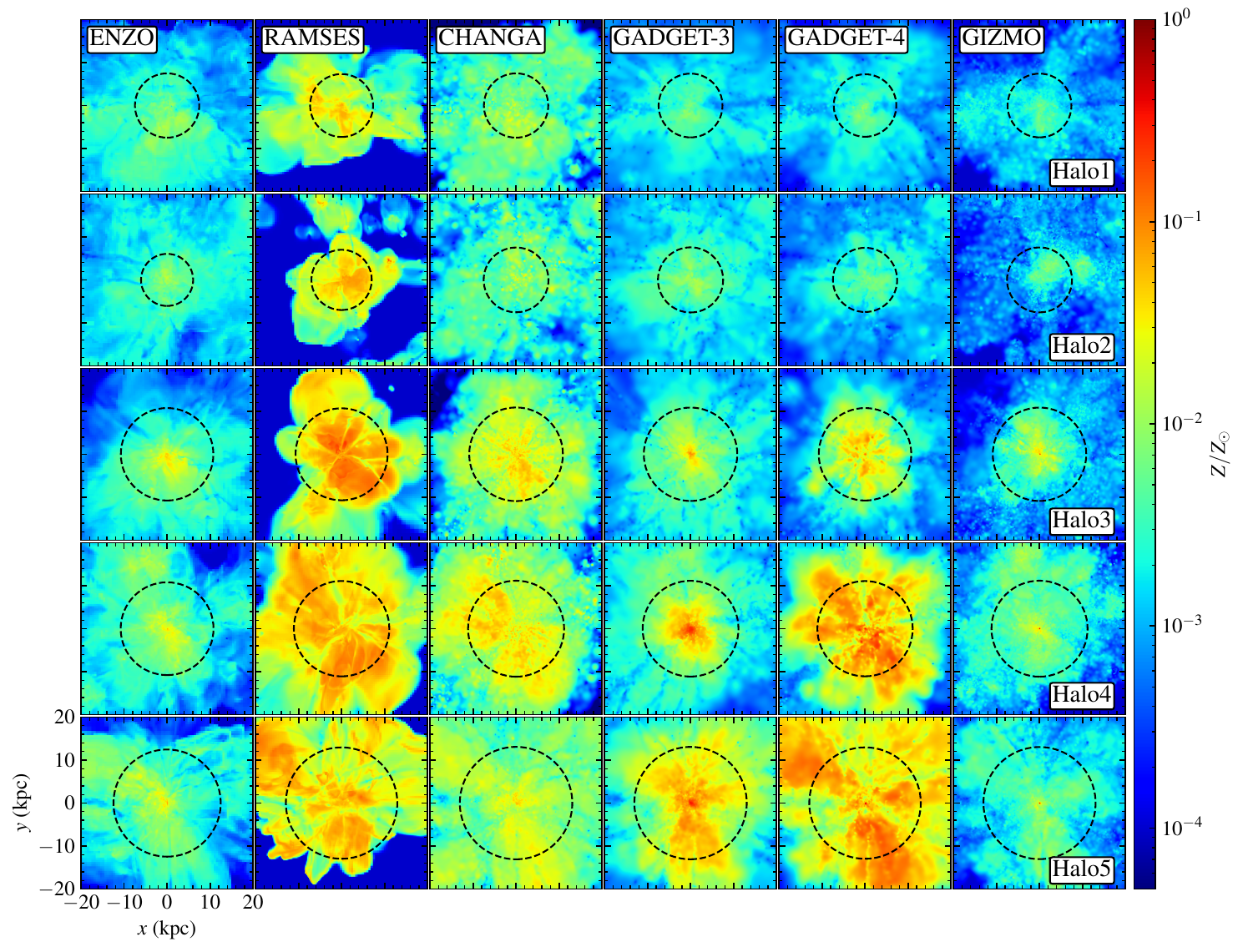}
    \vspace{-5mm}
    \caption{Density-weighted projections of gas metallicity at $z=10$ from the \textit{High-z Run} for the five target halos. As in Figures \ref{fig:density_projection} and \ref{fig:temperature_projeciton}, the black circles indicate the virial radius of the halos, and the color scale represents gas metallicity in units of ${Z}_{\odot}$. Although the total gas mass within the virial radius is comparable across codes, metal distributions differ significantly. See Section \ref{sec:metal} for more information.} 
    \vspace{2mm}

    \label{fig:metallicity_projection}
\end{figure*}

\begin{figure*}
    \vspace{0mm}
    \centering
    \includegraphics[width = 1\linewidth]{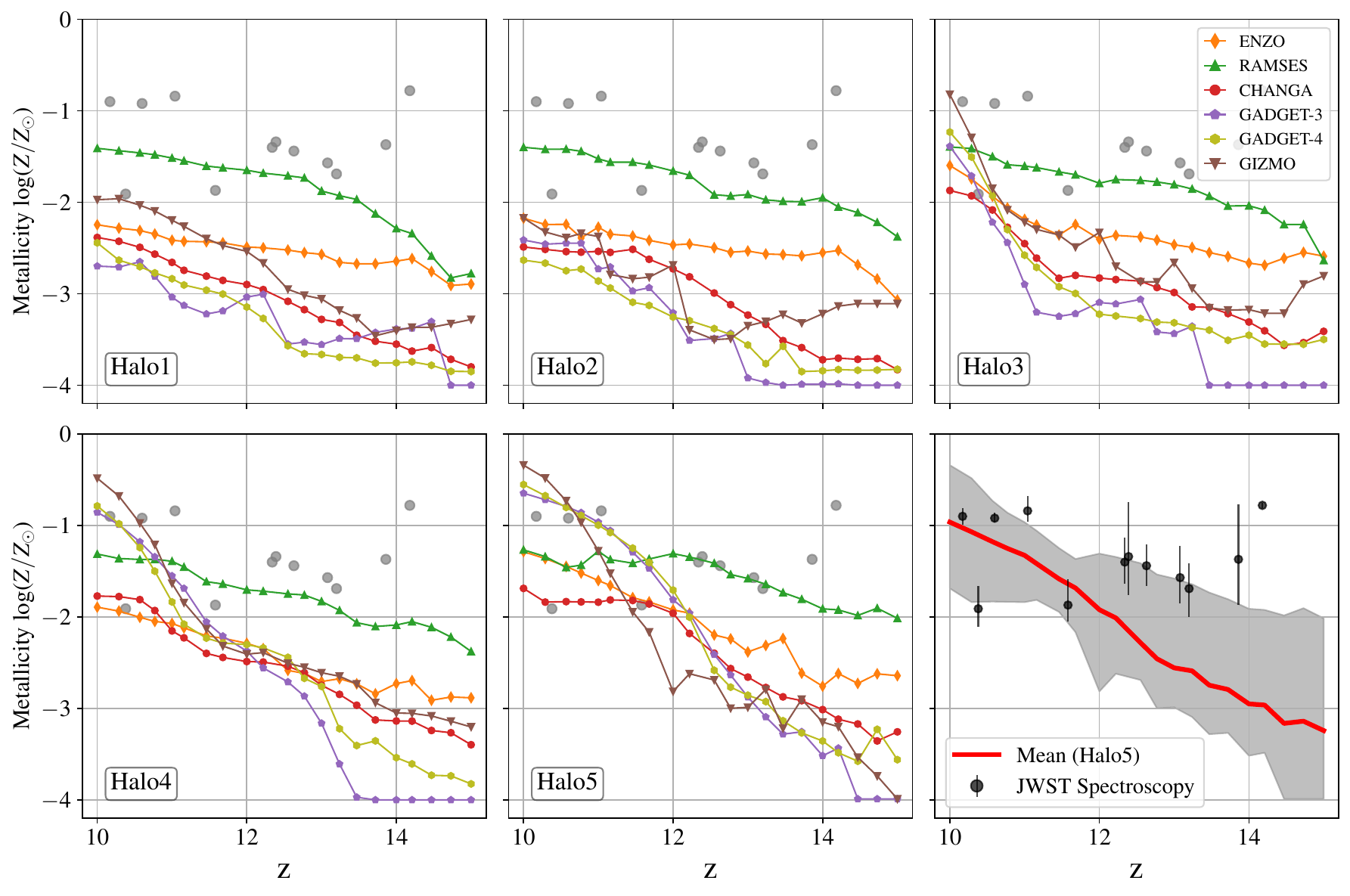}
    \vspace{-5mm}
    \caption{Metallicity evolution histories in the \textit{High-z Run} as a function of redshift. Metallicity is defined as the mass-weighted metallicity of star particles younger than 100 Myr, assuming young star particles share the metallicity of their star-forming regions. Metallicity exhibits larger inter-code variations than stellar mass. 
    To compare with observations, in the bottom right panel we present the mean value (thick red line) and the minimum–maximum range (gray shaded region) of Halo 5 and the JWST observation data (black dots with error bars) from the following works: \cite{2023A&A...677A..88B, 2023NatAs...7..622C, 2023ApJ...957L..34W,  2024ApJ...973....8H, 2024ApJ...972..143C, 2025A&A...696A..87C, 2025ApJ...992..212W}.  The observation data points are replicated in other panels as translucent gray dots. The \textit{High-z Run} reproduces the observed metallicity at $z<12$, but underpredicts at higher redshift, with the exception of \textsc{Ramses}. See Section \ref{sec:metal} for more information.}
    \vspace{2mm}
    \label{fig:metallicity}
\end{figure*}

\begin{figure}[t] 
    \centering
    \includegraphics[width=\columnwidth]{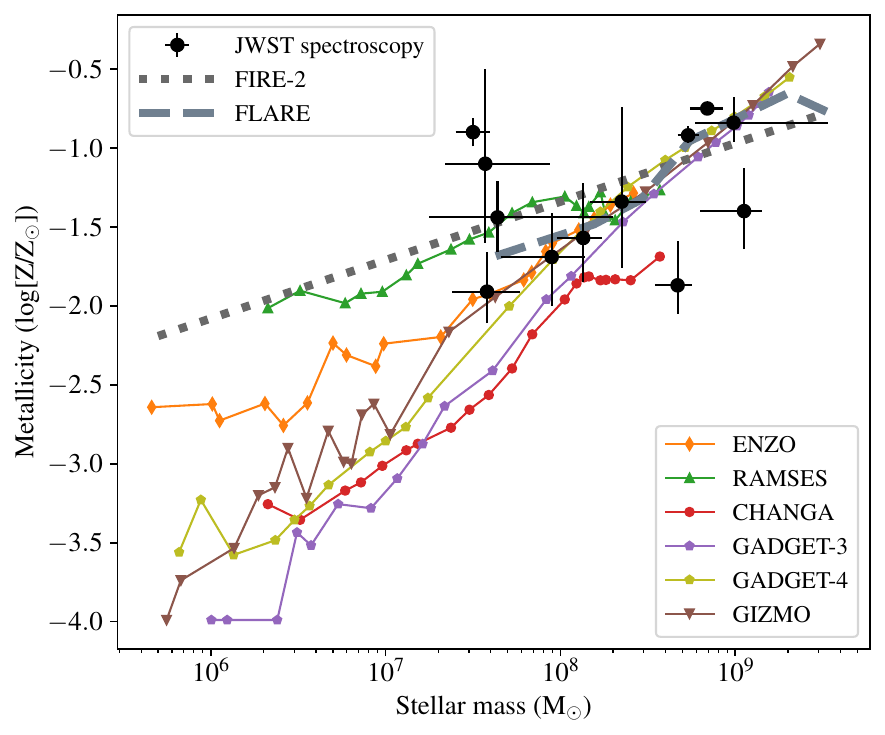}
    \caption{The stellar mass–metallicity relation (MZR) from the \textit{High-z Run}. We present MZR evolution of Halo~5, alongside JWST observational data points \citep[black dots with error bars,][]{2023A&A...677A..88B, 2023NatAs...7..622C, 2023ApJ...957L..34W,  2024ApJ...973....8H, 2024ApJ...972..143C, 2025A&A...696A..87C, 2025ApJ...992..212W}. We also include the gas-phase mass-metallicity relation from the FIRE-2 simulation \citep[gray dotted line,][]{2024ApJ...967L..41M} and FLARE simulation \citep[gray dashed line][]{2023MNRAS.519.3118W}. The participating codes diverge at the low-mass end, but converge well at higher mass, showing good agreement with observations and other cosmological simulations. See Section \ref{sec:metal} for more information.}
    \label{fig:mzr}
\end{figure}

\section{RESULTS}

\subsection{Gas Properties}\label{sec:gas_properties}

Figure~\ref{fig:density_projection} presents the projected gas density of each halo in the \textit{High-z Run}. Each panel spans 40 kpc in physical scale, with dashed white circles marking the virial radii. While the participating codes broadly agree on the overall gas distribution, inter-code differences emerge in the small scale, especially in massive halos. In Halos 3, 4 and 5, \textsc{Gadget-3}, \textsc{Gadget-4}, and \textsc{Gizmo} show highly compact cores ($\sim 500 \ \rm pc, 10\ \rm g/cm^2$) at the halo center, whereas mesh-based codes and maintain more diffuse gas.

Figure~\ref{fig:temperature_projeciton} shows the corresponding density-weighted temperature maps for the same regions. Here, the differences are even more pronounced. Particle-based codes generate hotter central regions and extended, smooth temperature gradients, while mesh-based codes produce sharper contrasts between cool, dense clumps and the hot, low-density CGM. This result is consistent with findings in Paper~VI. Mesh-based codes inject energy into a single cell and then spread it into the surrounding area, whereas particle-based codes distribute feedback energy to neighboring particles, producing a more extended hot region than mesh-based codes.

These figures highlight that, unlike the dark matter distribution (Figure~\ref{fig:darkmatter_plot}), the gas density and thermal structure exhibit clear inter-code variations that arise from the different numerical solvers of each simulation framework.

To investigate the details of gas properties, Figure \ref{fig:pdf} presents the density–temperature probability distribution function (PDF) at $z=10$ for gas within $R_{\rm vir}$ of halos, with the star formation density threshold ($n_{\rm H}=1\ \rm cm^{-3}$, vertical black dashed line) and density threshold ($n_{\rm H}=3000\ \rm cm^{-3}$, vertical red dashed line) proposed from FFB scenario \citep{2023MNRAS.523.3201D}. First, we can see that the physics of the \textit{CosmoRun} is clearly reflected in the \textit{High-z Run}. (For the details, see Figure 14 of Paper III). (1) the black bins at low-density and high-temperature regions in mesh-based codes, which come from the small mass gas cells in the CGM. (2) hot gas above the density threshold for the star formation due to delayed-cooling in \textsc{Ramses}, \textsc{Gadget-3}, and \textsc{Gadget-4}. (3) cooling curve at $\sim10^4 \ \rm K$, where gas cooling rate balances the heating rate in the \textsc{Cloudy} table.  Nevertheless, there are also differences between the \textit{High-z Run} and \textit{CosmoRun}. Since the target halos of the \textit{High-z Run} have very massive halo mass already at extremely high-redshift, gas is concentrated rapidly, which results in a very dense gas region above the star formation threshold. Especially, Halos 3, 4, and 5 have very dense and cold gas ($\sim10^{-21}{\rm g/cm^{3}}$, $10^{2}\ \rm K$), which is not shown in Halos 1 and 2. Also, \textsc{Enzo, Gadget-3, Gadget-4}, and \textsc{Gizmo} show significant amount ($\sim 10^{9} \ {\rm M}_{\odot}$) of cold dense gas. At high redshift, accretion in halos is dominated by cold streams from dense filaments despite the presence of a hot surrounding CGM. In these flows, the cooling time is shorter than the compression heating time from the virial shock \citep{2006MNRAS.368....2D}, preventing the formation of a stable shock. This allows cold gas to penetrate into the hot CGM and sustain inflowing streams into the halos. Moreover, the host masses of Halos 3, 4, and 5 exceed the mass threshold ($M_{\rm halo} = (1+z)^{-6.2}10^{17}\ {\rm M}_{\odot}$, $3.5\times10^{10}\ {\rm M}_{\odot}$ at $z=10$) proposed by \cite{2023MNRAS.523.3201D}, leading to the very dense gas reservoirs ($n_{\rm H}\sim3000\ \rm cm^{-3}$), where the stellar feedback becomes inefficient.

\textsc{Ramses} consistently produces such dense gas cells regardless of halo mass. This feature comes from the refinement strategy of \textsc{Ramses} implemented in the \textit{CosmoRun} suite. \textsc{Ramses} extends the high-resolution region around the dense cells to avoid discontinuity of resolution between the high density region and the nearby lower density regions. As a consequence, \textsc{Ramses} has a more extended and refined region, compared to the other codes.

\begin{figure*}
    \vspace{0mm}
    \centering
    \includegraphics[width = 1\linewidth]{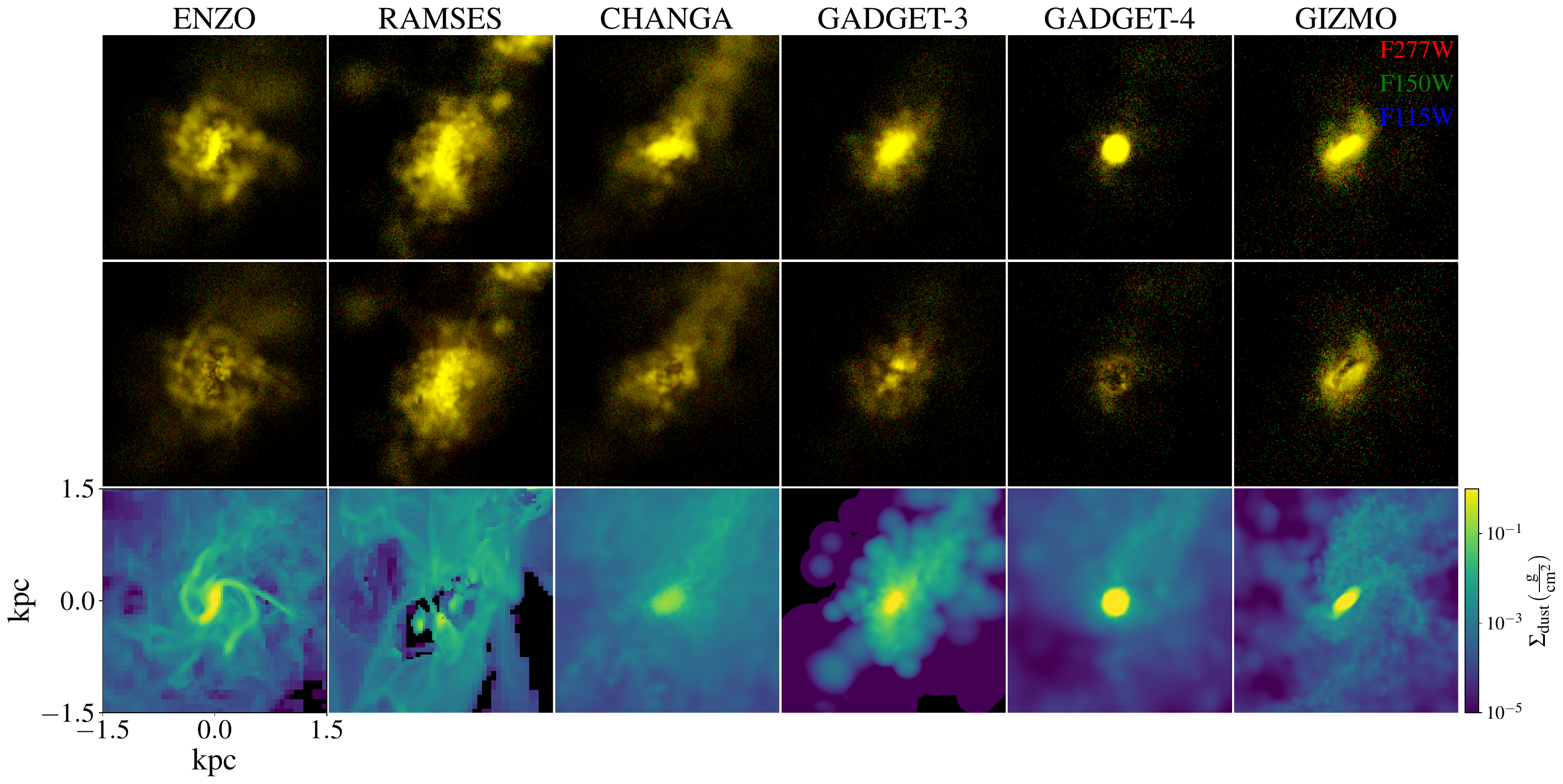}
    \vspace{-5mm}
    \caption{SKIRT-generated mock JWST NIRCam images and dust density projections of Halo 5 at $z=10$. Three filters (R: F277W, G: F150W, B: F115W) are used as RGB composites. The images are generated from the gas and stellar components within the virial radius $R_{\rm vir}$. \textit{Upper panels}: Mock image without gas and dust medium. \textit{Middle panels}: Mock image with gas and dust medium. \textit{Bottom panels}: Density projections of dust used to generate mock images. Depending on the dust distribution, dust makes a huge difference in the mock image. See Section \ref{sec:mock} for more information.}
    \vspace{2mm}
    \label{fig:mock}
\end{figure*}

\subsection{Stellar Mass Evolution}\label{sec:stellar}
Figure \ref{fig:stellar_mass} shows the stellar mass growth histories of halos as functions of redshift. Stellar mass is measured within the virial radius ($R_{\mathrm{vir}}$), centered on each halo. The difference between participating codes tends to broaden as the host halo mass increases. Codes show very good agreement within less than 0.5 dex in Halo 1 except \textsc{Ramses}, but inter-code differences grow to $\sim1$ dex in Halo 5. This trend is expected, as halo mass reflects the depth of the gravitational potential well at the galaxy center. The deep potential well of Halo 5 enables more efficient gas accretion from the IGM and sustains higher star formation, which results in various star formation histories from different stellar feedback across codes. Previous studies \citep{2017MNRAS.465.1682H, 2025A&A...693A.149K} show that different stellar feedback recipes can affect star formation at high redshift. Looking at some individual codes, \textsc{Ramses} consistently produces higher stellar masses in the early stages regardless of the halo mass. These results come from its refinement strategy, which achieves a more extended maximum spatial resolution than other codes. For example, \textsc{Enzo} and \textsc{Ramses} are both mesh-based code, the number of the finest cells differs at $z=15$ by a factor of 4 in the center of the galaxy. Despite this early disparity, \textsc{Ramses} converges to comparable stellar masses at the final redshift when the structures collapse and the star forming region becomes dense enough for star formation in all codes.\footnote{We note that the star formation of \textsc{Ramses} at higher redshift ($z \sim 15$) is highly sensitive to the refinement strategy choice, but reaches similar stellar mass at $z \sim 10$. In the future study, we will investigate the effect of refinement at high redshift, when the first stars are formed.} \textsc{Gizmo} exhibits distinctive sensitivity to merger events. While Halos 3 and 4 have similar final halo masses, Halo 3 starts a major merger (mass ratio $\sim 1:1$) around $z\sim12$, leading to a pronounced increase in stellar mass for \textsc{Gizmo} afterward. This is also observed in the other codes, but \textsc{Gizmo} shows the rapid stellar mass increase.

Overall, the tendency of stellar mass growth is similar across the simulation codes, but $\sim1$ dex of inter-code differences exist throughout the redshift range. However, compared to the stellar mass assembly history (SMAH) presented in Paper III, the \textit{High-z Run} shows better convergence at $z=10$ with a similar stellar mass range ($\sim$1.5 dex in the \textit{CosmoRun} and $\sim$1 dex in the \textit{High-z Run}). We also conduct an additional run with different feedback strengths and found that convergence is driven more by differences in feedback models between codes than by variations in feedback energy strength within a single code (see Appendix \ref{sec:convergence} for detail). From these results, we conclude that the simulation codes converge well, although they are calibrated at a relatively low redshift ($z=4$).

We also compared our results to the spectroscopically confirmed JWST observation data \citep{2023A&A...677A..88B, 2023ApJ...952...74T, 2023ApJ...957L..34W, 2024ApJ...960...56H,2024ApJ...973....8H,2024ApJ...970...31R, 2025A&A...696A..87C, 2025ApJ...992..212W}. In the bottom-right panel, we present the high-redshift galaxies from JWST observation as black dots with error bars, and the red line and gray shaded region represent the mean values and minimum-maximum values of the six participating codes from Halo 5. Without changing the subgrid physics for low redshift, the \textit{High-z Run} can reproduce the stellar mass at $z<12$ from JWST observation data, but it is insufficient for redshifts above that.

The stellar mass–halo mass (SMHM)  relation is a fundamental relation of galaxy evolution that quantifies the efficiency of star formation in the host halos \citep{2010ApJ...710..903M}. Figure \ref{fig:shmr} presents the evolution of the SMHM relation of the \textit{High-z Run} during $z=15\ \textendash{}\ 10$. For comparison, we also compare our work to other theoretical models and simulations. These include the semi-empirical model UniverseMachine \citep[solid black line,][]{2020MNRAS.499.5702B}, the semi-analytic model Santa Cruz SAM \citep[blue dashed line,][]{2019MNRAS.490.2855Y}, and simulations from $\rm FIREbox^{\it HR}$ \citep[red dashed line,][]{2025MNRAS.536..988F}, FirstLight \citep[orange dashed line,][]{2024A&A...689A.244C}, and THESAN-ZOOM \citep[green dots,][]{2025OJAp....8E.153K}. The \textit{High-z Run} results are marked with open symbols, and we represent the median (Santa Cruz SAM) and average values ($\rm FIREbox^{\it HR}$ and FirstLight) of previous studies as dashed lines for comparison. Across the redshift range, our results are consistent with previous studies. Except for \textsc{Ramses}, the participating codes show good convergence below $M_{\rm halo} \sim 10^{10}M_{\odot}$, but begin to diverge toward lower redshifts, with differences of up to about 1 dex. Nevertheless, the overall trends in the \textit{High-z Run} remain in broad agreement with earlier works.
 
\subsection{Metallicity Properties}\label{sec:metal}
Figure \ref{fig:metallicity_projection} shows the density-weighted gas metallicity projections for the five halos across six simulation codes. While the overall gas distributions are broadly similar, the metallicity patterns exhibit clear differences. In Halos 1 and 2, only \textsc{Ramses} reaches significantly higher metallicity than the other codes due to its higher star formation in low-mass halos, which enhances early metal enrichment. For Halos 3, 4, and 5, the divergence among codes becomes more pronounced, reflecting the distinct outflow behaviors discussed in Paper VI. In Paper VI, \textsc{Ramses, Changa}, and \textsc{Gadget-3} have strong outflows that mix chemical elements around the galaxy center, while \textsc{Enzo} (very narrow outflow) and \textsc{Gizmo} (weak outflow) do not. Therefore, in the \textit{High-z Run}, the CGM region of the former three codes has higher metallicity, the rest show less enrichment (\textsc{Enzo}), or are concentrated in the center (\textsc{Gizmo}). \textsc{Gadget-4}, which is newly added, shows the most efficient metal enrichment in the CGM. In this study, we do not examine the inflow and outflow characteristics of the \textit{High-z Run} galaxies, this analysis will be conducted in the future \textit{AGORA} project. These results highlight how differences in the implementation of the star formation and feedback models affect the metal enrichment processes.

Metallicity is an important tracer for estimating galaxy properties such as stellar masses, ages, star formation rates, and redshifts. Figure \ref{fig:metallicity} illustrates metallicity histories in the \textit{High-z Run} as a function of redshift. We follow the definition of galaxy metallicity from \cite{2020MNRAS.494.1988L}, calculating it as the mass-weighted average metallicity of star particles younger than 100 Myr within the virial radius ($R_{\mathrm{vir}}$), under the assumption that these particles share the metallicity of unresolved nebular regions, allowing comparison of our results with observational data. Generally, the metallicity evolution follows the star formation history of each halo, although it shows larger inter-code differences. For example, while \textsc{Gadget-3} exhibits efficient metal diffusion at $z=10$ (as shown in Figure \ref{fig:metallicity_projection}), but its diffusion is weak at $z\ {\geq}\ 13$ because it does not implement an explicit metal diffusion scheme. Metals spread only via kernel-weighted injection in supernova events, and at very high redshift, metal enrichment is local and the turbulent mixing time can exceed the collapse time, resulting in a distinct plateau around a metallicity floor of $\sim$ $10^{-4}{\ \rm Z}_{\odot}$. In contrast, \textsc{Ramses} consistently produces higher metallicities at earlier epochs due to its efficient star formation. The other codes (\textsc{Enzo, Changa, Gadget-4}, and \textsc{Gizmo}) broadly show agreement, although \textsc{Gizmo} exhibits a rapid increase at $z\ {\leq}\ 11$ in more massive halos.

Observational metallicity data from JWST spectroscopy \citep{2023A&A...677A..88B, 2023NatAs...7..622C, 2023ApJ...957L..34W,  2024ApJ...973....8H, 2024ApJ...972..143C, 2025A&A...696A..87C, 2025ApJ...992..212W} are also included for comparison in bottom right panel as black dots with the red mean line and the minimum-maximum gray shaded region of Halo 5. Interestingly, \textsc{Ramses} is exactly in line with the observation data at $z\sim13$. We cannot directly compare the \textit{High-z Run} result and observation since the halo mass of the observation dataset is not known, but this result suggests the possibility of reproducibility of the current galaxy formation model to high-redshift galaxies. At the same time, the rest of \textit{High-z Run} galaxies tend to underpredict metallicity in this regime but generally show good agreement with JWST observations at $z \leq 12$, following similar trends to the star formation histories.

The stellar mass–metallicity relation is known to have a strong positive correlation between stellar mass and metallicity, indicating that more massive galaxies tend to be more metal-enriched \citep{2004ApJ...613..898T}. Coupled with the stellar mass, Figure \ref{fig:mzr} presents the MZR for Halo 5 in the \textit{High-z Run}, compared with JWST spectroscopic measurements and gas-phase MZR of previous cosmological simulations at z = 10, FIRE-2 \citep[gray dotted line,][]{2024ApJ...967L..41M} and FLARE \citep[gray dashed line,][]{2023MNRAS.519.3118W}.  The participating codes show a broad divergence at the low-mass end ($M_\star \leq 10^7 M_\odot$), mesh-based codes tend to have higher metallicity than particle-based codes at the low-mass end. However, \textit{High-z Run} galaxies have much better convergence above $10^8 \ \rm{M}_\odot$. Although not shown here, this trend is also observed in the other halos and in Paper V. The convergence at higher mass suggests that the dominant chemical enrichment pathways are robust across different simulation methods, while low-mass galaxies remain more susceptible to code-specific differences in the metal diffusion scheme. Compared with previous simulations, the \textit{High-z Run} galaxies have lower metallicity at the low-mass end and a steeper slope, which comes from different metal enrichment of stellar-phase and gas-phase metallicity at the early stage of galaxy evolution \citep[][]{2018MNRAS.479.1180M}. However, at the high-mass end where metal enrichment is sufficiently processed, the differences get smaller and both the \textit{High-z Run} and previous studies show good agreement with JWST spectroscopic observations.

\begin{figure*}
    \vspace{0mm}
    \centering
    \includegraphics[width = 1\linewidth]{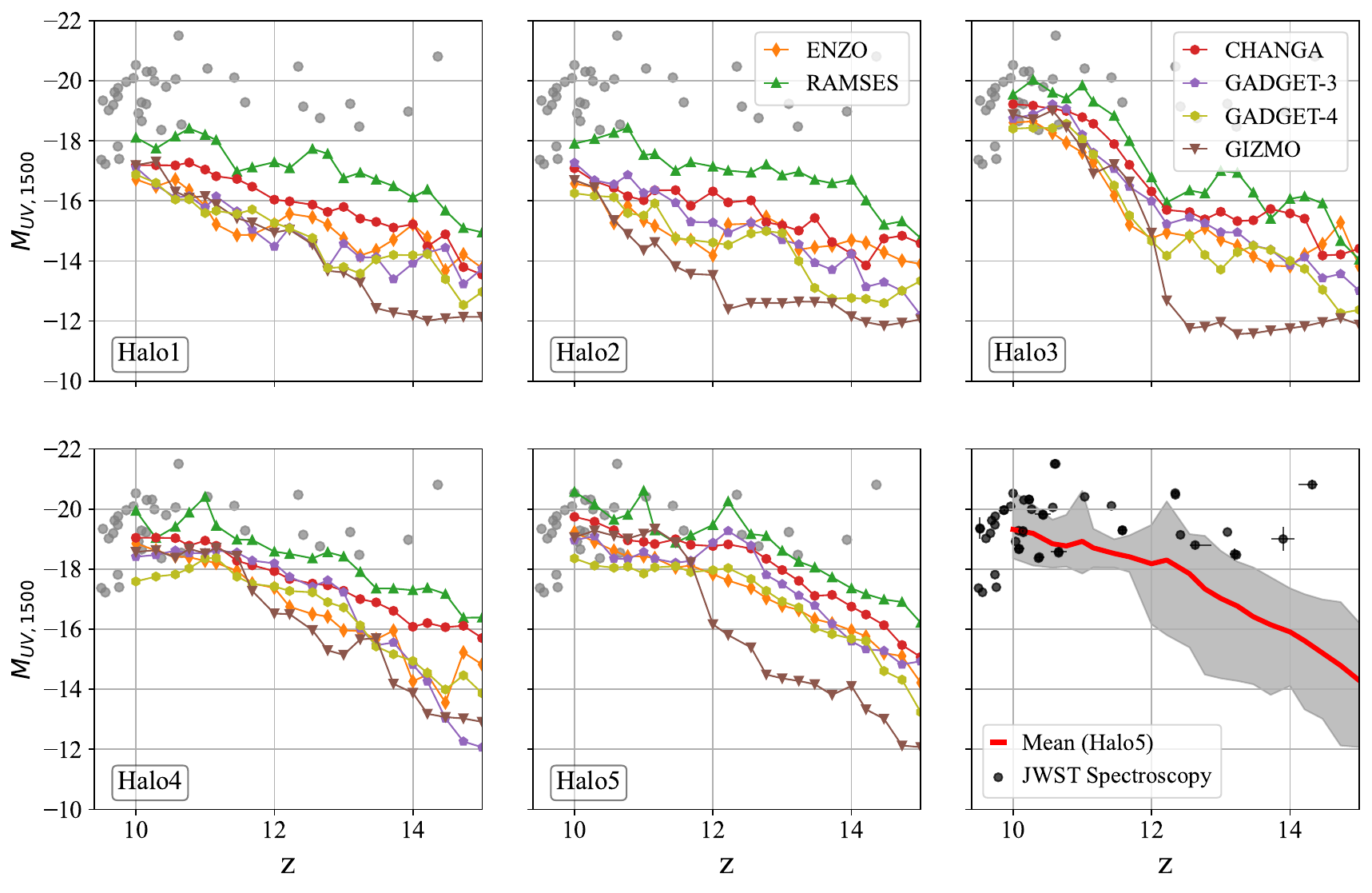}
    \vspace{-5mm}
    \caption{The evolution of UV magnitudes of the \textit{High-z Run} galaxies as a function of the redshift. The UV magnitudes are estimated using the results from the dust-radiative transfer code, SKIRT. To compare with observations, in the bottom right panel we present the mean value (thick red line) and the minimum–maximum range (gray shaded region) of Halo 5 and the JWST observation data --- black dots with error bars: \cite{2023NatAs...7..611R, 2023A&A...677A..88B, 2024Natur.633..318C, 2024ApJ...972..143C, 2025A&A...693A..50N}, black dots without error bars: \cite{2025ApJ...980..138H}. The observation data points are replicated in other panels as translucent gray dots. See Section \ref{sec:mock} for more information.}
    \vspace{2mm}
    \label{fig:luminosity}
\end{figure*}

\subsection{Mock JWST Image and UV luminosity}\label{sec:mock}
Using SKIRT, we generate the mock observation data of simulated galaxies in the \textit{High-z Run}.
Figure \ref{fig:mock} presents mock JWST NIRCam images of Halo 5 at $z=10$, generated using the F277W (R), F150W (G), and F115W (B) filters as an RGB composite. The upper panels display images generated from star particles only, while the middle panels additionally include the effects of gas and dust within the virial radius. The lower panels show the projected dust density used in SKIRT. Since we ignore the flux shorter than the Lyman-$\alpha$ wavelength ($\sim1.3 \ {\mu m}$), the color of galaxies appears yellow. Each pixel corresponds to 17 pc, equivalent to an angular size of about 0.004" at $z=10$ under the same cosmology adopted in the simulations. This resolution is about eight times better than the actual JWST NIRCam resolution (0.031"). Images degraded to the JWST resolution, along with clearly visible dust lanes in the rest frame, are presented in the Appendix \ref{sec:mock_compare}. Since the \textit{High-z Run} galaxies do not exhibit a well-defined disk structure, we select the x-axis of the simulation box as the viewing angle. The effect of the medium is evident. While the upper panels illustrate only the stellar distribution, the middle panels display pronounced dark dust lanes resulting from attenuation. The degree of attenuation differs across codes, reflecting variations in the dust density and spatial distribution. For instance, \textsc{Gadget-4} exhibits a highly concentrated dust core that significantly dims the central region, whereas \textsc{Ramses} shows weaker attenuation due to its relatively low dust content in the center. Since we impose a temperature threshold for generating dust in SKIRT, hot gas elements in \textsc{Ramses} are excluded and produce a cavity-like feature. As dust distribution depends on the gas density, metallicity and temperature (Section 2.3), this result further highlights the sensitivity of dust physics to subgrid prescriptions in different simulation codes.

Figure~\ref{fig:luminosity} shows the rest-frame UV magnitude at 1500~\AA{} as a function of redshift, based on our mock observations.  The UV flux $f_{\rm UV, 1500}$, averaged over the three viewing axes (x, y, z), is converted to absolute magnitude using the relation $M_{\rm UV, 1500} = -2.5 \log f_{\rm UV, 1500} - 48.6$ \citep{1983ApJ...266..713O}, and we confirm that the orientation does not make a big difference in flux. Each panel presents the evolution of UV magnitude for individual halos from $z=15$ to $z=10$. The overall trends closely follow those seen in Figure~\ref{fig:stellar_mass}. \textsc{Ramses} consistently produces brighter magnitudes than the other codes, while \textsc{Gizmo} shows a sharp increase in luminosity following a merger event. To verify the SKIRT results, we compare the UV luminosity histories to the stellar mass histories using the empirical $ \log_{10}(\rm M_{\star}/M_{\odot})$ – $M_{\rm UV}$ relation at high redshift \citep{2019MNRAS.486.3805B}. The best-fit slope from the \textit{High-z Run} is -0.3, which is a similar value of -0.4 in \cite{2019MNRAS.486.3805B}. Although the relation does not perfectly match the empirical relation, the UV magnitudes from SKIRT simulations show good agreement with the observed scaling relation.

\begin{figure*}
    \vspace{0mm}
    \centering
    \includegraphics[width = 1\linewidth]{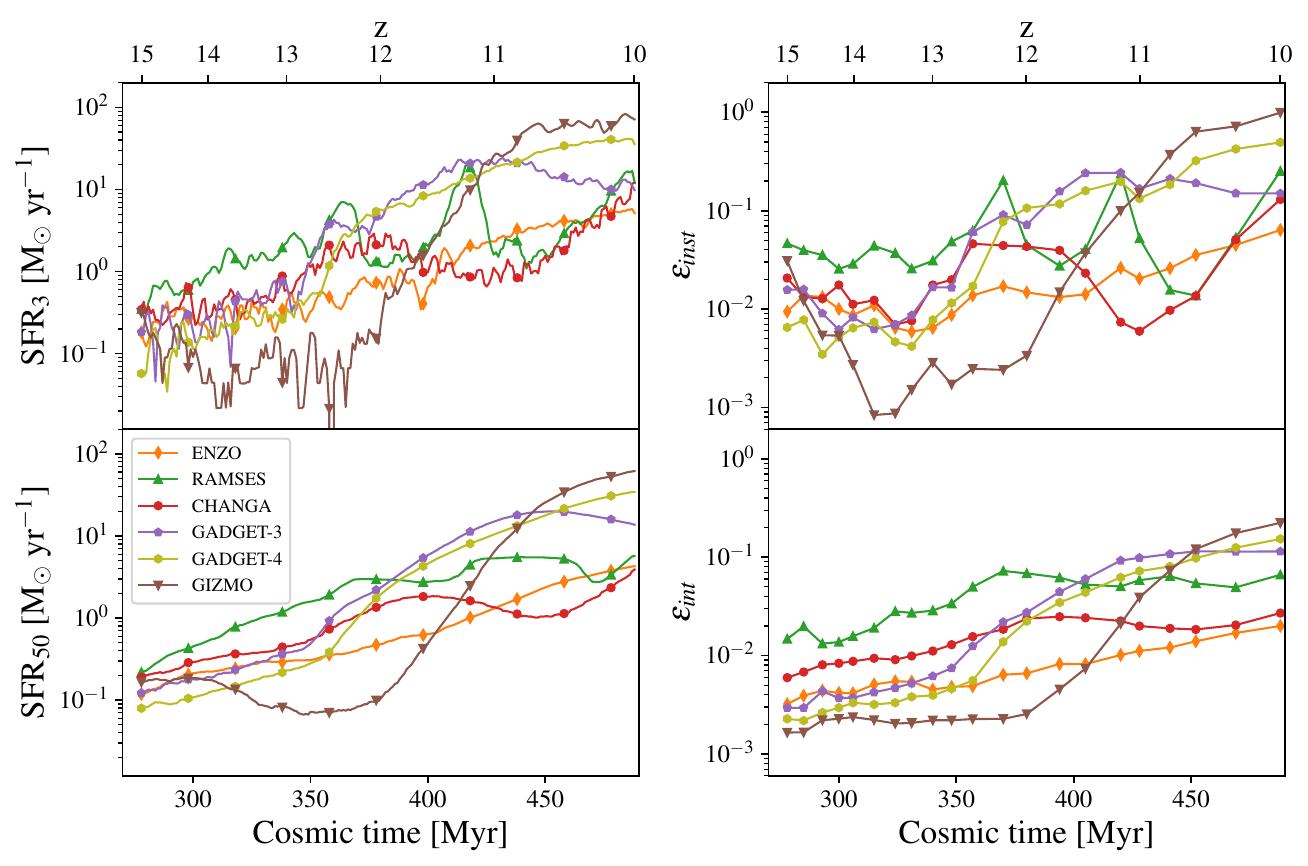}
    \vspace{-5mm}
\caption{\textit{Upper left panel}: Star formation rate (SFR) of Halo 5 as a function of time, averaged over 3 Myr. \textit{Bottom left panel}: Same as the upper left panel, but averaged over 50 Myr. \textit{Upper right panel}: Instantaneous star formation efficiency, $\epsilon_{\rm inst} =\dot{M}_\star/(\dot{M}_{\rm vir} f_{\rm b})$ at multiple redshifts. \textit{Bottom right panel}: Integrated star formation efficiency, $ \epsilon_{int} \equiv {M}_\star/({M}_{\rm vir} f_{\rm b})$. All codes exhibit an overall increasing trend in all panels. See Section \ref{sec:sfr} for more information.} 
    \vspace{2mm}
    \label{fig:SFR}
\end{figure*}

For observational comparison, we include black and gray dots representing spectroscopically confirmed galaxies from JWST observations \citep{2023NatAs...7..611R, 2023A&A...677A..88B, 2024Natur.633..318C, 2024ApJ...972..143C, 2025A&A...693A..50N, 2025ApJ...980..138H}. The red mean line and gray minimum-maximum region of Halo 5 are also included, as in Figures \ref{fig:stellar_mass} and \ref{fig:metallicity}. Our analysis indicates that a halo mass exceeding $5 \times 10^{10} {\ \rm M}_{\odot}$ is generally required to match the observed UV brightness. None of the codes in Halos 1 and 2 reproduce the luminosity of observed galaxies, whereas Halos 3, 4, and 5 partially achieve comparable UV magnitudes.

\section{DISCUSSION}

\subsection{Star Formation at Cosmic Dawn}\label{sec:sfr}
In this section, we investigate star formation in the \textit{High-z Run} using the most massive target halo.
The upper left panel of Figure \ref{fig:SFR} presents the $\rm SFR_3$ of Halo 5 as a function of cosmic time, star formation rate averaged over 3 Myr to capture the rapid variability of the SFR. All participating codes exhibit fluctuations on timescales from a few to tens of Myr, showing bursty star formation. The bottom left panel shows the same SFRs averaged over 50~Myr ($\rm SFR_{50}$), comparable to observationally inferred timescales. This smoothing reduces short-term burstiness and inter-code scatter, indicating that bursty star formation features are not easily captured from observation.

The systematic differences among codes are also evident. Particle-based codes generally show higher SFRs than mesh-based codes, reflecting differences in the underlying numerical solvers. When the Jeans length becomes unresolved, particle-based codes allow gas to collapse to much higher densities, whereas mesh-based codes tend to stabilize collapse earlier by enforcing a minimum cell size \citep{2023ApJ...950..132H}. Also, particle-based codes are effective for shock capturing, but they could broaden contact discontinuities and smear pressure gradients, allowing cold and dense gas to remain coherent and form long-lived clumps. By contrast, mesh codes with Riemann solvers naturally resolve shocks and shear flows more sharply, leading to faster clump disruption. As a consequence, illustrated in Figure \ref{fig:pdf}, particle-based codes reach gas number densities of $n_{\rm H} \sim  3000{\ \rm cm}^{-3}$, with some even exceeding $10000{\rm \ cm}^{-3}$, potentially driving the bursty star formation characteristic of Cosmic Dawn. Even within the same hydrodynamic solver, differences emerge. \textsc{Enzo} exhibits a relatively smooth rise in SFR, while \textsc{Ramses} shows stronger fluctuations, likely due to its feedback prescription. \textsc{Ramses} employs a delayed-cooling model, which suppresses star formation more effectively on short timescales than purely thermal injection schemes by generating a strong dense outflow \citep{2017MNRAS.466...11R}.

The growth of galaxies is mainly dominated by the mass growth of the host dark matter halo. In the right panel, we compute two star formation efficiencies, instantaneous star formation efficiency $\epsilon_{\rm inst} = \dot{M}_{\rm\star}/(\dot{M}_{\rm vir} f_{\rm b})$ and integrated star formation efficiency $\epsilon_{\rm int} \equiv {M}_\star/({M}_{\rm vir} f_{\rm b})$, where $f_b=0.165$ is the baryon fraction of the universe adopted in the \textit{High-z Run}. The definitions capture complementary aspects of the baryon-to-star conversion process. The instantaneous SFE traces the ratio of current star formation rate to the halo accretion rate, whereas the integrated SFE quantifies the cumulative efficiency of assembling the stellar mass relative to the integrated baryon inflow. 

The overall behavior mirrors the SFR histories as the halo mass growth is similar across the codes. Since $\epsilon_{\rm inst}$ responds to the instantaneous balance of gas supply and feedback, it can reach higher values and exhibit larger dispersion than $\epsilon_{\rm int}$. Compared with the FirstLight \citep{2024A&A...689A.244C} simulation, the value and scatter of $\epsilon_{\rm inst}$ in the \textit{High-z Run} are comparable. In contrast, the scatter of $\epsilon_{\rm int}$ is much smaller in FirstLight but remains large in the \textit{High-z Run}. This difference reflects the single feedback model used in FirstLight, which regulates star formation across the galaxy population toward similar cumulative efficiencies, whereas the \textit{High-z Run} compares multiple codes with distinct feedback implementations. The diversity in feedback prescriptions leads to a broader range of $\epsilon_{\rm int}$, indicating that star formation in the high-redshift is very sensitive to the feedback description. In our setup, we impose a fixed star formation efficiency of 1\%. \textsc{Enzo} and \textsc{Changa} yield $\epsilon_{\rm int}$ values consistent with this assumption, but the other codes have substantially higher efficiencies. 

\textsc{Gizmo} reaches $\epsilon_{\rm inst}\sim 1$ at $z=10$, corresponding to a stellar mass of $3\times10^{9}\ {\rm M}_\odot$ and gas densities of $\sim 10000 \ {\rm cm}^{-3}$. Such extreme values resemble the predictions of \citet{2023MNRAS.523.3201D}, suggesting that feedback regulation in \textsc{Gizmo} is inefficient under these conditions. Indeed, \citet{2021MNRAS.506.5512F} shows that feedback can be inefficient in compact and massive clouds, resulting in a very high SFE. However, it should be emphasized that the spatial resolution of particle-based runs is $\sim 80$ pc, which is insufficient to resolve and examine the internal structure of galaxies to verify the FFB model or other mechanism for the \textit{High-z Run}. In the future, a more extensive analysis with a higher resolution simulation will be conducted to investigate the formation of high-redshift galaxies.

\begin{figure}[t] 
    \centering
    \includegraphics[width=\columnwidth]{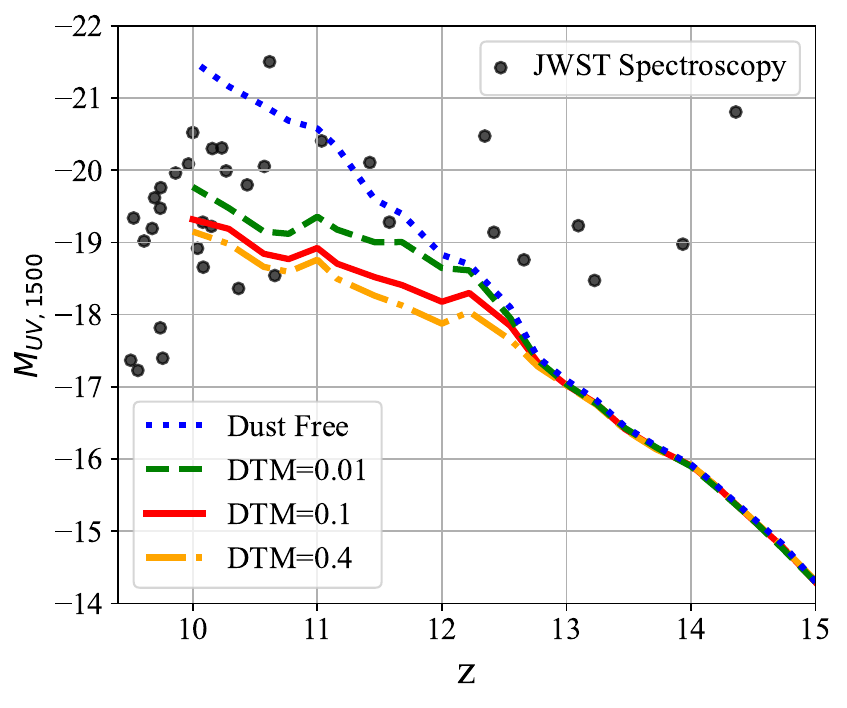}
    \caption{Same as the bottom right panel of Figure \ref{fig:luminosity}, but with different dust-to-metal ratios (DTM ratio= 0.01, 0.1, and 0.4) and a dust-free model. Variations in the DTM ratio produce mild differences, whereas the absence of dust decreases the UV magnitude by nearly 2 magnitudes at $z=10$. See Section \ref{sec:dust} for more information.}
    \label{fig:dtm}
\end{figure}

\subsection{Effect of Dust Attenuation}\label{sec:dust}
Dust physics is especially uncertain above the reionization era due to the faint signal of high-redshift galaxies and limited wavelength. While recent studies \citep[e.g.,][]{2025MNRAS.540.3693S} have attempted to constrain the dust properties of high-redshift galaxies, there are still ongoing degeneracies with dust temperature, which makes it difficult to estimate the dust mass. 

To investigate this uncertainty, we vary the DTM ratio over a range of 0.01, 0.1, 0.4, and a dust-free (only consider star particles) model using SKIRT. Figure \ref{fig:dtm} illustrates the effect of dust attenuation on the average of the rest-frame UV magnitude of Halo 5 in the \textit{High-z Run}, compared against JWST spectroscopy data. At $z > 13$, all models converge closely, indicating that dust might have only a minor effect on the brightest high-redshift galaxies similar to the ones studied here. Since the dust density is proportional to the gas density and metallicity, the dust mass in this regime is very small compared to the stellar mass due to low density and metallicity.\footnote{\textsc{Ramses} exhibits a very high metallicity due to active star formation even at $z>13$, but at the same time, stellar feedback lowers the gas density, limiting the dust mass.} However, after $z \sim 13$, the dust-free model clearly increases, producing UV magnitudes nearly two magnitudes brighter than dust-attenuated cases, while the three dusty models remain relatively close to each other. 

The comparison highlights that the presence of dust, rather than the precise DTM ratio, is the dominant factor in suppression of UV luminosity. Even the DTM ratio of 0.01 reduces the UV brightness significantly compared to the dust-free case, while increasing the DTM ratio to 0.4 only modestly enhances attenuation. This suggests that uncertainties in the exact efficiency of dust formation at these redshifts may be less important than whether dust is present at all. Observationally, the JWST spectroscopic points lie between the dust-free and dusty models, showing that dust extinction is already non-negligible at $z \sim 10-12$, although its precise abundance remains difficult to constrain. The sensitivity to dust, however, differs across codes. While varying the DTM ratio between 0.01 and 0.4 produces a small change in most cases, the contrast between dusty and dust-free models varies considerably. Due to the large reservoir of cold and dense gas in the center of particle-based codes, they also produce more dust mass, which leads to stronger attenuation than in mesh-based codes. \textsc{Gizmo} shows the largest differences between the DTM ratio = 0.1 model and the dust-free model at $z=10$, \textsc{Ramses} has the smallest difference. Differences of all participating codes are provided in Appendix \ref{sec:dtm_ratio}.

\subsection{Limitations \& Caveats} \label{sec:limits}
Despite its success in reproducing some JWST observation data, the \textit{High-z Run} suite has several limitations.

\begin{itemize}[itemsep=1\baselineskip]

\item 
Our target resolution of a few tens of parsecs does not resolve the internal structure of galaxies such as giant molecular clouds, which have a scale length of the Jeans length $\lambda_J=c_s/\sqrt{G\rho/\pi}$.  
For gas at the threshold density for the FFB scenario (\(\sim 3\times10^{3}\,{\rm cm^{-3}}\)) and \(T \sim 300\,{\rm K}\), the Jeans length falls below 10 pc and this is well beneath our resolution. Consequently, we cannot directly capture the fragmentation of dense clumps or validate small-scale star formation conditions. Instead, we can only test whether our simulation results with coarse resolution is broadly consistent with the proposed galaxy/star formation scenarios.

\item 
Recent studies strongly suggest the presence of AGN even in galaxies at \(z \sim 10\) \citep[e.g.,][]{2025A&A...693A..50N}. The \textit{High-z Run}, however, does not include black hole seeding or AGN feedback processes. As mentioned before, AGN may not dominate the UV luminosity budget \citep[e.g.,][]{2023ApJ...951...72O,2024MNRAS.529.3563T}, but their mechanical and radiative feedback could have significant effects on star formation, CGM heating, and the observed spectral energy distribution (SED). A more complete model of galaxy formation at high redshift should therefore include AGN physics.

\item 
We do not follow primordial non-equilibrium $\rm H, H_2$, and He chemistry. Instead, we impose a metallicity floor of $10^{-4}\ Z_{\odot}$ to approximate early enrichment from Pop III stars. Because our target halos already reach \(M_{\rm vir} \sim 10^{9}\,M_\odot\) by \(z=15\), cooling is expected to be dominated by atomic processes rather than molecular hydrogen \citep{2003ApJ...592..645Y}, we therefore expect this choice to have little impact on the overall galaxy properties. However, primordial chemistry can be significant in low-metallicity gas at Cosmic Dawn, potentially altering the thermal balance, fragmentation, and star formation in the earliest systems.

\item 
Finally, simulated galaxies should be compared only with observed systems whose halo masses match those of our simulated halos at the same redshift. However, because most high-redshift observational data lack any halo mass estimates, we can only suggest that the \textit{High-z Run} galaxies fall broadly within the range of observed properties. To fully and accurately validate our results, future work will require refined semi-analytic models and dedicated survey data that probes dynamical tracers of halo masses.

\end{itemize}

\section{Summary and Conclusion}\label{sec:conclusion}
In this paper, we presented results from the \textit{AGORA High-z Run}, a suite of cosmological zoom-in simulations targeting massive galaxies at Cosmic Dawn ($z\ge10$). Using six state-of-the-art hydrodynamical codes, we study inter-code agreements and differences in the high-redshift environment, and compare the simulated galaxies against JWST observation data. Our main findings are summarized as follows:

\begin{itemize}[itemsep=1\baselineskip]

  \item 
  Despite differences in feedback implementations among the simulation codes, the stellar mass histories exhibit a broad convergence across codes (Figure \ref{fig:stellar_mass}). In contrast, the metallicity evolution shows substantial inter-code variation, reflecting the diversity in feedback prescriptions (Figure \ref{fig:metallicity}). The mass–metallicity relation displays a remarkably tight positive correlation in most codes (Figure \ref{fig:mzr}), demonstrating the overall robustness of galaxy formation models irrespective of the specific numerical schemes adopted.

  \item 
  Without additional subgrid physics, the \textit{High-z Run} galaxies in massive halos succeed in reproducing galactic properties (Figures \ref{fig:stellar_mass},  \ref{fig:metallicity} and  \ref{fig:luminosity}) that are comparable to JWST observation data at $z\leq12$. Although the \textit{High-z Run} galaxies underpredict galaxy properties $z\sim13-14$, these results suggest that ``vanilla'' $\Lambda$CDM galaxy formation framework can sufficiently account for galaxies at Cosmic Dawn if the host halo is massive enough.

  \item 
  Some galaxies in the \textit{High-z Run} exhibit the signatures that are broadly consistent with the feedback-free starburst model ($n_{\rm H}>3000~\rm cm^{-3}, ~\rm SFR \sim 60 ~M_{\odot}/yr, ~\epsilon_{inst}>0.2$; Figures \ref{fig:pdf} and \ref{fig:SFR}). Although we impose a fixed local star formation efficiency of 1\% in all our simulations, these galaxies show significantly higher SFE, reflecting the rapid conversion of gas into stars in dense environments. However, we cannot readily verify this scenario since the resolution of the \textit{High-z Run} is not sufficient to resolve internal galactic structure. 

  \item The presence of dust plays a crucial role in determining the UV luminosities of the simulated galaxies in the \textit{High-z Run}. At $z > 13$, the difference between dust-free and dust-included models remains minimal, suggesting that dust has little impact on the earliest galaxies. However, at later times, the dust-free model predicts UV luminosities up to $\sim2$ magnitudes brighter than the dust-included cases, highlighting the strong attenuation effect once dust exists. In contrast, variations in the dust-to-metal ratio introduce only a minor difference, indicating that the key factor is the presence of dust itself rather than its precise abundance (Figure \ref{fig:dtm}).  
\end{itemize}

The \textit{AGORA High-z Run} is the first step toward understanding the formation and evolution of galaxies at Cosmic Dawn, as well as the effect of different gravity and hydrodynamics schemes in the high-redshift environment. Despite its limitations such as insufficient resolution and simplified subgrid physics, this suite of simulations offers valuable insights that pave the way for more refined, computationally expensive future simulations. Indeed, we plan to carry out the new ``\textit{High-res High-z Run}'' suite with higher spatial resolution (effective resolution of $\leq 10$ pc) and sophisticated subgrid physics.  In the \textit{High-z Run}, we are not able to measure some observational properties (e.g., galaxy size) due to limited resolution. However, the \textit{High-res High-z Run} will allow us to investigate physical processes in greater detail and compare a wider range of observational properties with JWST, enabling a more realistic exploration of galaxy formation in the early universe.

\vspace{4.5mm}

J.-H.K.’s work was supported by the National Research Foundation of Korea (NRF) grant funded by the Korea government (MSIT) (No. 2022M3K3A1093827 and No. 2023R1A2C1003244). His work was also supported by the National Institute of Supercomputing and Network/Korea Institute of Science and Technology Information with supercomputing resources including technical support, grants KSC-2021-CRE-0442, KSC-2022-CRE-0355 and KSC-2024-CRE-0232. His work was also supported by the GlobalLAMP Program of the NRF grant funded by the Ministry of Education (No. RS-2023-00301976). D.C. is supported by research grant PID2021-122603NB-C21 funded by the Ministerio de Ciencia, Innovaci\'{o}n y Universidades (MI-CIU/FEDER) and the research grant CNS2024-154550 funded by MI-CIU/AEI/10.13039/501100011033. \textsc{Gadget-4} simulations were performed using the SQUID at the D3 Center, the University of Osaka, through the HPCI System Research Project (hp230089, hp240141, hp250119). KN is partly supported by the MEXT/JSPS KAKENHI Grant Numbers JP20H00180, JP22K21349, 24H00002, 24H00241 and 25K01032 (K.N.). HV was supported by DGAPA-PAPIIT-UNAM project IN111425. \textsc{ChaNGa} runs were carried out in the CNS-IPICYT-SECIHTI, LNS-SECIHTI, and LAMOD-UNAM. R.R.C. acknowledges financial support from the Spanish Ministry of Science and Innovation through the research grant PID2021-123417OB-I00, funded by MCIN/AEI/10.13039/501100011033/FEDER, EU. and from IND2022/TIC-23643 project funded by Comunidad de Madrid.

\vspace{5mm}

\bibliography{main}{}

@ARTICLE{2025A&A...704A.158C,
       author = {{Castellano}, M. and {Fontana}, A. and {Merlin}, E. and {Santini}, P. and {Napolitano}, L. and {Menci}, N. and {P{\'e}rez-Gonz{\'a}lez}, P.~G. and {Calabr{\`o}}, A. and {Paris}, D. and {Pentericci}, L. and {Zavala}, J.~A. and {Dickinson}, M. and {Finkelstein}, S.~L. and {Treu}, T. and {Amorin}, R.~O. and {Arrabal Haro}, P. and {Bergamini}, P. and {Bisigello}, L. and {Catone}, M. and {Daddi}, E. and {Dayal}, P. and {Dekel}, A. and {Ferrara}, A. and {Fortuni}, F. and {Gandolfi}, G. and {Giavalisco}, M. and {Grillo}, C. and {Guida}, S.~T. and {Hathi}, N.~P. and {Holwerda}, B.~W. and {Koekemoer}, A.~M. and {Kokorev}, V. and {Li}, Z. and {Llerena}, M. and {Lucas}, R.~A. and {Mascia}, S. and {Metha}, B. and {Morishita}, T. and {Nanayakkara}, T. and {Pacucci}, F. and {Roberts-Borsani}, G. and {Rodighiero}, G. and {Rosati}, P. and {Salazar}, V. and {Schneider}, R. and {Somerville}, R.~S. and {Taylor}, A. and {Trenti}, M. and {Trinca}, A. and {Wang}, X. and {Watson}, P.~J. and {Yang}, L. and {Yung}, L.~Y.~A.},
        title = "{Pushing JWST to the extremes: Search and scrutiny of bright galaxy candidates at z ≃ 15─30}",
      journal = {\aap},
         year = 2025,
        month = dec,
       volume = {704},
          eid = {A158},
        pages = {A158},
          doi = {10.1051/0004-6361/202555082},
archivePrefix = {arXiv},
       eprint = {2504.05893},
 primaryClass = {astro-ph.GA},
       adsurl = {https://ui.adsabs.harvard.edu/abs/2025A&A...704A.158C}
}

@ARTICLE{2025ApJ...994..245J,
       author = {{Jung}, Minyong and {Kim}, Ji-Hoon and {Nguy\~{\^e}n}, Th\`inh{\~{u}}'u. and {Rodr{\'\i}guez-Cardoso}, Ram{\'o}n and {Roca-F{\`a}brega}, Santi and {Primack}, Joel R. and {Barrow}, Kirk S.~S. and {Genina}, Anna and {Granizo}, Pablo and {Kim}, Hyeonyong and {Nagamine}, Kentaro and {Oku}, Yuri and {Powell}, Johnny W. and {Revaz}, Yves and {Vel{\'a}zquez}, H{\'e}ctor and {Lupi}, Alessandro and {Shimizu}, Ikkoh and {Abel}, Tom and {Agertz}, Oscar and {Cen}, Renyue and {Ceverino}, Daniel and {Dekel}, Avishai and {Jeong}, Chaerin and {Mayer}, Lucio and {Oh}, Boon Kiat and {Quinn}, Thomas R. and {Song}, Hyunmi and {Agora Collaboration}},
        title = "{The AGORA High-resolution Galaxy Simulations Comparison Project. VIII. Disk Formation and Evolution of Simulated Milky Way Mass Galaxy Progenitors at 1 < z < 5}",
      journal = {\apj},
         year = 2025,
        month = dec,
       volume = {994},
       number = {2},
          eid = {245},
        pages = {245},
          doi = {10.3847/1538-4357/ae112d},
archivePrefix = {arXiv},
       eprint = {2505.05720},
 primaryClass = {astro-ph.GA},
       adsurl = {https://ui.adsabs.harvard.edu/abs/2025ApJ...994..245J}
}

@ARTICLE{2025MNRAS.543.2760V,
       author = {{van Donkelaar}, Floor and {Capelo}, Pedro R. and {Mayer}, Lucio and {Reed}, Darren S. and {Quinn}, Thomas R.},
        title = "{Introducing the PHOEBOS simulation: galaxy properties at the dawn of galaxy formation}",
      journal = {\mnras},
         year = 2025,
        month = nov,
       volume = {543},
       number = {3},
        pages = {2760-2780},
          doi = {10.1093/mnras/staf1638},
archivePrefix = {arXiv},
       eprint = {2507.04927},
 primaryClass = {astro-ph.GA},
       adsurl = {https://ui.adsabs.harvard.edu/abs/2025MNRAS.543.2760V}
}

@ARTICLE{2025ApJ...991..179P,
       author = {{P{\'e}rez-Gonz{\'a}lez}, Pablo G. and {{\"O}stlin}, G{\"o}ran and {Costantin}, Luca and {Melinder}, Jens and {Finkelstein}, Steven L. and {Somerville}, Rachel S. and {Annunziatella}, Marianna and {{\'A}lvarez-M{\'a}rquez}, Javier and {Colina}, Luis and {Dekel}, Avishai and {Ferguson}, Henry C. and {Li}, Zhaozhou and {Yung}, L.~Y. Aaron and {Bagley}, Micaela B. and {Boogaard}, Leindert A. and {Burgarella}, Denis and {Calabr{\`o}}, Antonello and {Caputi}, Karina I. and {Cheng}, Yingjie and {Dickinson}, Mark and {Eckart}, Andreas and {Giavalisco}, Mauro and {Gillman}, Steven and {Greve}, Thomas R. and {Hamed}, Mahmoud and {Hathi}, Nimish P. and {Hjorth}, Jens and {Huertas-Company}, Marc and {Kartaltepe}, Jeyhan S. and {Koekemoer}, Anton M. and {Kokorev}, Vasily and {Labiano}, {\'A}lvaro and {Langeroodi}, Danial and {Leung}, Gene C.~K. and {Natarajan}, Priyamvada and {Papovich}, Casey and {Peissker}, Florian and {Pentericci}, Laura and {Pirzkal}, Nor and {Rinaldi}, Pierluigi and {van der Werf}, Paul and {Walter}, Fabian},
        title = "{The Rise of the Galactic Empire: Ultraviolet Luminosity Functions at z {\ensuremath{\sim}} 17 and z {\ensuremath{\sim}} 25 Estimated with the MIDIS+NGDEEP Ultra-deep JWST/NIRCam Data Set}",
      journal = {\apj},
         year = 2025,
        month = oct,
       volume = {991},
       number = {2},
          eid = {179},
        pages = {179},
          doi = {10.3847/1538-4357/adf8c9},
archivePrefix = {arXiv},
       eprint = {2503.15594},
 primaryClass = {astro-ph.GA},
       adsurl = {https://ui.adsabs.harvard.edu/abs/2025ApJ...991..179P}
}

@ARTICLE{2025ApJ...992..212W,
       author = {{Wu}, Zihao and {Eisenstein}, Daniel J. and {Johnson}, Benjamin D. and {Jakobsen}, Peter and {Alberts}, Stacey and {Arribas}, Santiago and {Baker}, William M. and {Bunker}, Andrew J. and {Carniani}, Stefano and {Charlot}, St{\'e}phane and {Chevallard}, Jacopo and {Curti}, Mirko and {Curtis-Lake}, Emma and {D'Eugenio}, Francesco and {Hainline}, Kevin and {Helton}, Jakob M. and {Hsiao}, Tiger Yu-Yang and {Ji}, Xihan and {Ji}, Zhiyuan and {Looser}, Tobias J. and {Rieke}, George and {Rinaldi}, Pierluigi and {Robertson}, Brant and {Scholtz}, Jan and {Sun}, Fengwu and {Tacchella}, Sandro and {Trussler}, James A.~A. and {Williams}, Christina C. and {Willmer}, Christopher N.~A. and {Willott}, Chris and {Witstok}, Joris and {Zhu}, Yongda},
        title = "{JADES-GS-z14-1: A Compact, Faint Galaxy at z {\ensuremath{\approx}} 14 with Weak Metal Lines from Extremely Deep JWST MIRI, NIRCam, and NIRSpec Observations}",
      journal = {\apj},
         year = 2025,
        month = oct,
       volume = {992},
       number = {2},
          eid = {212},
        pages = {212},
          doi = {10.3847/1538-4357/ae01a1},
archivePrefix = {arXiv},
       eprint = {2507.22858},
 primaryClass = {astro-ph.GA},
       adsurl = {https://ui.adsabs.harvard.edu/abs/2025ApJ...992..212W}
}

@ARTICLE{2025ApJ...992...63W,
       author = {{Whitler}, Lily and {Stark}, Daniel P. and {Topping}, Michael W. and {Robertson}, Brant and {Rieke}, Marcia and {Hainline}, Kevin N. and {Endsley}, Ryan and {Chen}, Zuyi and {Baker}, William M. and {Bhatawdekar}, Rachana and {Bunker}, Andrew J. and {Carniani}, Stefano and {Charlot}, St{\'e}phane and {Chevallard}, Jacopo and {Curtis-Lake}, Emma and {Egami}, Eiichi and {Eisenstein}, Daniel J. and {Helton}, Jakob M. and {Ji}, Zhiyuan and {Johnson}, Benjamin D. and {P{\'e}rez-Gonz{\'a}lez}, Pablo G. and {Rinaldi}, Pierluigi and {Tacchella}, Sandro and {Williams}, Christina C. and {Willmer}, Christopher N.~A. and {Willott}, Chris and {Witstok}, Joris},
        title = "{The z {\ensuremath{\gtrsim}} 9 Galaxy UV Luminosity Function from the JWST Advanced Deep Extragalactic Survey: Insights into Early Galaxy Evolution and Reionization}",
      journal = {\apj},
         year = 2025,
        month = oct,
       volume = {992},
       number = {1},
          eid = {63},
        pages = {63},
          doi = {10.3847/1538-4357/adfddc},
archivePrefix = {arXiv},
       eprint = {2501.00984},
 primaryClass = {astro-ph.GA},
       adsurl = {https://ui.adsabs.harvard.edu/abs/2025ApJ...992...63W}
}

@ARTICLE{2025OJAp....8E.153K,
       author = {{Kannan}, Rahul and {Puchwein}, Ewald and {Smith}, Aaron and {Borrow}, Josh and {Garaldi}, Enrico and {Keating}, Laura and {Vogelsberger}, Mark and {Zier}, Oliver and {McClymont}, William and {Shen}, Xuejian and {Popovic}, Filip and {Tacchella}, Sandro and {Hernquist}, Lars and {Springel}, Volker},
        title = "{Introducing the THESAN-ZOOM project: radiation-hydrodynamic simulations of high-redshift galaxies with a multi-phase interstellar medium}",
      journal = {The Open Journal of Astrophysics},
         year = 2025,
        month = oct,
       volume = {8},
          eid = {153},
        pages = {153},
          doi = {10.33232/001c.145804},
archivePrefix = {arXiv},
       eprint = {2502.20437},
 primaryClass = {astro-ph.GA},
       adsurl = {https://ui.adsabs.harvard.edu/abs/2025OJAp....8E.153K}
}

@ARTICLE{2025MNRAS.540.3350A,
       author = {{Andalman}, Zachary L. and {Teyssier}, Romain and {Dekel}, Avishai},
        title = "{On the origin of the high star formation efficiency in massive galaxies at Cosmic Dawn}",
      journal = {\mnras},
         year = 2025,
        month = jul,
       volume = {540},
       number = {4},
        pages = {3350-3383},
          doi = {10.1093/mnras/staf930},
archivePrefix = {arXiv},
       eprint = {2410.20530},
 primaryClass = {astro-ph.GA},
       adsurl = {https://ui.adsabs.harvard.edu/abs/2025MNRAS.540.3350A}
}

@ARTICLE{2025MNRAS.540.3693S,
       author = {{Sommovigo}, Laura and {Algera}, Hiddo},
        title = "{Realistic multitemperature dust: how well can we constrain the dust properties of high-redshift galaxies?}",
      journal = {\mnras},
         year = 2025,
        month = jul,
       volume = {540},
       number = {4},
        pages = {3693-3708},
          doi = {10.1093/mnras/staf897},
archivePrefix = {arXiv},
       eprint = {2505.20105},
 primaryClass = {astro-ph.GA},
       adsurl = {https://ui.adsabs.harvard.edu/abs/2025MNRAS.540.3693S}
}

@ARTICLE{2025ApJ...988...19S,
       author = {{Schouws}, Sander and {Bouwens}, Rychard J. and {Ormerod}, Katherine and {Smit}, Renske and {Algera}, Hiddo and {Sommovigo}, Laura and {Hodge}, Jacqueline and {Ferrara}, Andrea and {Oesch}, Pascal A. and {Rowland}, Lucie E. and {van Leeuwen}, Ivana and {Stefanon}, Mauro and {Herard-Demanche}, Thomas and {Fudamoto}, Yoshinobu and {R{\"o}ttgering}, Huub and {van der Werf}, Paul},
        title = "{Detection of [O III]$_{88 {\ensuremath{\mu}}m}$ in JADES-GS-z14-0 at z = 14.1793}",
      journal = {\apj},
         year = 2025,
        month = jul,
       volume = {988},
       number = {1},
          eid = {19},
        pages = {19},
          doi = {10.3847/1538-4357/adbf1b},
archivePrefix = {arXiv},
       eprint = {2409.20549},
 primaryClass = {astro-ph.GA},
       adsurl = {https://ui.adsabs.harvard.edu/abs/2025ApJ...988...19S}
}

@ARTICLE{2025A&A...698A.303R,
       author = {{Rodr{\'\i}guez-Cardoso}, Ram{\'o}n and {Roca-F{\`a}brega}, Santi and {Jung}, Minyong and {Nguy\~{\^e}n}, Th\`inh. and {Kim}, Ji-Hoon and {Primack}, Joel and {Agertz}, Oscar and {Barrow}, Kirk S.~S. and {Gallego}, Jesus and {Nagamine}, Kentaro and {Powell}, Johnny W. and {Revaz}, Yves and {Vel{\'a}zquez}, Hector and {Genina}, Anna and {Kim}, Hyeonyong and {Lupi}, Alessandro and {Abel}, Tom and {Cen}, Renyue and {Ceverino}, Daniel and {Dekel}, Avishai and {Oh}, Boon Kiat and {Quinn}, Thomas R. and {The Agora Collaboration}},
        title = "{The AGORA High-Resolution Galaxy Simulations Comparison Project: VII. Satellite quenching in zoom-in simulation of a Milky Way-mass halo}",
      journal = {\aap},
         year = 2025,
        month = jun,
       volume = {698},
          eid = {A303},
        pages = {A303},
          doi = {10.1051/0004-6361/202453639},
archivePrefix = {arXiv},
       eprint = {2505.05844},
 primaryClass = {astro-ph.GA},
       adsurl = {https://ui.adsabs.harvard.edu/abs/2025A&A...698A.303R}
}

@ARTICLE{2025A&A...696A..87C,
       author = {{Carniani}, Stefano and {D'Eugenio}, Francesco and {Ji}, Xihan and {Parlanti}, Eleonora and {Scholtz}, Jan and {Sun}, Fengwu and {Venturi}, Giacomo and {Bakx}, Tom J.~L.~C. and {Curti}, Mirko and {Maiolino}, Roberto and {Tacchella}, Sandro and {Zavala}, Jorge A. and {Hainline}, Kevin and {Witstok}, Joris and {Johnson}, Benjamin D. and {Alberts}, Stacey and {Bunker}, Andrew J. and {Charlot}, St{\'e}phane and {Eisenstein}, Daniel J. and {Helton}, Jakob M. and {Jakobsen}, Peter and {Kumari}, Nimisha and {Robertson}, Brant and {Saxena}, Aayush and {{\"U}bler}, Hannah and {Williams}, Christina C. and {Willmer}, Christopher N.~A. and {Willott}, Chris},
        title = "{The eventful life of a luminous galaxy at z = 14: metal enrichment, feedback, and low gas fraction?}",
      journal = {\aap},
         year = 2025,
        month = apr,
       volume = {696},
          eid = {A87},
        pages = {A87},
          doi = {10.1051/0004-6361/202452451},
archivePrefix = {arXiv},
       eprint = {2409.20533},
 primaryClass = {astro-ph.GA},
       adsurl = {https://ui.adsabs.harvard.edu/abs/2025A&A...696A..87C}
}

@ARTICLE{2025ApJ...980...10J,
       author = {{Jeong}, Tae Bong and {Jeon}, Myoungwon and {Song}, Hyunmi and {Bromm}, Volker},
        title = "{Simulating High-redshift Galaxies: Enhancing UV Luminosity with Star Formation Efficiency and a Top-heavy IMF}",
      journal = {\apj},
         year = 2025,
        month = feb,
       volume = {980},
       number = {1},
          eid = {10},
        pages = {10},
          doi = {10.3847/1538-4357/ada27d},
archivePrefix = {arXiv},
       eprint = {2411.17007},
 primaryClass = {astro-ph.GA},
       adsurl = {https://ui.adsabs.harvard.edu/abs/2025ApJ...980...10J}
}

@ARTICLE{2025ApJ...980..138H,
       author = {{Harikane}, Yuichi and {Inoue}, Akio K. and {Ellis}, Richard S. and {Ouchi}, Masami and {Nakazato}, Yurina and {Yoshida}, Naoki and {Ono}, Yoshiaki and {Sun}, Fengwu and {Sato}, Riku A. and {Ferrami}, Giovanni and {Fujimoto}, Seiji and {Kashikawa}, Nobunari and {McLeod}, Derek J. and {P{\'e}rez-Gonz{\'a}lez}, Pablo G. and {Sawicki}, Marcin and {Sugahara}, Yuma and {Xu}, Yi and {Yamanaka}, Satoshi and {Carnall}, Adam C. and {Cullen}, Fergus and {Dunlop}, James S. and {Egami}, Eiichi and {Grogin}, Norman and {Isobe}, Yuki and {Koekemoer}, Anton M. and {Laporte}, Nicolas and {Lee}, Chien-Hsiu and {Magee}, Dan and {Matsuo}, Hiroshi and {Matsuoka}, Yoshiki and {Mawatari}, Ken and {Nakajima}, Kimihiko and {Nakane}, Minami and {Tamura}, Yoichi and {Umeda}, Hiroya and {Yanagisawa}, Hiroto},
        title = "{JWST, ALMA, and Keck Spectroscopic Constraints on the UV Luminosity Functions at z {\ensuremath{\sim}} 7─14: Clumpiness and Compactness of the Brightest Galaxies in the Early Universe}",
      journal = {\apj},
         year = 2025,
        month = feb,
       volume = {980},
       number = {1},
          eid = {138},
        pages = {138},
          doi = {10.3847/1538-4357/ad9b2c},
archivePrefix = {arXiv},
       eprint = {2406.18352},
 primaryClass = {astro-ph.GA},
       adsurl = {https://ui.adsabs.harvard.edu/abs/2025ApJ...980..138H}
}

@ARTICLE{2025A&A...694A.286F,
       author = {{Ferrara}, A. and {Pallottini}, A. and {Sommovigo}, L.},
        title = "{Blue monsters at z > 10: Where all their dust has gone}",
      journal = {\aap},
         year = 2025,
        month = feb,
       volume = {694},
          eid = {A286},
        pages = {A286},
          doi = {10.1051/0004-6361/202452707},
archivePrefix = {arXiv},
       eprint = {2410.19042},
 primaryClass = {astro-ph.GA},
       adsurl = {https://ui.adsabs.harvard.edu/abs/2025A&A...694A.286F}
}

@ARTICLE{2025A&A...693A..50N,
       author = {{Napolitano}, L. and {Castellano}, M. and {Pentericci}, L. and {Arrabal Haro}, P. and {Fontana}, A. and {Treu}, T. and {Bergamini}, P. and {Calabr{\`o}}, A. and {Mascia}, S. and {Morishita}, T. and {Roberts-Borsani}, G. and {Santini}, P. and {Vanzella}, E. and {Vulcani}, B. and {Zakharova}, D. and {Bakx}, T. and {Dickinson}, M. and {Grillo}, C. and {Leethochawalit}, N. and {Llerena}, M. and {Merlin}, E. and {Paris}, D. and {Rojas-Ruiz}, S. and {Rosati}, P. and {Wang}, X. and {Yoon}, I. and {Zavala}, J.},
        title = "{Seven wonders of Cosmic Dawn: JWST confirms a high abundance of galaxies and AGN at z ≃ 9─11 in the GLASS field}",
      journal = {\aap},
         year = 2025,
        month = jan,
       volume = {693},
          eid = {A50},
        pages = {A50},
          doi = {10.1051/0004-6361/202452090},
archivePrefix = {arXiv},
       eprint = {2410.10967},
 primaryClass = {astro-ph.GA},
       adsurl = {https://ui.adsabs.harvard.edu/abs/2025A&A...693A..50N}
}

@ARTICLE{2025MNRAS.536..988F,
       author = {{Feldmann}, Robert and {Boylan-Kolchin}, Michael and {Bullock}, James S. and {{\c{C}}atmabacak}, Onur and {Faucher-Gigu{\`e}re}, Claude-Andr{\'e} and {Hayward}, Christopher C. and {Kere{\v{s}}}, Du{\v{s}}an and {Lazar}, Alexandres and {Liang}, Lichen and {Moreno}, Jorge and {Oesch}, Pascal A. and {Quataert}, Eliot and {Shen}, Xuejian and {Sun}, Guochao},
        title = "{Elevated UV luminosity density at Cosmic Dawn explained by non-evolving, weakly mass-dependent star formation efficiency}",
      journal = {\mnras},
         year = 2025,
        month = jan,
       volume = {536},
       number = {1},
        pages = {988-1016},
          doi = {10.1093/mnras/stae2633},
archivePrefix = {arXiv},
       eprint = {2407.02674},
 primaryClass = {astro-ph.CO},
       adsurl = {https://ui.adsabs.harvard.edu/abs/2025MNRAS.536..988F}
}

@ARTICLE{2025A&A...693A.149K,
       author = {{Kang}, Cheonsu and {Kimm}, Taysun and {Han}, Daniel and {Katz}, Harley and {Devriendt}, Julien and {Slyz}, Adrianne and {Teyssier}, Romain},
        title = "{Impact of star formation models on the growth of simulated galaxies at high redshifts}",
      journal = {\aap},
         year = 2025,
        month = jan,
       volume = {693},
          eid = {A149},
        pages = {A149},
          doi = {10.1051/0004-6361/202451502},
archivePrefix = {arXiv},
       eprint = {2407.12090},
 primaryClass = {astro-ph.GA},
       adsurl = {https://ui.adsabs.harvard.edu/abs/2025A&A...693A.149K}
}

@ARTICLE{2024ApJ...975..183O,
       author = {{Oku}, Yuri and {Nagamine}, Kentaro},
        title = "{Osaka Feedback Model. III. Cosmological Simulation CROCODILE}",
      journal = {\apj},
         year = 2024,
        month = nov,
       volume = {975},
       number = {2},
          eid = {183},
        pages = {183},
          doi = {10.3847/1538-4357/ad77d3},
archivePrefix = {arXiv},
       eprint = {2401.06324},
 primaryClass = {astro-ph.GA},
       adsurl = {https://ui.adsabs.harvard.edu/abs/2024ApJ...975..183O}
}

@ARTICLE{2024A&A...690A.108L,
       author = {{Li}, Zhaozhou and {Dekel}, Avishai and {Sarkar}, Kartick C. and {Aung}, Han and {Giavalisco}, Mauro and {Mandelker}, Nir and {Tacchella}, Sandro},
        title = "{Feedback-free starbursts at cosmic dawn: Observable predictions for JWST}",
      journal = {\aap},
         year = 2024,
        month = oct,
       volume = {690},
          eid = {A108},
        pages = {A108},
          doi = {10.1051/0004-6361/202348727},
archivePrefix = {arXiv},
       eprint = {2311.14662},
 primaryClass = {astro-ph.GA},
       adsurl = {https://ui.adsabs.harvard.edu/abs/2024A&A...690A.108L}
}

@ARTICLE{2024ApJ...972..143C,
       author = {{Castellano}, Marco and {Napolitano}, Lorenzo and {Fontana}, Adriano and {Roberts-Borsani}, Guido and {Treu}, Tommaso and {Vanzella}, Eros and {Zavala}, Jorge A. and {Arrabal Haro}, Pablo and {Calabr{\`o}}, Antonello and {Llerena}, Mario and {Mascia}, Sara and {Merlin}, Emiliano and {Paris}, Diego and {Pentericci}, Laura and {Santini}, Paola and {Bakx}, Tom J.~L.~C. and {Bergamini}, Pietro and {Cupani}, Guido and {Dickinson}, Mark and {Filippenko}, Alexei V. and {Glazebrook}, Karl and {Grillo}, Claudio and {Kelly}, Patrick L. and {Malkan}, Matthew A. and {Mason}, Charlotte A. and {Morishita}, Takahiro and {Nanayakkara}, Themiya and {Rosati}, Piero and {Sani}, Eleonora and {Wang}, Xin and {Yoon}, Ilsang},
        title = "{JWST NIRSpec Spectroscopy of the Remarkable Bright Galaxy GHZ2/GLASS-z12 at Redshift 12.34}",
      journal = {\apj},
         year = 2024,
        month = sep,
       volume = {972},
       number = {2},
          eid = {143},
        pages = {143},
          doi = {10.3847/1538-4357/ad5f88},
archivePrefix = {arXiv},
       eprint = {2403.10238},
 primaryClass = {astro-ph.GA},
       adsurl = {https://ui.adsabs.harvard.edu/abs/2024ApJ...972..143C}
}

@ARTICLE{2024A&A...689A.244C,
       author = {{Ceverino}, D. and {Nakazato}, Y. and {Yoshida}, N. and {Klessen}, R.~S. and {Glover}, S.~C.~O.},
        title = "{Redshift-dependent galaxy formation efficiency at z = 5 {\ensuremath{-}} 13 in the FirstLight Simulations}",
      journal = {\aap},
         year = 2024,
        month = sep,
       volume = {689},
          eid = {A244},
        pages = {A244},
          doi = {10.1051/0004-6361/202450224},
archivePrefix = {arXiv},
       eprint = {2404.02537},
 primaryClass = {astro-ph.GA},
       adsurl = {https://ui.adsabs.harvard.edu/abs/2024A&A...689A.244C}
}

@ARTICLE{2024ApJ...973....8H,
       author = {{Hsiao}, Tiger Yu-Yang and {Abdurro'uf} and {Coe}, Dan and {Larson}, Rebecca L. and {Jung}, Intae and {Mingozzi}, Matilde and {Dayal}, Pratika and {Kumari}, Nimisha and {Kokorev}, Vasily and {Vikaeus}, Anton and {Brammer}, Gabriel and {Furtak}, Lukas J. and {Adamo}, Angela and {Andrade-Santos}, Felipe and {Antwi-Danso}, Jacqueline and {Brada{\v{c}}}, Maru{\v{s}}a and {Bradley}, Larry D. and {Broadhurst}, Tom and {Carnall}, Adam C. and {Conselice}, Christopher J. and {Diego}, Jose M. and {Donahue}, Megan and {Eldridge}, Jan J. and {Fujimoto}, Seiji and {Henry}, Alaina and {Hernandez}, Svea and {Hutchison}, Taylor A. and {James}, Bethan L. and {Norman}, Colin and {Park}, Hyunbae and {Pirzkal}, Norbert and {Postman}, Marc and {Ricotti}, Massimo and {Rigby}, Jane R. and {Vanzella}, Eros and {Welch}, Brian and {Wilkins}, Stephen M. and {Windhorst}, Rogier A. and {Xu}, Xinfeng and {Zackrisson}, Erik and {Zitrin}, Adi},
        title = "{JWST NIRSpec Spectroscopy of the Triply Lensed z = 10.17 Galaxy MACS0647─JD}",
      journal = {\apj},
         year = 2024,
        month = sep,
       volume = {973},
       number = {1},
          eid = {8},
        pages = {8},
          doi = {10.3847/1538-4357/ad5da8},
archivePrefix = {arXiv},
       eprint = {2305.03042},
 primaryClass = {astro-ph.GA},
       adsurl = {https://ui.adsabs.harvard.edu/abs/2024ApJ...973....8H}
}

@ARTICLE{2024Natur.633..318C,
       author = {{Carniani}, Stefano and {Hainline}, Kevin and {D'Eugenio}, Francesco and {Eisenstein}, Daniel J. and {Jakobsen}, Peter and {Witstok}, Joris and {Johnson}, Benjamin D. and {Chevallard}, Jacopo and {Maiolino}, Roberto and {Helton}, Jakob M. and {Willott}, Chris and {Robertson}, Brant and {Alberts}, Stacey and {Arribas}, Santiago and {Baker}, William M. and {Bhatawdekar}, Rachana and {Boyett}, Kristan and {Bunker}, Andrew J. and {Cameron}, Alex J. and {Cargile}, Phillip A. and {Charlot}, St{\'e}phane and {Curti}, Mirko and {Curtis-Lake}, Emma and {Egami}, Eiichi and {Giardino}, Giovanna and {Isaak}, Kate and {Ji}, Zhiyuan and {Jones}, Gareth C. and {Kumari}, Nimisha and {Maseda}, Michael V. and {Parlanti}, Eleonora and {P{\'e}rez-Gonz{\'a}lez}, Pablo G. and {Rawle}, Tim and {Rieke}, George and {Rieke}, Marcia and {Del Pino}, Bruno Rodr{\'\i}guez and {Saxena}, Aayush and {Scholtz}, Jan and {Smit}, Renske and {Sun}, Fengwu and {Tacchella}, Sandro and {{\"U}bler}, Hannah and {Venturi}, Giacomo and {Williams}, Christina C. and {Willmer}, Christopher N.~A.},
        title = "{Spectroscopic confirmation of two luminous galaxies at a redshift of 14}",
      journal = {\nat},
         year = 2024,
        month = sep,
       volume = {633},
       number = {8029},
        pages = {318-322},
          doi = {10.1038/s41586-024-07860-9},
archivePrefix = {arXiv},
       eprint = {2405.18485},
 primaryClass = {astro-ph.GA},
       adsurl = {https://ui.adsabs.harvard.edu/abs/2024Natur.633..318C}
}

@ARTICLE{2024ApJ...970...31R,
       author = {{Robertson}, Brant and {Johnson}, Benjamin D. and {Tacchella}, Sandro and {Eisenstein}, Daniel J. and {Hainline}, Kevin and {Arribas}, Santiago and {Baker}, William M. and {Bunker}, Andrew J. and {Carniani}, Stefano and {Cargile}, Phillip A. and {Carreira}, Courtney and {Charlot}, Stephane and {Chevallard}, Jacopo and {Curti}, Mirko and {Curtis-Lake}, Emma and {D'Eugenio}, Francesco and {Egami}, Eiichi and {Hausen}, Ryan and {Helton}, Jakob M. and {Jakobsen}, Peter and {Ji}, Zhiyuan and {Jones}, Gareth C. and {Maiolino}, Roberto and {Maseda}, Michael V. and {Nelson}, Erica and {P{\'e}rez-Gonz{\'a}lez}, Pablo G. and {Pusk{\'a}s}, D{\'a}vid and {Rieke}, Marcia and {Smit}, Renske and {Sun}, Fengwu and {{\"U}bler}, Hannah and {Whitler}, Lily and {Williams}, Christina C. and {Willmer}, Christopher N.~A. and {Willott}, Chris and {Witstok}, Joris},
        title = "{Earliest Galaxies in the JADES Origins Field: Luminosity Function and Cosmic Star Formation Rate Density 300 Myr after the Big Bang}",
      journal = {\apj},
         year = 2024,
        month = jul,
       volume = {970},
       number = {1},
          eid = {31},
        pages = {31},
          doi = {10.3847/1538-4357/ad463d},
archivePrefix = {arXiv},
       eprint = {2312.10033},
 primaryClass = {astro-ph.GA},
       adsurl = {https://ui.adsabs.harvard.edu/abs/2024ApJ...970...31R}
}

@ARTICLE{2024ApJ...969L...2F,
       author = {{Finkelstein}, Steven L. and {Leung}, Gene C.~K. and {Bagley}, Micaela B. and {Dickinson}, Mark and {Ferguson}, Henry C. and {Papovich}, Casey and {Akins}, Hollis B. and {Arrabal Haro}, Pablo and {Dav{\'e}}, Romeel and {Dekel}, Avishai and {Kartaltepe}, Jeyhan S. and {Kocevski}, Dale D. and {Koekemoer}, Anton M. and {Pirzkal}, Nor and {Somerville}, Rachel S. and {Yung}, L.~Y. Aaron and {Amor{\'\i}n}, Ricardo O. and {Backhaus}, Bren E. and {Behroozi}, Peter and {Bisigello}, Laura and {Bromm}, Volker and {Casey}, Caitlin M. and {Ch{\'a}vez Ortiz}, {\'O}scar A. and {Cheng}, Yingjie and {Chworowsky}, Katherine and {Cleri}, Nikko J. and {Cooper}, M.~C. and {Davis}, Kelcey and {de la Vega}, Alexander and {Elbaz}, David and {Franco}, Maximilien and {Fontana}, Adriano and {Fujimoto}, Seiji and {Giavalisco}, Mauro and {Grogin}, Norman A. and {Holwerda}, Benne W. and {Huertas-Company}, Marc and {Hirschmann}, Michaela and {Iyer}, Kartheik G. and {Jogee}, Shardha and {Jung}, Intae and {Larson}, Rebecca L. and {Lucas}, Ray A. and {Mobasher}, Bahram and {Morales}, Alexa M. and {Morley}, Caroline V. and {Mukherjee}, Sagnick and {P{\'e}rez-Gonz{\'a}lez}, Pablo G. and {Ravindranath}, Swara and {Rodighiero}, Giulia and {Rowland}, Melanie J. and {Tacchella}, Sandro and {Taylor}, Anthony J. and {Trump}, Jonathan R. and {Wilkins}, Stephen M.},
        title = "{The Complete CEERS Early Universe Galaxy Sample: A Surprisingly Slow Evolution of the Space Density of Bright Galaxies at z {\ensuremath{\sim}} 8.5─14.5}",
      journal = {\apjl},
         year = 2024,
        month = jul,
       volume = {969},
       number = {1},
          eid = {L2},
        pages = {L2},
          doi = {10.3847/2041-8213/ad4495},
archivePrefix = {arXiv},
       eprint = {2311.04279},
 primaryClass = {astro-ph.GA},
       adsurl = {https://ui.adsabs.harvard.edu/abs/2024ApJ...969L...2F}
}

@ARTICLE{2024ApJ...967L..41M,
       author = {{Marszewski}, Andrew and {Sun}, Guochao and {Faucher-Gigu{\`e}re}, Claude-Andr{\'e} and {Hayward}, Christopher C. and {Feldmann}, Robert},
        title = "{The High-Redshift Gas-Phase Mass─Metallicity Relation in FIRE-2}",
      journal = {\apjl},
         year = 2024,
        month = jun,
       volume = {967},
       number = {2},
          eid = {L41},
        pages = {L41},
          doi = {10.3847/2041-8213/ad4cee},
archivePrefix = {arXiv},
       eprint = {2403.08853},
 primaryClass = {astro-ph.GA},
       adsurl = {https://ui.adsabs.harvard.edu/abs/2024ApJ...967L..41M}
}

@ARTICLE{2024ApJ...968..125R,
       author = {{Roca-F{\`a}brega}, Santi and {Kim}, Ji-Hoon and {Primack}, Joel R. and {Jung}, Minyong and {Genina}, Anna and {Hausammann}, Loic and {Kim}, Hyeonyong and {Lupi}, Alessandro and {Nagamine}, Kentaro and {Powell}, Johnny W. and {Revaz}, Yves and {Shimizu}, Ikkoh and {Strawn}, Clayton and {Vel{\'a}zquez}, H{\'e}ctor and {Abel}, Tom and {Ceverino}, Daniel and {Dong}, Bili and {Quinn}, Thomas R. and {Shin}, Eun-Jin and {Segovia-Otero}, Alvaro and {Agertz}, Oscar and {Barrow}, Kirk S.~S. and {Cadiou}, Corentin and {Dekel}, Avishai and {Hummels}, Cameron and {Oh}, Boon Kiat and {Teyssier}, Romain and {AGORA Collaboration}},
        title = "{The AGORA High-resolution Galaxy Simulations Comparison Project. IV. Halo and Galaxy Mass Assembly in a Cosmological Zoom-in Simulation at z {\ensuremath{\leq}} 2}",
      journal = {\apj},
         year = 2024,
        month = jun,
       volume = {968},
       number = {2},
          eid = {125},
        pages = {125},
          doi = {10.3847/1538-4357/ad43de},
archivePrefix = {arXiv},
       eprint = {2402.06202},
 primaryClass = {astro-ph.CO},
       adsurl = {https://ui.adsabs.harvard.edu/abs/2024ApJ...968..125R}
}

@ARTICLE{2024MNRAS.529.3563T,
       author = {{Trinca}, Alessandro and {Schneider}, Raffaella and {Valiante}, Rosa and {Graziani}, Luca and {Ferrotti}, Arianna and {Omukai}, Kazuyuki and {Chon}, Sunmyon},
        title = "{Exploring the nature of UV-bright z {\ensuremath{\gtrsim}} 10 galaxies detected by JWST: star formation, black hole accretion, or a non-universal IMF?}",
      journal = {\mnras},
         year = 2024,
        month = apr,
       volume = {529},
       number = {4},
        pages = {3563-3581},
          doi = {10.1093/mnras/stae651},
archivePrefix = {arXiv},
       eprint = {2305.04944},
 primaryClass = {astro-ph.GA},
       adsurl = {https://ui.adsabs.harvard.edu/abs/2024MNRAS.529.3563T}
}

@ARTICLE{2024ApJ...964..123J,
       author = {{Jung}, Minyong and {Roca-F{\`a}brega}, Santi and {Kim}, Ji-Hoon and {Genina}, Anna and {Hausammann}, Loic and {Kim}, Hyeonyong and {Lupi}, Alessandro and {Nagamine}, Kentaro and {Powell}, Johnny W. and {Revaz}, Yves and {Shimizu}, Ikkoh and {Vel{\'a}zquez}, H{\'e}ctor and {Ceverino}, Daniel and {Primack}, Joel R. and {Quinn}, Thomas R. and {Strawn}, Clayton and {Abel}, Tom and {Dekel}, Avishai and {Dong}, Bili and {Oh}, Boon Kiat and {Teyssier}, Romain and {AGORA Collaboration}},
        title = "{The AGORA High-resolution Galaxy Simulations Comparison Project. V. Satellite Galaxy Populations in a Cosmological Zoom-in Simulation of a Milky Way─Mass Halo}",
      journal = {\apj},
         year = 2024,
        month = apr,
       volume = {964},
       number = {2},
          eid = {123},
        pages = {123},
          doi = {10.3847/1538-4357/ad245b},
archivePrefix = {arXiv},
       eprint = {2402.05392},
 primaryClass = {astro-ph.GA},
       adsurl = {https://ui.adsabs.harvard.edu/abs/2024ApJ...964..123J}
}

@ARTICLE{2024ApJ...962...29S,
       author = {{Strawn}, Clayton and {Roca-F{\`a}brega}, Santi and {Primack}, Joel R. and {Kim}, Ji-Hoon and {Genina}, Anna and {Hausammann}, Loic and {Kim}, Hyeonyong and {Lupi}, Alessandro and {Nagamine}, Kentaro and {Powell}, Johnny W. and {Revaz}, Yves and {Shimizu}, Ikkoh and {Vel{\'a}zquez}, H{\'e}ctor and {Abel}, Tom and {Ceverino}, Daniel and {Dong}, Bili and {Jung}, Minyong and {Quinn}, Thomas R. and {Shin}, Eun-Jin and {Barrow}, Kirk S.~S. and {Dekel}, Avishai and {Oh}, Boon Kiat and {Mandelker}, Nir and {Teyssier}, Romain and {Hummels}, Cameron and {Maji}, Soumily and {Man}, Antonio and {Mayerhofer}, Paul and {The Agora Collaboration}},
        title = "{The AGORA High-resolution Galaxy Simulations Comparison Project. VI. Similarities and Differences in the Circumgalactic Medium}",
      journal = {\apj},
         year = 2024,
        month = feb,
       volume = {962},
       number = {1},
          eid = {29},
        pages = {29},
          doi = {10.3847/1538-4357/ad12cb},
archivePrefix = {arXiv},
       eprint = {2402.05246},
 primaryClass = {astro-ph.GA},
       adsurl = {https://ui.adsabs.harvard.edu/abs/2024ApJ...962...29S}
}

@ARTICLE{2024MNRAS.527.5929Y,
       author = {{Yung}, L.~Y. Aaron and {Somerville}, Rachel S. and {Finkelstein}, Steven L. and {Wilkins}, Stephen M. and {Gardner}, Jonathan P.},
        title = "{Are the ultra-high-redshift galaxies at z > 10 surprising in the context of standard galaxy formation models?}",
      journal = {\mnras},
         year = 2024,
        month = jan,
       volume = {527},
       number = {3},
        pages = {5929-5948},
          doi = {10.1093/mnras/stad3484},
archivePrefix = {arXiv},
       eprint = {2304.04348},
 primaryClass = {astro-ph.GA},
       adsurl = {https://ui.adsabs.harvard.edu/abs/2024MNRAS.527.5929Y}
}

@ARTICLE{2024A&A...681A..64K,
       author = {{Konstantopoulou}, Christina and {De Cia}, Annalisa and {Ledoux}, C{\'e}dric and {Krogager}, Jens-Kristian and {Mattsson}, Lars and {Watson}, Darach and {Heintz}, Kasper E. and {P{\'e}roux}, C{\'e}line and {Noterdaeme}, Pasquier and {Andersen}, Anja C. and {Fynbo}, Johan P.~U. and {Jermann}, Iris and {Ramburuth-Hurt}, Tanita},
        title = "{Dust depletion of metals from local to distant galaxies. II. Cosmic dust-to-metal ratio and dust composition}",
      journal = {\aap},
         year = 2024,
        month = jan,
       volume = {681},
          eid = {A64},
        pages = {A64},
          doi = {10.1051/0004-6361/202347171},
archivePrefix = {arXiv},
       eprint = {2310.07709},
 primaryClass = {astro-ph.GA},
       adsurl = {https://ui.adsabs.harvard.edu/abs/2024A&A...681A..64K}
}

@ARTICLE{2024ApJ...960...56H,
       author = {{Harikane}, Yuichi and {Nakajima}, Kimihiko and {Ouchi}, Masami and {Umeda}, Hiroya and {Isobe}, Yuki and {Ono}, Yoshiaki and {Xu}, Yi and {Zhang}, Yechi},
        title = "{Pure Spectroscopic Constraints on UV Luminosity Functions and Cosmic Star Formation History from 25 Galaxies at z $_{spec}$ = 8.61-13.20 Confirmed with JWST/NIRSpec}",
      journal = {\apj},
         year = 2024,
        month = jan,
       volume = {960},
       number = {1},
          eid = {56},
        pages = {56},
          doi = {10.3847/1538-4357/ad0b7e},
archivePrefix = {arXiv},
       eprint = {2304.06658},
 primaryClass = {astro-ph.GA},
       adsurl = {https://ui.adsabs.harvard.edu/abs/2024ApJ...960...56H}
}

@ARTICLE{2023MNRAS.526.4801T,
       author = {{Tsuna}, Daichi and {Nakazato}, Yurina and {Hartwig}, Tilman},
        title = "{A photon burst clears the earliest dusty galaxies: modelling dust in high-redshift galaxies from ALMA to JWST}",
      journal = {\mnras},
         year = 2023,
        month = dec,
       volume = {526},
       number = {4},
        pages = {4801-4813},
          doi = {10.1093/mnras/stad3043},
archivePrefix = {arXiv},
       eprint = {2309.02415},
 primaryClass = {astro-ph.GA},
       adsurl = {https://ui.adsabs.harvard.edu/abs/2023MNRAS.526.4801T}
}

@ARTICLE{2023MNRAS.526.3871K,
       author = {{Kapoor}, Anand Utsav and {Baes}, Maarten and {van der Wel}, Arjen and {Gebek}, Andrea and {Camps}, Peter and {Nersesian}, Angelos and {Meidt}, Sharon E. and {Smith}, Aaron and {Vicens}, Sebastien and {D'Eugenio}, Francesco and {Martorano}, Marco and {Barrientos}, Daniela and {Sartorio}, Nina Sanches},
        title = "{TODDLERS: a new UV-mm emission library for star-forming regions - I. Integration with SKIRT and public release}",
      journal = {\mnras},
         year = 2023,
        month = dec,
       volume = {526},
       number = {3},
        pages = {3871-3901},
          doi = {10.1093/mnras/stad2977},
archivePrefix = {arXiv},
       eprint = {2310.00388},
 primaryClass = {astro-ph.GA},
       adsurl = {https://ui.adsabs.harvard.edu/abs/2023MNRAS.526.3871K}
}

@ARTICLE{2023MNRAS.526.2665S,
       author = {{Sun}, Guochao and {Faucher-Gigu{\`e}re}, Claude-Andr{\'e} and {Hayward}, Christopher C. and {Shen}, Xuejian},
        title = "{Seen and unseen: bursty star formation and its implications for observations of high-redshift galaxies with JWST}",
      journal = {\mnras},
         year = 2023,
        month = dec,
       volume = {526},
       number = {2},
        pages = {2665-2672},
          doi = {10.1093/mnras/stad2902},
archivePrefix = {arXiv},
       eprint = {2305.02713},
 primaryClass = {astro-ph.GA},
       adsurl = {https://ui.adsabs.harvard.edu/abs/2023MNRAS.526.2665S}
}

@ARTICLE{2023MNRAS.525.4976M,
       author = {{Mushtaq}, Muzammil and {Ceverino}, Daniel and {Klessen}, Ralf S. and {Reissl}, Stefan and {Puttasiddappa}, Prajwal Hassan},
        title = "{Dust attenuation in galaxies at cosmic dawn from the FirstLight simulations}",
      journal = {\mnras},
         year = 2023,
        month = nov,
       volume = {525},
       number = {4},
        pages = {4976-4984},
          doi = {10.1093/mnras/stad2602},
archivePrefix = {arXiv},
       eprint = {2304.10150},
 primaryClass = {astro-ph.GA},
       adsurl = {https://ui.adsabs.harvard.edu/abs/2023MNRAS.525.4976M}
}

@ARTICLE{2023ApJ...957L..34W,
       author = {{Wang}, Bingjie and {Fujimoto}, Seiji and {Labb{\'e}}, Ivo and {Furtak}, Lukas J. and {Miller}, Tim B. and {Setton}, David J. and {Zitrin}, Adi and {Atek}, Hakim and {Bezanson}, Rachel and {Brammer}, Gabriel and {Leja}, Joel and {Oesch}, Pascal A. and {Price}, Sedona H. and {Chemerynska}, Iryna and {Cutler}, Sam E. and {Dayal}, Pratika and {van Dokkum}, Pieter and {Goulding}, Andy D. and {Greene}, Jenny E. and {Fudamoto}, Y. and {Khullar}, Gourav and {Kokorev}, Vasily and {Marchesini}, Danilo and {Pan}, Richard and {Weaver}, John R. and {Whitaker}, Katherine E. and {Williams}, Christina C.},
        title = "{UNCOVER: Illuminating the Early Universe-JWST/NIRSpec Confirmation of z > 12 Galaxies}",
      journal = {\apjl},
         year = 2023,
        month = nov,
       volume = {957},
       number = {2},
          eid = {L34},
        pages = {L34},
          doi = {10.3847/2041-8213/acfe07},
archivePrefix = {arXiv},
       eprint = {2308.03745},
 primaryClass = {astro-ph.GA},
       adsurl = {https://ui.adsabs.harvard.edu/abs/2023ApJ...957L..34W}
}

@ARTICLE{2023A&A...679A..91H,
       author = {{Heintz}, K.~E. and {De Cia}, A. and {Th{\"o}ne}, C.~C. and {Krogager}, J.-K. and {Yates}, R.~M. and {Vejlgaard}, S. and {Konstantopoulou}, C. and {Fynbo}, J.~P.~U. and {Watson}, D. and {Narayanan}, D. and {Wilson}, S.~N. and {Arabsalmani}, M. and {Campana}, S. and {D'Elia}, V. and {De Pasquale}, M. and {Hartmann}, D.~H. and {Izzo}, L. and {Jakobsson}, P. and {Kouveliotou}, C. and {Levan}, A. and {Li}, Q. and {Malesani}, D.~B. and {Melandri}, A. and {Milvang-Jensen}, B. and {M{\o}ller}, P. and {Palazzi}, E. and {Palmerio}, J. and {Petitjean}, P. and {Pugliese}, G. and {Rossi}, A. and {Saccardi}, A. and {Salvaterra}, R. and {Savaglio}, S. and {Schady}, P. and {Stratta}, G. and {Tanvir}, N.~R. and {de Ugarte Postigo}, A. and {Vergani}, S.~D. and {Wiersema}, K. and {Wijers}, R.~A.~M.~J. and {Zafar}, T.},
        title = "{The cosmic buildup of dust and metals. Accurate abundances from GRB-selected star-forming galaxies at 1.7 < z < 6.3}",
      journal = {\aap},
         year = 2023,
        month = nov,
       volume = {679},
          eid = {A91},
        pages = {A91},
          doi = {10.1051/0004-6361/202347418},
archivePrefix = {arXiv},
       eprint = {2308.14812},
 primaryClass = {astro-ph.GA},
       adsurl = {https://ui.adsabs.harvard.edu/abs/2023A&A...679A..91H}
}

@ARTICLE{2023Natur.622..707A,
       author = {{Arrabal Haro}, Pablo and {Dickinson}, Mark and {Finkelstein}, Steven L. and {Kartaltepe}, Jeyhan S. and {Donnan}, Callum T. and {Burgarella}, Denis and {Carnall}, Adam C. and {Cullen}, Fergus and {Dunlop}, James S. and {Fern{\'a}ndez}, Vital and {Fujimoto}, Seiji and {Jung}, Intae and {Krips}, Melanie and {Larson}, Rebecca L. and {Papovich}, Casey and {P{\'e}rez-Gonz{\'a}lez}, Pablo G. and {Amor{\'\i}n}, Ricardo O. and {Bagley}, Micaela B. and {Buat}, V{\'e}ronique and {Casey}, Caitlin M. and {Chworowsky}, Katherine and {Cohen}, Seth H. and {Ferguson}, Henry C. and {Giavalisco}, Mauro and {Huertas-Company}, Marc and {Hutchison}, Taylor A. and {Kocevski}, Dale D. and {Koekemoer}, Anton M. and {Lucas}, Ray A. and {McLeod}, Derek J. and {McLure}, Ross J. and {Pirzkal}, Norbert and {Seill{\'e}}, Lise-Marie and {Trump}, Jonathan R. and {Weiner}, Benjamin J. and {Wilkins}, Stephen M. and {Zavala}, Jorge A.},
        title = "{Confirmation and refutation of very luminous galaxies in the early Universe}",
      journal = {\nat},
         year = 2023,
        month = oct,
       volume = {622},
       number = {7984},
        pages = {707-711},
          doi = {10.1038/s41586-023-06521-7},
archivePrefix = {arXiv},
       eprint = {2303.15431},
 primaryClass = {astro-ph.GA},
       adsurl = {https://ui.adsabs.harvard.edu/abs/2023Natur.622..707A}
}

@ARTICLE{2023OJAp....6E..47M,
       author = {{McCaffrey}, Joe and {Hardin}, Samantha and {Wise}, John H. and {Regan}, John A.},
        title = "{No Tension: JWST Galaxies at z > 10 Consistent with Cosmological Simulations}",
      journal = {The Open Journal of Astrophysics},
         year = 2023,
        month = sep,
       volume = {6},
          eid = {47},
        pages = {47},
          doi = {10.21105/astro.2304.13755},
archivePrefix = {arXiv},
       eprint = {2304.13755},
 primaryClass = {astro-ph.GA},
       adsurl = {https://ui.adsabs.harvard.edu/abs/2023OJAp....6E..47M}
}

@ARTICLE{2023A&A...677A..88B,
       author = {{Bunker}, Andrew J. and {Saxena}, Aayush and {Cameron}, Alex J. and {Willott}, Chris J. and {Curtis-Lake}, Emma and {Jakobsen}, Peter and {Carniani}, Stefano and {Smit}, Renske and {Maiolino}, Roberto and {Witstok}, Joris and {Curti}, Mirko and {D'Eugenio}, Francesco and {Jones}, Gareth C. and {Ferruit}, Pierre and {Arribas}, Santiago and {Charlot}, Stephane and {Chevallard}, Jacopo and {Giardino}, Giovanna and {de Graaff}, Anna and {Looser}, Tobias J. and {L{\"u}tzgendorf}, Nora and {Maseda}, Michael V. and {Rawle}, Tim and {Rix}, Hans-Walter and {Del Pino}, Bruno Rodr{\'\i}guez and {Alberts}, Stacey and {Egami}, Eiichi and {Eisenstein}, Daniel J. and {Endsley}, Ryan and {Hainline}, Kevin and {Hausen}, Ryan and {Johnson}, Benjamin D. and {Rieke}, George and {Rieke}, Marcia and {Robertson}, Brant E. and {Shivaei}, Irene and {Stark}, Daniel P. and {Sun}, Fengwu and {Tacchella}, Sandro and {Tang}, Mengtao and {Williams}, Christina C. and {Willmer}, Christopher N.~A. and {Baker}, William M. and {Baum}, Stefi and {Bhatawdekar}, Rachana and {Bowler}, Rebecca and {Boyett}, Kristan and {Chen}, Zuyi and {Circosta}, Chiara and {Helton}, Jakob M. and {Ji}, Zhiyuan and {Kumari}, Nimisha and {Lyu}, Jianwei and {Nelson}, Erica and {Parlanti}, Eleonora and {Perna}, Michele and {Sandles}, Lester and {Scholtz}, Jan and {Suess}, Katherine A. and {Topping}, Michael W. and {{\"U}bler}, Hannah and {Wallace}, Imaan E.~B. and {Whitler}, Lily},
        title = "{JADES NIRSpec Spectroscopy of GN-z11: Lyman-{\ensuremath{\alpha}} emission and possible enhanced nitrogen abundance in a z = 10.60 luminous galaxy}",
      journal = {\aap},
         year = 2023,
        month = sep,
       volume = {677},
          eid = {A88},
        pages = {A88},
          doi = {10.1051/0004-6361/202346159},
archivePrefix = {arXiv},
       eprint = {2302.07256},
 primaryClass = {astro-ph.GA},
       adsurl = {https://ui.adsabs.harvard.edu/abs/2023A&A...677A..88B}
}

@ARTICLE{2023ApJ...954L..46L,
       author = {{Leung}, Gene C.~K. and {Bagley}, Micaela B. and {Finkelstein}, Steven L. and {Ferguson}, Henry C. and {Koekemoer}, Anton M. and {P{\'e}rez-Gonz{\'a}lez}, Pablo G. and {Morales}, Alexa and {Kocevski}, Dale D. and {Yang}, Guang and {Somerville}, Rachel S. and {Wilkins}, Stephen M. and {Yung}, L.~Y. Aaron and {Fujimoto}, Seiji and {Larson}, Rebecca L. and {Papovich}, Casey and {Pirzkal}, Nor and {Berg}, Danielle A. and {Lotz}, Jennifer M. and {Castellano}, Marco and {Ch{\'a}vez Ortiz}, {\'O}scar A. and {Cheng}, Yingjie and {Dickinson}, Mark and {Giavalisco}, Mauro and {Hathi}, Nimish P. and {Hutchison}, Taylor A. and {Jung}, Intae and {Kartaltepe}, Jeyhan S. and {Natarajan}, Priyamvada and {Rothberg}, Barry},
        title = "{NGDEEP Epoch 1: The Faint End of the Luminosity Function at z   9-12 from Ultradeep JWST Imaging}",
      journal = {\apjl},
         year = 2023,
        month = sep,
       volume = {954},
       number = {2},
          eid = {L46},
        pages = {L46},
          doi = {10.3847/2041-8213/acf365},
archivePrefix = {arXiv},
       eprint = {2306.06244},
 primaryClass = {astro-ph.GA},
       adsurl = {https://ui.adsabs.harvard.edu/abs/2023ApJ...954L..46L}
}

@ARTICLE{2023MNRAS.523.3201D,
       author = {{Dekel}, Avishai and {Sarkar}, Kartick C. and {Birnboim}, Yuval and {Mandelker}, Nir and {Li}, Zhaozhou},
        title = "{Efficient formation of massive galaxies at cosmic dawn by feedback-free starbursts}",
      journal = {\mnras},
         year = 2023,
        month = aug,
       volume = {523},
       number = {3},
        pages = {3201-3218},
          doi = {10.1093/mnras/stad1557},
archivePrefix = {arXiv},
       eprint = {2303.04827},
 primaryClass = {astro-ph.GA},
       adsurl = {https://ui.adsabs.harvard.edu/abs/2023MNRAS.523.3201D}
}

@ARTICLE{2023ApJ...952...74T,
       author = {{Tacchella}, Sandro and {Eisenstein}, Daniel J. and {Hainline}, Kevin and {Johnson}, Benjamin D. and {Baker}, William M. and {Helton}, Jakob M. and {Robertson}, Brant and {Suess}, Katherine A. and {Chen}, Zuyi and {Nelson}, Erica and {Pusk{\'a}s}, D{\'a}vid and {Sun}, Fengwu and {Alberts}, Stacey and {Egami}, Eiichi and {Hausen}, Ryan and {Rieke}, George and {Rieke}, Marcia and {Shivaei}, Irene and {Williams}, Christina C. and {Willmer}, Christopher N.~A. and {Bunker}, Andrew and {Cameron}, Alex J. and {Carniani}, Stefano and {Charlot}, Stephane and {Curti}, Mirko and {Curtis-Lake}, Emma and {Looser}, Tobias J. and {Maiolino}, Roberto and {Maseda}, Michael V. and {Rawle}, Tim and {Rix}, Hans-Walter and {Smit}, Renske and {{\"U}bler}, Hannah and {Willott}, Chris and {Witstok}, Joris and {Baum}, Stefi and {Bhatawdekar}, Rachana and {Boyett}, Kristan and {Danhaive}, A. Lola and {de Graaff}, Anna and {Endsley}, Ryan and {Ji}, Zhiyuan and {Lyu}, Jianwei and {Sandles}, Lester and {Saxena}, Aayush and {Scholtz}, Jan and {Topping}, Michael W. and {Whitler}, Lily},
        title = "{JADES Imaging of GN-z11: Revealing the Morphology and Environment of a Luminous Galaxy 430 Myr after the Big Bang}",
      journal = {\apj},
         year = 2023,
        month = jul,
       volume = {952},
       number = {1},
          eid = {74},
        pages = {74},
          doi = {10.3847/1538-4357/acdbc6},
archivePrefix = {arXiv},
       eprint = {2302.07234},
 primaryClass = {astro-ph.GA},
       adsurl = {https://ui.adsabs.harvard.edu/abs/2023ApJ...952...74T}
}

@ARTICLE{2023MNRAS.523.1036B,
       author = {{Bouwens}, Rychard J. and {Stefanon}, Mauro and {Brammer}, Gabriel and {Oesch}, Pascal A. and {Herard-Demanche}, Thomas and {Illingworth}, Garth D. and {Matthee}, Jorryt and {Naidu}, Rohan P. and {van Dokkum}, Pieter G. and {van Leeuwen}, Ivana F.},
        title = "{Evolution of the UV LF from z   15 to z   8 using new JWST NIRCam medium-band observations over the HUDF/XDF}",
      journal = {\mnras},
         year = 2023,
        month = jul,
       volume = {523},
       number = {1},
        pages = {1036-1055},
          doi = {10.1093/mnras/stad1145},
archivePrefix = {arXiv},
       eprint = {2211.02607},
 primaryClass = {astro-ph.GA},
       adsurl = {https://ui.adsabs.harvard.edu/abs/2023MNRAS.523.1036B}
}

@ARTICLE{2023ApJ...951...72O,
       author = {{Ono}, Yoshiaki and {Harikane}, Yuichi and {Ouchi}, Masami and {Yajima}, Hidenobu and {Abe}, Makito and {Isobe}, Yuki and {Shibuya}, Takatoshi and {Wise}, John H. and {Zhang}, Yechi and {Nakajima}, Kimihiko and {Umeda}, Hiroya},
        title = "{Morphologies of Galaxies at z {\ensuremath{\gtrsim}} 9 Uncovered by JWST/NIRCam Imaging: Cosmic Size Evolution and an Identification of an Extremely Compact Bright Galaxy at z 12}",
      journal = {\apj},
         year = 2023,
        month = jul,
       volume = {951},
       number = {1},
          eid = {72},
        pages = {72},
          doi = {10.3847/1538-4357/acd44a},
archivePrefix = {arXiv},
       eprint = {2208.13582},
 primaryClass = {astro-ph.GA},
       adsurl = {https://ui.adsabs.harvard.edu/abs/2023ApJ...951...72O}
}

@ARTICLE{2023ApJ...950..132H,
       author = {{Hu}, Chia-Yu and {Smith}, Matthew C. and {Teyssier}, Romain and {Bryan}, Greg L. and {Verbeke}, Robbert and {Emerick}, Andrew and {Somerville}, Rachel S. and {Burkhart}, Blakesley and {Li}, Yuan and {Forbes}, John C. and {Starkenburg}, Tjitske},
        title = "{Code Comparison in Galaxy-scale Simulations with Resolved Supernova Feedback: Lagrangian versus Eulerian Methods}",
      journal = {\apj},
         year = 2023,
        month = jun,
       volume = {950},
       number = {2},
          eid = {132},
        pages = {132},
          doi = {10.3847/1538-4357/accf9e},
archivePrefix = {arXiv},
       eprint = {2208.10528},
 primaryClass = {astro-ph.GA},
       adsurl = {https://ui.adsabs.harvard.edu/abs/2023ApJ...950..132H}
}

@ARTICLE{2023NatAs...7..731B,
       author = {{Boylan-Kolchin}, Michael},
        title = "{Stress testing {\ensuremath{\Lambda}}CDM with high-redshift galaxy candidates}",
      journal = {Nature Astronomy},
         year = 2023,
        month = jun,
       volume = {7},
        pages = {731-735},
          doi = {10.1038/s41550-023-01937-7},
archivePrefix = {arXiv},
       eprint = {2208.01611},
 primaryClass = {astro-ph.CO},
       adsurl = {https://ui.adsabs.harvard.edu/abs/2023NatAs...7..731B}
}

@ARTICLE{2023NatAs...7..611R,
       author = {{Robertson}, B.~E. and {Tacchella}, S. and {Johnson}, B.~D. and {Hainline}, K. and {Whitler}, L. and {Eisenstein}, D.~J. and {Endsley}, R. and {Rieke}, M. and {Stark}, D.~P. and {Alberts}, S. and {Dressler}, A. and {Egami}, E. and {Hausen}, R. and {Rieke}, G. and {Shivaei}, I. and {Williams}, C.~C. and {Willmer}, C.~N.~A. and {Arribas}, S. and {Bonaventura}, N. and {Bunker}, A. and {Cameron}, A.~J. and {Carniani}, S. and {Charlot}, S. and {Chevallard}, J. and {Curti}, M. and {Curtis-Lake}, E. and {D'Eugenio}, F. and {Jakobsen}, P. and {Looser}, T.~J. and {L{\"u}tzgendorf}, N. and {Maiolino}, R. and {Maseda}, M.~V. and {Rawle}, T. and {Rix}, H.-W. and {Smit}, R. and {{\"U}bler}, H. and {Willott}, C. and {Witstok}, J. and {Baum}, S. and {Bhatawdekar}, R. and {Boyett}, K. and {Chen}, Z. and {de Graaff}, A. and {Florian}, M. and {Helton}, J.~M. and {Hviding}, R.~E. and {Ji}, Z. and {Kumari}, N. and {Lyu}, J. and {Nelson}, E. and {Sandles}, L. and {Saxena}, A. and {Suess}, K.~A. and {Sun}, F. and {Topping}, M. and {Wallace}, I.~E.~B.},
        title = "{Identification and properties of intense star-forming galaxies at redshifts z > 10}",
      journal = {Nature Astronomy},
         year = 2023,
        month = may,
       volume = {7},
        pages = {611-621},
          doi = {10.1038/s41550-023-01921-1},
archivePrefix = {arXiv},
       eprint = {2212.04480},
 primaryClass = {astro-ph.GA},
       adsurl = {https://ui.adsabs.harvard.edu/abs/2023NatAs...7..611R}
}

@ARTICLE{2023NatAs...7..622C,
       author = {{Curtis-Lake}, Emma and {Carniani}, Stefano and {Cameron}, Alex and {Charlot}, Stephane and {Jakobsen}, Peter and {Maiolino}, Roberto and {Bunker}, Andrew and {Witstok}, Joris and {Smit}, Renske and {Chevallard}, Jacopo and {Willott}, Chris and {Ferruit}, Pierre and {Arribas}, Santiago and {Bonaventura}, Nina and {Curti}, Mirko and {D'Eugenio}, Francesco and {Franx}, Marijn and {Giardino}, Giovanna and {Looser}, Tobias J. and {L{\"u}tzgendorf}, Nora and {Maseda}, Michael V. and {Rawle}, Tim and {Rix}, Hans-Walter and {Rodr{\'\i}guez del Pino}, Bruno and {{\"U}bler}, Hannah and {Sirianni}, Marco and {Dressler}, Alan and {Egami}, Eiichi and {Eisenstein}, Daniel J. and {Endsley}, Ryan and {Hainline}, Kevin and {Hausen}, Ryan and {Johnson}, Benjamin D. and {Rieke}, Marcia and {Robertson}, Brant and {Shivaei}, Irene and {Stark}, Daniel P. and {Tacchella}, Sandro and {Williams}, Christina C. and {Willmer}, Christopher N.~A. and {Bhatawdekar}, Rachana and {Bowler}, Rebecca and {Boyett}, Kristan and {Chen}, Zuyi and {de Graaff}, Anna and {Helton}, Jakob M. and {Hviding}, Raphael E. and {Jones}, Gareth C. and {Kumari}, Nimisha and {Lyu}, Jianwei and {Nelson}, Erica and {Perna}, Michele and {Sandles}, Lester and {Saxena}, Aayush and {Suess}, Katherine A. and {Sun}, Fengwu and {Topping}, Michael W. and {Wallace}, Imaan E.~B. and {Whitler}, Lily},
        title = "{Spectroscopic confirmation of four metal-poor galaxies at z = 10.3-13.2}",
      journal = {Nature Astronomy},
         year = 2023,
        month = may,
       volume = {7},
        pages = {622-632},
          doi = {10.1038/s41550-023-01918-w},
archivePrefix = {arXiv},
       eprint = {2212.04568},
 primaryClass = {astro-ph.GA},
       adsurl = {https://ui.adsabs.harvard.edu/abs/2023NatAs...7..622C}
}

@ARTICLE{2023Natur.616..266L,
       author = {{Labb{\'e}}, Ivo and {van Dokkum}, Pieter and {Nelson}, Erica and {Bezanson}, Rachel and {Suess}, Katherine A. and {Leja}, Joel and {Brammer}, Gabriel and {Whitaker}, Katherine and {Mathews}, Elijah and {Stefanon}, Mauro and {Wang}, Bingjie},
        title = "{A population of red candidate massive galaxies  600 Myr after the Big Bang}",
      journal = {\nat},
         year = 2023,
        month = apr,
       volume = {616},
       number = {7956},
        pages = {266-269},
          doi = {10.1038/s41586-023-05786-2},
archivePrefix = {arXiv},
       eprint = {2207.12446},
 primaryClass = {astro-ph.GA},
       adsurl = {https://ui.adsabs.harvard.edu/abs/2023Natur.616..266L}
}

@ARTICLE{2023ApJ...946L..13F,
       author = {{Finkelstein}, Steven L. and {Bagley}, Micaela B. and {Ferguson}, Henry C. and {Wilkins}, Stephen M. and {Kartaltepe}, Jeyhan S. and {Papovich}, Casey and {Yung}, L.~Y. Aaron and {Arrabal Haro}, Pablo and {Behroozi}, Peter and {Dickinson}, Mark and {Kocevski}, Dale D. and {Koekemoer}, Anton M. and {Larson}, Rebecca L. and {Le Bail}, Aur{\'e}lien and {Morales}, Alexa M. and {P{\'e}rez-Gonz{\'a}lez}, Pablo G. and {Burgarella}, Denis and {Dav{\'e}}, Romeel and {Hirschmann}, Michaela and {Somerville}, Rachel S. and {Wuyts}, Stijn and {Bromm}, Volker and {Casey}, Caitlin M. and {Fontana}, Adriano and {Fujimoto}, Seiji and {Gardner}, Jonathan P. and {Giavalisco}, Mauro and {Grazian}, Andrea and {Grogin}, Norman A. and {Hathi}, Nimish P. and {Hutchison}, Taylor A. and {Jha}, Saurabh W. and {Jogee}, Shardha and {Kewley}, Lisa J. and {Kirkpatrick}, Allison and {Long}, Arianna S. and {Lotz}, Jennifer M. and {Pentericci}, Laura and {Pierel}, Justin D.~R. and {Pirzkal}, Nor and {Ravindranath}, Swara and {Ryan}, Russell E. and {Trump}, Jonathan R. and {Yang}, Guang and {Bhatawdekar}, Rachana and {Bisigello}, Laura and {Buat}, V{\'e}ronique and {Calabr{\`o}}, Antonello and {Castellano}, Marco and {Cleri}, Nikko J. and {Cooper}, M.~C. and {Croton}, Darren and {Daddi}, Emanuele and {Dekel}, Avishai and {Elbaz}, David and {Franco}, Maximilien and {Gawiser}, Eric and {Holwerda}, Benne W. and {Huertas-Company}, Marc and {Jaskot}, Anne E. and {Leung}, Gene C.~K. and {Lucas}, Ray A. and {Mobasher}, Bahram and {Pandya}, Viraj and {Tacchella}, Sandro and {Weiner}, Benjamin J. and {Zavala}, Jorge A.},
        title = "{CEERS Key Paper. I. An Early Look into the First 500 Myr of Galaxy Formation with JWST}",
      journal = {\apjl},
         year = 2023,
        month = mar,
       volume = {946},
       number = {1},
          eid = {L13},
        pages = {L13},
          doi = {10.3847/2041-8213/acade4},
archivePrefix = {arXiv},
       eprint = {2211.05792},
 primaryClass = {astro-ph.GA},
       adsurl = {https://ui.adsabs.harvard.edu/abs/2023ApJ...946L..13F}
}

@ARTICLE{2023MNRAS.518.6011D,
       author = {{Donnan}, C.~T. and {McLeod}, D.~J. and {Dunlop}, J.~S. and {McLure}, R.~J. and {Carnall}, A.~C. and {Begley}, R. and {Cullen}, F. and {Hamadouche}, M.~L. and {Bowler}, R.~A.~A. and {Magee}, D. and {McCracken}, H.~J. and {Milvang-Jensen}, B. and {Moneti}, A. and {Targett}, T.},
        title = "{The evolution of the galaxy UV luminosity function at redshifts z ≃ 8 - 15 from deep JWST and ground-based near-infrared imaging}",
      journal = {\mnras},
         year = 2023,
        month = feb,
       volume = {518},
       number = {4},
        pages = {6011-6040},
          doi = {10.1093/mnras/stac3472},
archivePrefix = {arXiv},
       eprint = {2207.12356},
 primaryClass = {astro-ph.GA},
       adsurl = {https://ui.adsabs.harvard.edu/abs/2023MNRAS.518.6011D}
}

@ARTICLE{2023MNRAS.519.3118W,
       author = {{Wilkins}, Stephen M. and {Vijayan}, Aswin P. and {Lovell}, Christopher C. and {Roper}, William J. and {Irodotou}, Dimitrios and {Caruana}, Joseph and {Seeyave}, Louise T.~C. and {Kuusisto}, Jussi K. and {Thomas}, Peter A. and {Parris}, Shedeur A.~K.},
        title = "{First light and reionization epoch simulations (FLARES) V: the redshift frontier}",
      journal = {\mnras},
         year = 2023,
        month = feb,
       volume = {519},
       number = {2},
        pages = {3118-3128},
          doi = {10.1093/mnras/stac3280},
archivePrefix = {arXiv},
       eprint = {2204.09431},
 primaryClass = {astro-ph.GA},
       adsurl = {https://ui.adsabs.harvard.edu/abs/2023MNRAS.519.3118W}
}

@ARTICLE{2023MNRAS.518.2511L,
       author = {{Lovell}, Christopher C. and {Harrison}, Ian and {Harikane}, Yuichi and {Tacchella}, Sandro and {Wilkins}, Stephen M.},
        title = "{Extreme value statistics of the halo and stellar mass distributions at high redshift: are JWST results in tension with {\ensuremath{\Lambda}}CDM?}",
      journal = {\mnras},
         year = 2023,
        month = jan,
       volume = {518},
       number = {2},
        pages = {2511-2520},
          doi = {10.1093/mnras/stac3224},
archivePrefix = {arXiv},
       eprint = {2208.10479},
 primaryClass = {astro-ph.GA},
       adsurl = {https://ui.adsabs.harvard.edu/abs/2023MNRAS.518.2511L}
}

@ARTICLE{2023MNRAS.518.4755A,
       author = {{Adams}, N.~J. and {Conselice}, C.~J. and {Ferreira}, L. and {Austin}, D. and {Trussler}, J.~A.~A. and {Juod{\v{z}}balis}, I. and {Wilkins}, S.~M. and {Caruana}, J. and {Dayal}, P. and {Verma}, A. and {Vijayan}, A.~P.},
        title = "{Discovery and properties of ultra-high redshift galaxies (9 < z < 12) in the JWST ERO SMACS 0723 Field}",
      journal = {\mnras},
         year = 2023,
        month = jan,
       volume = {518},
       number = {3},
        pages = {4755-4766},
          doi = {10.1093/mnras/stac3347},
archivePrefix = {arXiv},
       eprint = {2207.11217},
 primaryClass = {astro-ph.GA},
       adsurl = {https://ui.adsabs.harvard.edu/abs/2023MNRAS.518.4755A}
}

@ARTICLE{2022ApJ...938L..10I,
       author = {{Inayoshi}, Kohei and {Harikane}, Yuichi and {Inoue}, Akio K. and {Li}, Wenxiu and {Ho}, Luis C.},
        title = "{A Lower Bound of Star Formation Activity in Ultra-high-redshift Galaxies Detected with JWST: Implications for Stellar Populations and Radiation Sources}",
      journal = {\apjl},
         year = 2022,
        month = oct,
       volume = {938},
       number = {2},
          eid = {L10},
        pages = {L10},
          doi = {10.3847/2041-8213/ac9310},
archivePrefix = {arXiv},
       eprint = {2208.06872},
 primaryClass = {astro-ph.GA},
       adsurl = {https://ui.adsabs.harvard.edu/abs/2022ApJ...938L..10I}
}

@ARTICLE{2022ApJ...938L..15C,
       author = {{Castellano}, Marco and {Fontana}, Adriano and {Treu}, Tommaso and {Santini}, Paola and {Merlin}, Emiliano and {Leethochawalit}, Nicha and {Trenti}, Michele and {Vanzella}, Eros and {Mestric}, Uros and {Bonchi}, Andrea and {Belfiori}, Davide and {Nonino}, Mario and {Paris}, Diego and {Polenta}, Gianluca and {Roberts-Borsani}, Guido and {Boyett}, Kristan and {Brada{\v{c}}}, Maru{\v{s}}a and {Calabr{\`o}}, Antonello and {Glazebrook}, Karl and {Grillo}, Claudio and {Mascia}, Sara and {Mason}, Charlotte and {Mercurio}, Amata and {Morishita}, Takahiro and {Nanayakkara}, Themiya and {Pentericci}, Laura and {Rosati}, Piero and {Vulcani}, Benedetta and {Wang}, Xin and {Yang}, Lilan},
        title = "{Early Results from GLASS-JWST. III. Galaxy Candidates at z  9-15}",
      journal = {\apjl},
         year = 2022,
        month = oct,
       volume = {938},
       number = {2},
          eid = {L15},
        pages = {L15},
          doi = {10.3847/2041-8213/ac94d0},
archivePrefix = {arXiv},
       eprint = {2207.09436},
 primaryClass = {astro-ph.GA},
       adsurl = {https://ui.adsabs.harvard.edu/abs/2022ApJ...938L..15C}
}

@ARTICLE{2022ApJS..262....9O,
       author = {{Oku}, Yuri and {Tomida}, Kengo and {Nagamine}, Kentaro and {Shimizu}, Ikkoh and {Cen}, Renyue},
        title = "{Osaka Feedback Model. II. Modeling Supernova Feedback Based on High-resolution Simulations}",
      journal = {\apjs},
         year = 2022,
        month = sep,
       volume = {262},
       number = {1},
          eid = {9},
        pages = {9},
          doi = {10.3847/1538-4365/ac77ff},
archivePrefix = {arXiv},
       eprint = {2201.00970},
 primaryClass = {astro-ph.GA},
       adsurl = {https://ui.adsabs.harvard.edu/abs/2022ApJS..262....9O}
}

@ARTICLE{2022MNRAS.514.4639C,
       author = {{Chon}, Sunmyon and {Ono}, Haruka and {Omukai}, Kazuyuki and {Schneider}, Raffaella},
        title = "{Impact of the cosmic background radiation on the initial mass function of metal-poor stars}",
      journal = {\mnras},
         year = 2022,
        month = aug,
       volume = {514},
       number = {3},
        pages = {4639-4654},
          doi = {10.1093/mnras/stac1549},
archivePrefix = {arXiv},
       eprint = {2205.15328},
 primaryClass = {astro-ph.GA},
       adsurl = {https://ui.adsabs.harvard.edu/abs/2022MNRAS.514.4639C}
}

@ARTICLE{2022ApJ...928...52F,
       author = {{Finkelstein}, Steven L. and {Bagley}, Micaela and {Song}, Mimi and {Larson}, Rebecca and {Papovich}, Casey and {Dickinson}, Mark and {Finkelstein}, Keely D. and {Koekemoer}, Anton M. and {Pirzkal}, Norbert and {Somerville}, Rachel S. and {Yung}, L.~Y. Aaron and {Behroozi}, Peter and {Ferguson}, Harry and {Giavalisco}, Mauro and {Grogin}, Norman and {Hathi}, Nimish and {Hutchison}, Taylor A. and {Jung}, Intae and {Kocevski}, Dale and {Kawinwanichakij}, Lalitwadee and {Rojas-Ruiz}, Sof{\'\i}a and {Ryan}, Russell and {Snyder}, Gregory F. and {Tacchella}, Sandro},
        title = "{A Census of the Bright z = 8.5-11 Universe with the Hubble and Spitzer Space Telescopes in the CANDELS Fields}",
      journal = {\apj},
         year = 2022,
        month = mar,
       volume = {928},
       number = {1},
          eid = {52},
        pages = {52},
          doi = {10.3847/1538-4357/ac3aed},
archivePrefix = {arXiv},
       eprint = {2106.13813},
 primaryClass = {astro-ph.GA},
       adsurl = {https://ui.adsabs.harvard.edu/abs/2022ApJ...928...52F}
}

@ARTICLE{2021MNRAS.506.5512F,
       author = {{Fukushima}, Hajime and {Yajima}, Hidenobu},
        title = "{Radiation hydrodynamics simulations of massive star cluster formation in giant molecular clouds}",
      journal = {\mnras},
         year = 2021,
        month = oct,
       volume = {506},
       number = {4},
        pages = {5512-5539},
          doi = {10.1093/mnras/stab2099},
archivePrefix = {arXiv},
       eprint = {2104.10892},
 primaryClass = {astro-ph.GA},
       adsurl = {https://ui.adsabs.harvard.edu/abs/2021MNRAS.506.5512F}
}

@ARTICLE{2021MNRAS.506.2871S,
       author = {{Springel}, Volker and {Pakmor}, R{\"u}diger and {Zier}, Oliver and {Reinecke}, Martin},
        title = "{Simulating cosmic structure formation with the GADGET-4 code}",
      journal = {\mnras},
         year = 2021,
        month = sep,
       volume = {506},
       number = {2},
        pages = {2871-2949},
          doi = {10.1093/mnras/stab1855},
archivePrefix = {arXiv},
       eprint = {2010.03567},
 primaryClass = {astro-ph.IM},
       adsurl = {https://ui.adsabs.harvard.edu/abs/2021MNRAS.506.2871S}
}

@ARTICLE{2021ApJ...917...64R,
       author = {{Roca-F{\`a}brega}, Santi and {Kim}, Ji-Hoon and {Hausammann}, Loic and {Nagamine}, Kentaro and {Lupi}, Alessandro and {Powell}, Johnny W. and {Shimizu}, Ikkoh and {Ceverino}, Daniel and {Primack}, Joel R. and {Quinn}, Thomas R. and {Revaz}, Yves and {Vel{\'a}zquez}, H{\'e}ctor and {Abel}, Tom and {Buehlmann}, Michael and {Dekel}, Avishai and {Dong}, Bili and {Hahn}, Oliver and {Hummels}, Cameron and {Kim}, Ki-Won and {Smith}, Britton D. and {Strawn}, Clayton and {Teyssier}, Romain and {Turk}, Matthew J. and {AGORA Collaboration}},
        title = "{The AGORA High-resolution Galaxy Simulations Comparison Project. III. Cosmological Zoom-in Simulation of a Milky Way-mass Halo}",
      journal = {\apj},
         year = 2021,
        month = aug,
       volume = {917},
       number = {2},
          eid = {64},
        pages = {64},
          doi = {10.3847/1538-4357/ac088a},
archivePrefix = {arXiv},
       eprint = {2106.09738},
 primaryClass = {astro-ph.GA},
       adsurl = {https://ui.adsabs.harvard.edu/abs/2021ApJ...917...64R}
}

@ARTICLE{2020MNRAS.499.5702B,
       author = {{Behroozi}, Peter and {Conroy}, Charlie and {Wechsler}, Risa H. and {Hearin}, Andrew and {Williams}, Christina C. and {Moster}, Benjamin P. and {Yung}, L.~Y. Aaron and {Somerville}, Rachel S. and {Gottl{\"o}ber}, Stefan and {Yepes}, Gustavo and {Endsley}, Ryan},
        title = "{The Universe at z > 10: predictions for JWST from the UNIVERSEMACHINE DR1}",
      journal = {\mnras},
         year = 2020,
        month = dec,
       volume = {499},
       number = {4},
        pages = {5702-5718},
          doi = {10.1093/mnras/staa3164},
archivePrefix = {arXiv},
       eprint = {2007.04988},
 primaryClass = {astro-ph.GA},
       adsurl = {https://ui.adsabs.harvard.edu/abs/2020MNRAS.499.5702B}
}

@ARTICLE{2020MNRAS.494.1988L,
       author = {{Langan}, Ivanna and {Ceverino}, Daniel and {Finlator}, Kristian},
        title = "{Weak evolution of the mass-metallicity relation at cosmic dawn in the FirstLight simulations}",
      journal = {\mnras},
         year = 2020,
        month = may,
       volume = {494},
       number = {2},
        pages = {1988-1993},
          doi = {10.1093/mnras/staa880},
archivePrefix = {arXiv},
       eprint = {1910.11729},
 primaryClass = {astro-ph.GA},
       adsurl = {https://ui.adsabs.harvard.edu/abs/2020MNRAS.494.1988L}
}

@ARTICLE{2020A&C....3100381C,
       author = {{Camps}, P. and {Baes}, M.},
        title = "{SKIRT 9: Redesigning an advanced dust radiative transfer code to allow kinematics, line transfer and polarization by aligned dust grains}",
      journal = {Astronomy and Computing},
         year = 2020,
        month = apr,
       volume = {31},
          eid = {100381},
        pages = {100381},
          doi = {10.1016/j.ascom.2020.100381},
archivePrefix = {arXiv},
       eprint = {2003.00721},
 primaryClass = {astro-ph.GA},
       adsurl = {https://ui.adsabs.harvard.edu/abs/2020A&C....3100381C}
}

@ARTICLE{2019MNRAS.490.2855Y,
       author = {{Yung}, L.~Y. Aaron and {Somerville}, Rachel S. and {Popping}, Gerg{\"o} and {Finkelstein}, Steven L. and {Ferguson}, Harry C. and {Dav{\'e}}, Romeel},
        title = "{Semi-analytic forecasts for JWST - II. Physical properties and scaling relations for galaxies at z = 4-10}",
      journal = {\mnras},
         year = 2019,
        month = dec,
       volume = {490},
       number = {2},
        pages = {2855-2879},
          doi = {10.1093/mnras/stz2755},
archivePrefix = {arXiv},
       eprint = {1901.05964},
 primaryClass = {astro-ph.GA},
       adsurl = {https://ui.adsabs.harvard.edu/abs/2019MNRAS.490.2855Y}
}

@ARTICLE{2019MNRAS.486.3805B,
       author = {{Bhatawdekar}, Rachana and {Conselice}, Christopher J. and {Margalef-Bentabol}, Berta and {Duncan}, Kenneth},
        title = "{Evolution of the galaxy stellar mass functions and UV luminosity functions at z = 6-9 in the Hubble Frontier Fields}",
      journal = {\mnras},
         year = 2019,
        month = jul,
       volume = {486},
       number = {3},
        pages = {3805-3830},
          doi = {10.1093/mnras/stz866},
archivePrefix = {arXiv},
       eprint = {1807.07580},
 primaryClass = {astro-ph.GA},
       adsurl = {https://ui.adsabs.harvard.edu/abs/2019MNRAS.486.3805B}
}

@ARTICLE{2019MNRAS.484.2632S,
       author = {{Shimizu}, Ikkoh and {Todoroki}, Keita and {Yajima}, Hidenobu and {Nagamine}, Kentaro},
        title = "{Osaka feedback model: isolated disc galaxy simulations}",
      journal = {\mnras},
         year = 2019,
        month = apr,
       volume = {484},
       number = {2},
        pages = {2632-2655},
          doi = {10.1093/mnras/stz098},
archivePrefix = {arXiv},
       eprint = {1901.03815},
 primaryClass = {astro-ph.GA},
       adsurl = {https://ui.adsabs.harvard.edu/abs/2019MNRAS.484.2632S}
}

@ARTICLE{2019MNRAS.484.1366C,
       author = {{Ceverino}, Daniel and {Klessen}, Ralf S. and {Glover}, Simon C.~O.},
        title = "{FirstLight III: rest-frame UV-optical spectral energy distributions of simulated galaxies at cosmic dawn}",
      journal = {\mnras},
         year = 2019,
        month = mar,
       volume = {484},
       number = {1},
        pages = {1366-1377},
          doi = {10.1093/mnras/stz079},
archivePrefix = {arXiv},
       eprint = {1810.09754},
 primaryClass = {astro-ph.GA},
       adsurl = {https://ui.adsabs.harvard.edu/abs/2019MNRAS.484.1366C}
}

@ARTICLE{2019A&A...623A...5D,
       author = {{De Vis}, P. and {Jones}, A. and {Viaene}, S. and {Casasola}, V. and {Clark}, C.~J.~R. and {Baes}, M. and {Bianchi}, S. and {Cassara}, L.~P. and {Davies}, J.~I. and {De Looze}, I. and {Galametz}, M. and {Galliano}, F. and {Lianou}, S. and {Madden}, S. and {Manilla-Robles}, A. and {Mosenkov}, A.~V. and {Nersesian}, A. and {Roychowdhury}, S. and {Xilouris}, E.~M. and {Ysard}, N.},
        title = "{A systematic metallicity study of DustPedia galaxies reveals evolution in the dust-to-metal ratios}",
      journal = {\aap},
         year = 2019,
        month = mar,
       volume = {623},
          eid = {A5},
        pages = {A5},
          doi = {10.1051/0004-6361/201834444},
archivePrefix = {arXiv},
       eprint = {1901.09040},
 primaryClass = {astro-ph.GA},
       adsurl = {https://ui.adsabs.harvard.edu/abs/2019A&A...623A...5D}
}

@ARTICLE{2018MNRAS.479...75S,
       author = {{Stanway}, E.~R. and {Eldridge}, J.~J.},
        title = "{Re-evaluating old stellar populations}",
      journal = {\mnras},
         year = 2018,
        month = sep,
       volume = {479},
       number = {1},
        pages = {75-93},
          doi = {10.1093/mnras/sty1353},
archivePrefix = {arXiv},
       eprint = {1805.08784},
 primaryClass = {astro-ph.GA},
       adsurl = {https://ui.adsabs.harvard.edu/abs/2018MNRAS.479...75S}
}

@ARTICLE{2018MNRAS.479.1180M,
       author = {{Mainali}, Ramesh and {Zitrin}, Adi and {Stark}, Daniel P. and {Ellis}, Richard S. and {Richard}, Johan and {Tang}, Mengtao and {Laporte}, Nicolas and {Oesch}, Pascal and {McGreer}, Ian},
        title = "{Spectroscopic constraints on UV metal line emission at z ≃ 6-9: the nature of Ly{\ensuremath{\alpha}} emitting galaxies in the reionization era}",
      journal = {\mnras},
         year = 2018,
        month = sep,
       volume = {479},
       number = {1},
        pages = {1180-1193},
          doi = {10.1093/mnras/sty1640},
archivePrefix = {arXiv},
       eprint = {1804.00041},
 primaryClass = {astro-ph.GA},
       adsurl = {https://ui.adsabs.harvard.edu/abs/2018MNRAS.479.1180M}
}

@ARTICLE{2018MNRAS.477..552B,
       author = {{Behrens}, C. and {Pallottini}, A. and {Ferrara}, A. and {Gallerani}, S. and {Vallini}, L.},
        title = "{Dusty galaxies in the Epoch of Reionization: simulations}",
      journal = {\mnras},
         year = 2018,
        month = jun,
       volume = {477},
       number = {1},
        pages = {552-565},
          doi = {10.1093/mnras/sty552},
archivePrefix = {arXiv},
       eprint = {1802.07772},
 primaryClass = {astro-ph.GA},
       adsurl = {https://ui.adsabs.harvard.edu/abs/2018MNRAS.477..552B}
}

@ARTICLE{2018MNRAS.477.1578H,
       author = {{Hopkins}, Philip F. and {Wetzel}, Andrew and {Kere{\v{s}}}, Du{\v{s}}an and {Faucher-Gigu{\`e}re}, Claude-Andr{\'e} and {Quataert}, Eliot and {Boylan-Kolchin}, Michael and {Murray}, Norman and {Hayward}, Christopher C. and {El-Badry}, Kareem},
        title = "{How to model supernovae in simulations of star and galaxy formation}",
      journal = {\mnras},
         year = 2018,
        month = jun,
       volume = {477},
       number = {2},
        pages = {1578-1603},
          doi = {10.1093/mnras/sty674},
archivePrefix = {arXiv},
       eprint = {1707.07010},
 primaryClass = {astro-ph.GA},
       adsurl = {https://ui.adsabs.harvard.edu/abs/2018MNRAS.477.1578H}
}

@ARTICLE{2017MNRAS.471.2357W,
       author = {{Wadsley}, James W. and {Keller}, Benjamin W. and {Quinn}, Thomas R.},
        title = "{Gasoline2: a modern smoothed particle hydrodynamics code}",
      journal = {\mnras},
         year = 2017,
        month = oct,
       volume = {471},
       number = {2},
        pages = {2357-2369},
          doi = {10.1093/mnras/stx1643},
archivePrefix = {arXiv},
       eprint = {1707.03824},
 primaryClass = {astro-ph.IM},
       adsurl = {https://ui.adsabs.harvard.edu/abs/2017MNRAS.471.2357W}
}

@ARTICLE{2017MNRAS.470.2791C,
       author = {{Ceverino}, Daniel and {Glover}, Simon C.~O. and {Klessen}, Ralf S.},
        title = "{Introducing the FirstLight project: UV luminosity function and scaling relations of primeval galaxies}",
      journal = {\mnras},
         year = 2017,
        month = sep,
       volume = {470},
       number = {3},
        pages = {2791-2798},
          doi = {10.1093/mnras/stx1386},
archivePrefix = {arXiv},
       eprint = {1703.02913},
 primaryClass = {astro-ph.GA},
       adsurl = {https://ui.adsabs.harvard.edu/abs/2017MNRAS.470.2791C}
}

@ARTICLE{2017MNRAS.466..105A,
       author = {{Aoyama}, Shohei and {Hou}, Kuan-Chou and {Shimizu}, Ikkoh and {Hirashita}, Hiroyuki and {Todoroki}, Keita and {Choi}, Jun-Hwan and {Nagamine}, Kentaro},
        title = "{Galaxy simulation with dust formation and destruction}",
      journal = {\mnras},
         year = 2017,
        month = apr,
       volume = {466},
       number = {1},
        pages = {105-121},
          doi = {10.1093/mnras/stw3061},
archivePrefix = {arXiv},
       eprint = {1609.07547},
 primaryClass = {astro-ph.GA},
       adsurl = {https://ui.adsabs.harvard.edu/abs/2017MNRAS.466..105A}
}

@ARTICLE{2017MNRAS.466...11R,
       author = {{Rosdahl}, Joakim and {Schaye}, Joop and {Dubois}, Yohan and {Kimm}, Taysun and {Teyssier}, Romain},
        title = "{Snap, crackle, pop: sub-grid supernova feedback in AMR simulations of disc galaxies}",
      journal = {\mnras},
         year = 2017,
        month = apr,
       volume = {466},
       number = {1},
        pages = {11-33},
          doi = {10.1093/mnras/stw3034},
archivePrefix = {arXiv},
       eprint = {1609.01296},
 primaryClass = {astro-ph.GA},
       adsurl = {https://ui.adsabs.harvard.edu/abs/2017MNRAS.466...11R}
}

@ARTICLE{2017MNRAS.466.2217S,
       author = {{Smith}, Britton D. and {Bryan}, Greg L. and {Glover}, Simon C.~O. and {Goldbaum}, Nathan J. and {Turk}, Matthew J. and {Regan}, John and {Wise}, John H. and {Schive}, Hsi-Yu and {Abel}, Tom and {Emerick}, Andrew and {O'Shea}, Brian W. and {Anninos}, Peter and {Hummels}, Cameron B. and {Khochfar}, Sadegh},
        title = "{GRACKLE: a chemistry and cooling library for astrophysics}",
      journal = {\mnras},
         year = 2017,
        month = apr,
       volume = {466},
       number = {2},
        pages = {2217-2234},
          doi = {10.1093/mnras/stw3291},
archivePrefix = {arXiv},
       eprint = {1610.09591},
 primaryClass = {astro-ph.CO},
       adsurl = {https://ui.adsabs.harvard.edu/abs/2017MNRAS.466.2217S}
}

@ARTICLE{2017AJ....153...85S,
       author = {{Saitoh}, Takayuki R.},
        title = "{Chemical Evolution Library for Galaxy Formation Simulation}",
      journal = {\aj},
         year = 2017,
        month = feb,
       volume = {153},
       number = {2},
          eid = {85},
        pages = {85},
          doi = {10.3847/1538-3881/153/2/85},
archivePrefix = {arXiv},
       eprint = {1612.02260},
 primaryClass = {astro-ph.IM},
       adsurl = {https://ui.adsabs.harvard.edu/abs/2017AJ....153...85S}
}

@ARTICLE{2017MNRAS.465.1682H,
       author = {{Hayward}, Christopher C. and {Hopkins}, Philip F.},
        title = "{How stellar feedback simultaneously regulates star formation and drives outflows}",
      journal = {\mnras},
         year = 2017,
        month = feb,
       volume = {465},
       number = {2},
        pages = {1682-1698},
          doi = {10.1093/mnras/stw2888},
archivePrefix = {arXiv},
       eprint = {1510.05650},
 primaryClass = {astro-ph.GA},
       adsurl = {https://ui.adsabs.harvard.edu/abs/2017MNRAS.465.1682H}
}

@ARTICLE{2016ApJ...833..202K,
       author = {{Kim}, Ji-hoon and {Agertz}, Oscar and {Teyssier}, Romain and {Butler}, Michael J. and {Ceverino}, Daniel and {Choi}, Jun-Hwan and {Feldmann}, Robert and {Keller}, Ben W. and {Lupi}, Alessandro and {Quinn}, Thomas and {Revaz}, Yves and {Wallace}, Spencer and {Gnedin}, Nickolay Y. and {Leitner}, Samuel N. and {Shen}, Sijing and {Smith}, Britton D. and {Thompson}, Robert and {Turk}, Matthew J. and {Abel}, Tom and {Arraki}, Kenza S. and {Benincasa}, Samantha M. and {Chakrabarti}, Sukanya and {DeGraf}, Colin and {Dekel}, Avishai and {Goldbaum}, Nathan J. and {Hopkins}, Philip F. and {Hummels}, Cameron B. and {Klypin}, Anatoly and {Li}, Hui and {Madau}, Piero and {Mandelker}, Nir and {Mayer}, Lucio and {Nagamine}, Kentaro and {Nickerson}, Sarah and {O'Shea}, Brian W. and {Primack}, Joel R. and {Roca-F{\`a}brega}, Santi and {Semenov}, Vadim and {Shimizu}, Ikkoh and {Simpson}, Christine M. and {Todoroki}, Keita and {Wadsley}, James W. and {Wise}, John H. and {AGORA Collaboration}},
        title = "{The AGORA High-resolution Galaxy Simulations Comparison Project. II. Isolated Disk Test}",
      journal = {\apj},
         year = 2016,
        month = dec,
       volume = {833},
       number = {2},
          eid = {202},
        pages = {202},
          doi = {10.3847/1538-4357/833/2/202},
archivePrefix = {arXiv},
       eprint = {1610.03066},
 primaryClass = {astro-ph.GA},
       adsurl = {https://ui.adsabs.harvard.edu/abs/2016ApJ...833..202K}
}

@ARTICLE{2016ApJ...819..129O,
       author = {{Oesch}, P.~A. and {Brammer}, G. and {van Dokkum}, P.~G. and {Illingworth}, G.~D. and {Bouwens}, R.~J. and {Labb{\'e}}, I. and {Franx}, M. and {Momcheva}, I. and {Ashby}, M.~L.~N. and {Fazio}, G.~G. and {Gonzalez}, V. and {Holden}, B. and {Magee}, D. and {Skelton}, R.~E. and {Smit}, R. and {Spitler}, L.~R. and {Trenti}, M. and {Willner}, S.~P.},
        title = "{A Remarkably Luminous Galaxy at z=11.1 Measured with Hubble Space Telescope Grism Spectroscopy}",
      journal = {\apj},
         year = 2016,
        month = mar,
       volume = {819},
       number = {2},
          eid = {129},
        pages = {129},
          doi = {10.3847/0004-637X/819/2/129},
archivePrefix = {arXiv},
       eprint = {1603.00461},
 primaryClass = {astro-ph.GA},
       adsurl = {https://ui.adsabs.harvard.edu/abs/2016ApJ...819..129O}
}

@ARTICLE{2015MNRAS.452.1502D,
       author = {{Dubois}, Yohan and {Volonteri}, Marta and {Silk}, Joseph and {Devriendt}, Julien and {Slyz}, Adrianne and {Teyssier}, Romain},
        title = "{Black hole evolution - I. Supernova-regulated black hole growth}",
      journal = {\mnras},
         year = 2015,
        month = sep,
       volume = {452},
       number = {2},
        pages = {1502-1518},
          doi = {10.1093/mnras/stv1416},
archivePrefix = {arXiv},
       eprint = {1504.00018},
 primaryClass = {astro-ph.GA},
       adsurl = {https://ui.adsabs.harvard.edu/abs/2015MNRAS.452.1502D}
}

@ARTICLE{2015MNRAS.450...53H,
       author = {{Hopkins}, Philip F.},
        title = "{A new class of accurate, mesh-free hydrodynamic simulation methods}",
      journal = {\mnras},
         year = 2015,
        month = jun,
       volume = {450},
       number = {1},
        pages = {53-110},
          doi = {10.1093/mnras/stv195},
archivePrefix = {arXiv},
       eprint = {1409.7395},
 primaryClass = {astro-ph.CO},
       adsurl = {https://ui.adsabs.harvard.edu/abs/2015MNRAS.450...53H}
}

@ARTICLE{2015ComAC...2....1M,
       author = {{Menon}, Harshitha and {Wesolowski}, Lukasz and {Zheng}, Gengbin and {Jetley}, Pritish and {Kale}, Laxmikant and {Quinn}, Thomas and {Governato}, Fabio},
        title = "{Adaptive techniques for clustered N-body cosmological simulations}",
      journal = {Computational Astrophysics and Cosmology},
         year = 2015,
        month = mar,
       volume = {2},
          eid = {1},
        pages = {1},
          doi = {10.1186/s40668-015-0007-9},
archivePrefix = {arXiv},
       eprint = {1409.1929},
 primaryClass = {astro-ph.IM},
       adsurl = {https://ui.adsabs.harvard.edu/abs/2015ComAC...2....1M}
}

@ARTICLE{2014MNRAS.442.3013K,
       author = {{Keller}, B.~W. and {Wadsley}, J. and {Benincasa}, S.~M. and {Couchman}, H.~M.~P.},
        title = "{A superbubble feedback model for galaxy simulations}",
      journal = {\mnras},
         year = 2014,
        month = aug,
       volume = {442},
       number = {4},
        pages = {3013-3025},
          doi = {10.1093/mnras/stu1058},
archivePrefix = {arXiv},
       eprint = {1405.2625},
 primaryClass = {astro-ph.GA},
       adsurl = {https://ui.adsabs.harvard.edu/abs/2014MNRAS.442.3013K}
}

@ARTICLE{2014ApJS..211...19B,
       author = {{Bryan}, Greg L. and {Norman}, Michael L. and {O'Shea}, Brian W. and {Abel}, Tom and {Wise}, John H. and {Turk}, Matthew J. and {Reynolds}, Daniel R. and {Collins}, David C. and {Wang}, Peng and {Skillman}, Samuel W. and {Smith}, Britton and {Harkness}, Robert P. and {Bordner}, James and {Kim}, Ji-hoon and {Kuhlen}, Michael and {Xu}, Hao and {Goldbaum}, Nathan and {Hummels}, Cameron and {Kritsuk}, Alexei G. and {Tasker}, Elizabeth and {Skory}, Stephen and {Simpson}, Christine M. and {Hahn}, Oliver and {Oishi}, Jeffrey S. and {So}, Geoffrey C. and {Zhao}, Fen and {Cen}, Renyue and {Li}, Yuan and {Enzo Collaboration}},
        title = "{ENZO: An Adaptive Mesh Refinement Code for Astrophysics}",
      journal = {\apjs},
         year = 2014,
        month = apr,
       volume = {211},
       number = {2},
          eid = {19},
        pages = {19},
          doi = {10.1088/0067-0049/211/2/19},
archivePrefix = {arXiv},
       eprint = {1307.2265},
 primaryClass = {astro-ph.IM},
       adsurl = {https://ui.adsabs.harvard.edu/abs/2014ApJS..211...19B}
}

@ARTICLE{2014ApJS..210...14K,
       author = {{Kim}, Ji-hoon and {Abel}, Tom and {Agertz}, Oscar and {Bryan}, Greg L. and {Ceverino}, Daniel and {Christensen}, Charlotte and {Conroy}, Charlie and {Dekel}, Avishai and {Gnedin}, Nickolay Y. and {Goldbaum}, Nathan J. and {Guedes}, Javiera and {Hahn}, Oliver and {Hobbs}, Alexander and {Hopkins}, Philip F. and {Hummels}, Cameron B. and {Iannuzzi}, Francesca and {Keres}, Dusan and {Klypin}, Anatoly and {Kravtsov}, Andrey V. and {Krumholz}, Mark R. and {Kuhlen}, Michael and {Leitner}, Samuel N. and {Madau}, Piero and {Mayer}, Lucio and {Moody}, Christopher E. and {Nagamine}, Kentaro and {Norman}, Michael L. and {Onorbe}, Jose and {O'Shea}, Brian W. and {Pillepich}, Annalisa and {Primack}, Joel R. and {Quinn}, Thomas and {Read}, Justin I. and {Robertson}, Brant E. and {Rocha}, Miguel and {Rudd}, Douglas H. and {Shen}, Sijing and {Smith}, Britton D. and {Szalay}, Alexander S. and {Teyssier}, Romain and {Thompson}, Robert and {Todoroki}, Keita and {Turk}, Matthew J. and {Wadsley}, James W. and {Wise}, John H. and {Zolotov}, Adi and {AGORA Collaboration29},the},
        title = "{The AGORA High-resolution Galaxy Simulations Comparison Project}",
      journal = {\apjs},
         year = 2014,
        month = jan,
       volume = {210},
       number = {1},
          eid = {14},
        pages = {14},
          doi = {10.1088/0067-0049/210/1/14},
archivePrefix = {arXiv},
       eprint = {1308.2669},
 primaryClass = {astro-ph.GA},
       adsurl = {https://ui.adsabs.harvard.edu/abs/2014ApJS..210...14K}
}

@ARTICLE{2013MNRAS.433.1230W,
       author = {{Watson}, William A. and {Iliev}, Ilian T. and {D'Aloisio}, Anson and {Knebe}, Alexander and {Shapiro}, Paul R. and {Yepes}, Gustavo},
        title = "{The halo mass function through the cosmic ages}",
      journal = {\mnras},
         year = 2013,
        month = aug,
       volume = {433},
       number = {2},
        pages = {1230-1245},
          doi = {10.1093/mnras/stt791},
archivePrefix = {arXiv},
       eprint = {1212.0095},
 primaryClass = {astro-ph.CO},
       adsurl = {https://ui.adsabs.harvard.edu/abs/2013MNRAS.433.1230W}
}

@ARTICLE{2012A&A...538A..82R,
       author = {{Revaz}, Y. and {Jablonka}, P.},
        title = "{The dynamical and chemical evolution of dwarf spheroidal galaxies with GEAR}",
      journal = {\aap},
         year = 2012,
        month = feb,
       volume = {538},
          eid = {A82},
        pages = {A82},
          doi = {10.1051/0004-6361/201117402},
archivePrefix = {arXiv},
       eprint = {1109.0989},
 primaryClass = {astro-ph.CO},
       adsurl = {https://ui.adsabs.harvard.edu/abs/2012A&A...538A..82R}
}
\bibliographystyle{aasjournal}

\appendix

\section{Inter-code Convergence Test in the Stellar Mass}\label{sec:convergence}
\begin{figure}
    \centering
    \includegraphics[width=0.6\linewidth]{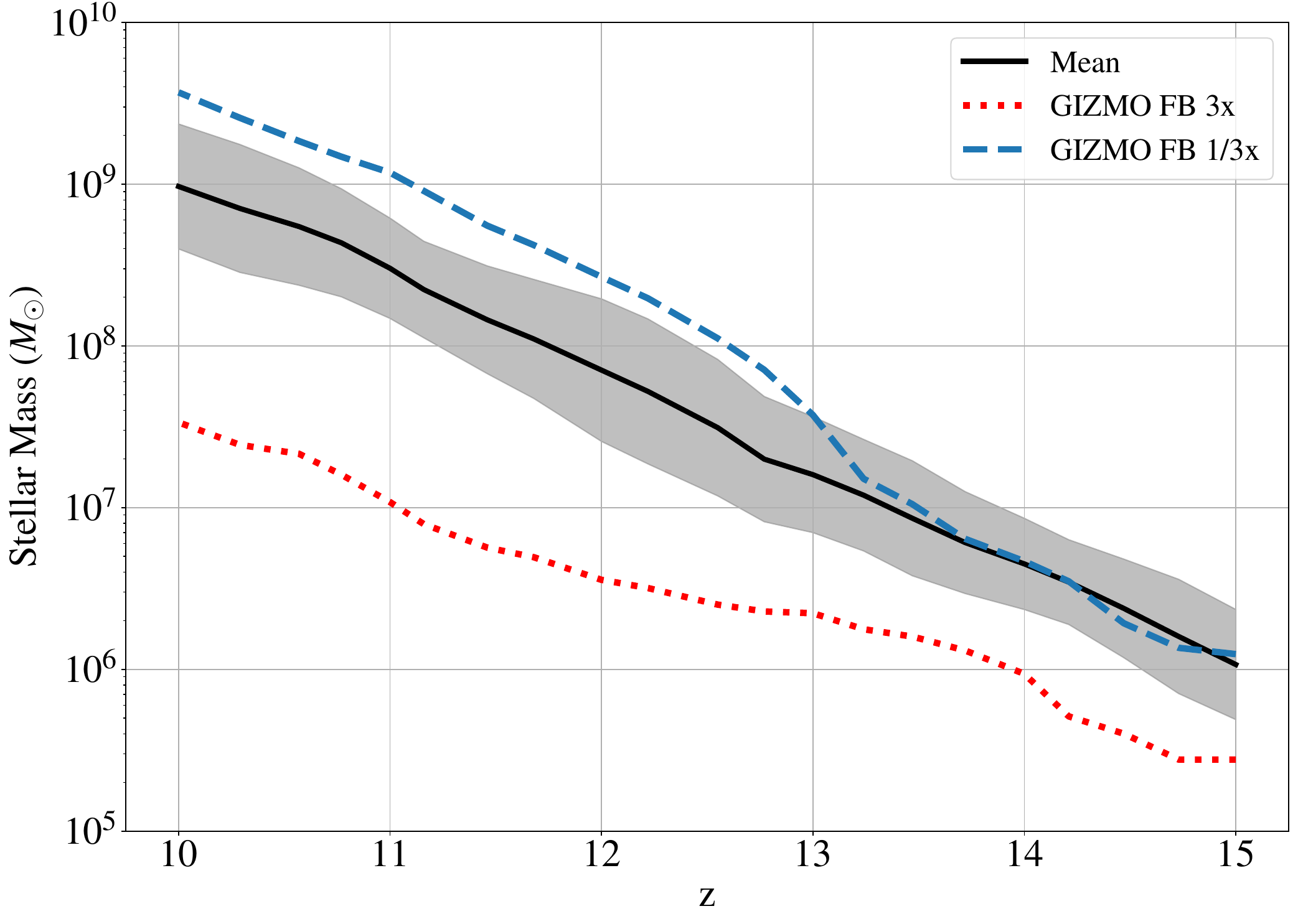}
    \caption{Same as Figure \ref{fig:stellar_mass}, but mean stellar mass evolution for Halo 4, with the black line indicating the average of participating codes and the gray shaded region representing the 1${\sigma}$ scatter. The colored lines show the results from additional feedback convergence tests using \textsc {Gizmo} --- red dotted line and blue dashed line correspond to \textsc{Gizmo} simulations with 3 times stronger and weaker supernova feedback energy, respectively. See Appendix \ref{sec:convergence} for more information.}
    \label{fig:convergence}
\end{figure}

Figure \ref{fig:convergence} shows the same stellar mass history of Halo 4 from Figure 5, but with the additional convergence test results. 
The black line indicates the mean stellar mass across all codes, and the shaded region represents the one standard deviation of participating codes. 
The red dotted line and blue dashed line indicate the evolution of stellar mass using \textsc{Gizmo} with feedback strength varied by a factor of 3. 
Each test line differs from the original simulation by nearly an order of magnitude. 
Although the difference between the minimum and maximum values among codes can reach 1 dex, the inter-code results generally show better convergence than the feedback variation lines, especially considering the substantial diversity in feedback implementations across the codes.

\begin{figure*}
    \centering
    \includegraphics[width=1\linewidth]{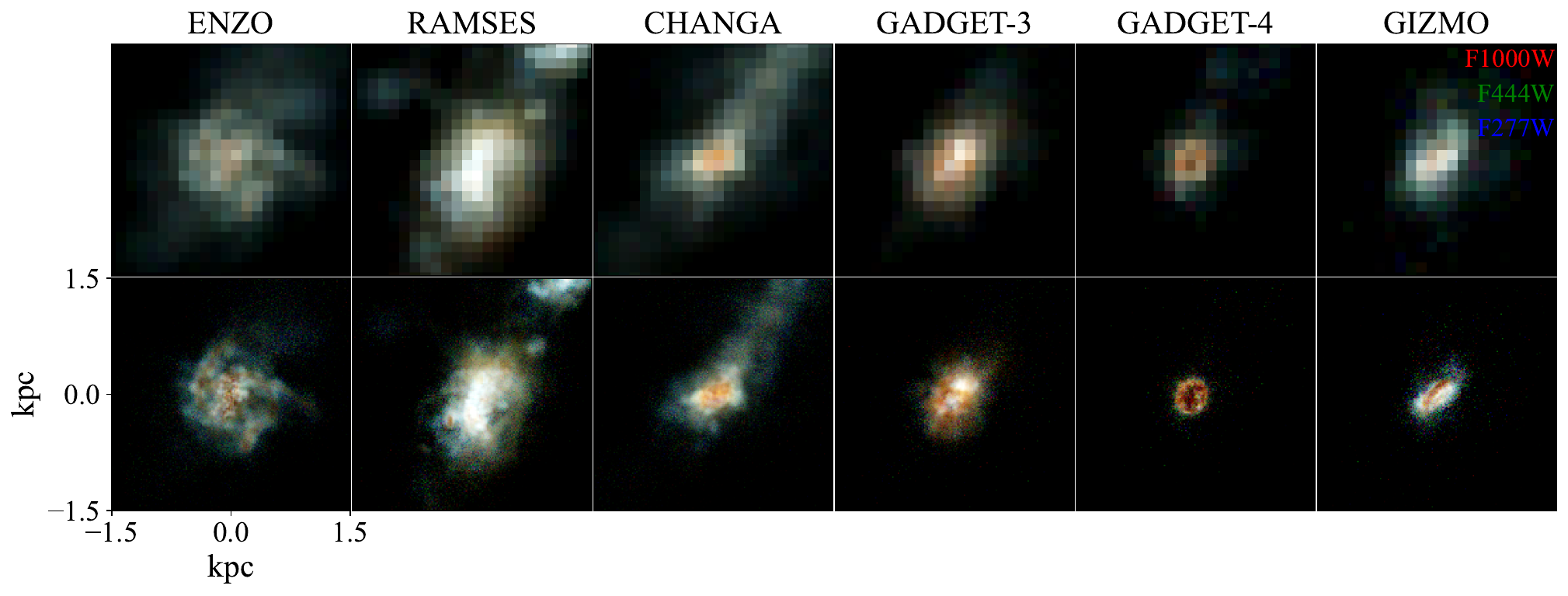}
    \caption{Comparison of mock images of a simulated galaxy at different resolutions and with different RGB composite filters (R: F1000W, G: F444W, B: F277W). The original mock image in the {\it bottom row} (${\rm W}_{\mathrm{pixel}} = 0.0039''$, equivalent to 17 pc at $z = 10$) is degraded to match the resolution of JWST NIRCam in the {\it top row} (${\rm W}_{\mathrm{pixel}} = 0.031''$, equivalent to 132 pc). See Appendix \ref{sec:mock_compare} for more information.}
    \label{fig:mock_compare}
\end{figure*}

\section{Mock images with different resolutions and RGB bands}\label{sec:mock_compare}
In Figure \ref{fig:mock_compare}, we compare mock images generated from SKIRT at two different resolutions. We also apply different RGB composite filters (R: F1000W, G: F444W, B: F277W) to examine the effects of dust on the mock images. The upper row corresponds to the JWST NIRCam resolution, with a pixel width of 0.031". 
The bottom row, reproduced from Figure \ref{fig:mock}, has 8 times higher spatial resolution. 
Although the higher-resolution images appear more compact and show finer structures, the average difference in the resulting SEDs is less than 2 percent, indicating that the resolution effect on integrated photometric properties is minimal. The impact of dust, which appears as dark lanes in Figure \ref{fig:mock}, is also shown more distinctly here as brownish lanes, making the attenuation features more evident.


\begin{figure*}
    \vspace{0mm}
    \centering
    \includegraphics[width = 0.6\linewidth]{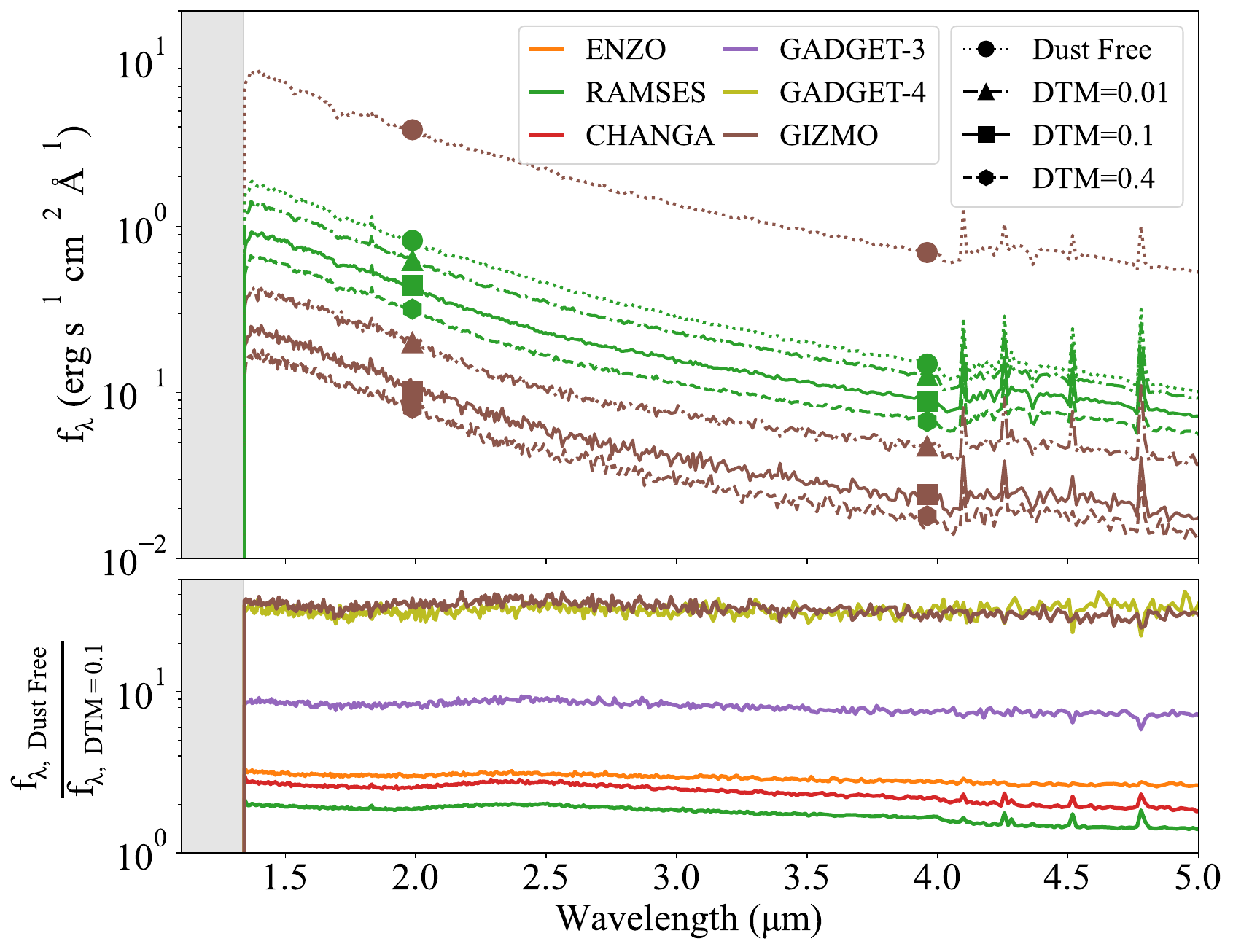}
\caption{\textit{Upper panel}: The simulated galaxies' spectral energy distributions (SEDs) with different dust-to-metal ratios (DTM ratio = 0.01, 0.1, and 0.4) and a dust-free model. Only two representative codes are shown here to illustrate the largest (\textsc{Gizmo}) and smallest (\textsc{Ramses}) deviations. \textit{Bottom panel}: The ratios of the dust-free model to the fiducial case (DTM ratio = 0.1) for all codes. See Appendix \ref{sec:dtm_ratio} for more information.}

    \label{fig:flux_ratio}
\end{figure*}

\section{Flux Differences between Different DTM Ratios}\label{sec:dtm_ratio}
Figure \ref{fig:flux_ratio} shows the SED of simulated galaxies with varying dust-to-metal ratios (DTM ratio = 0.01, 0.1, and 0.4) and a dust-free model. \footnote{SKIRT nebular emission library assumes a shell around star-forming regions, which produces emission lines even in the dust-free model. However, the differences are also observed in the non-nebular emission library.}  The upper panel demonstrates that the presence of dust substantially suppresses the UV continuum, with \textsc{Gizmo} exhibiting the strongest attenuation and \textsc{Ramses} the weakest. Variations in the DTM ratio produce only modest deviations from the fiducial model (DTM ratio = 0.1). The bottom panel presents the flux ratios of the dust-free model relative to the fiducial case, showing that dust attenuation can reduce the UV flux by factors ranging from about 2 to 30, depending on the simulation code.




\end{document}